\documentclass[12pt]{article}

\usepackage{makeidx}         

\usepackage{graphicx}        
\usepackage{multicol}        

\makeindex             
\input amssym

\def\oneone{\rlap 1\mkern4mu{\rm l}}
\def\coeff#1#2{\relax{\textstyle {#1 \over #2}}\displaystyle}
\def\Nb{\overline{N}}
\def\Pb{\overline{P}}
\def\Qb{\overline{Q}}
\def\IR{\Bbb{R}}
\def\ZZ{\Bbb{Z}}
\def\flux{\Pi}
\def\cF{{\cal F}}
\def\cN{{\cal N}}
\def\cQ{{\cal Q}}
\def\cS{{\cal S}}
\def\exbox#1{\bigskip \framebox{\parbox{5.6 in}{{\bf \underline{Exercise}:\ \ }{\it #1}}}\bigskip }
\newcommand{\be}{\begin{equation}}
\newcommand{\ee}{\end{equation}}

\newcommand{\bea}{\begin{eqnarray}}
\newcommand{\eea}{\end{eqnarray}}
\newcommand{\Tr}{{\rm Tr}}

\newcommand{\p}{\partial}

\def\rt{\tilde{r}}
\def\Rt{\widetilde{R}}

\def\psit{\tilde{\psi}}
\def\phit{\tilde{\phi}}

\def\thetat{\tilde{\theta}}
\def\Sigmat{\tilde{\Sigma}}

\def\Qb{\overline{Q}}
\def\tZ{ \tilde{Z}}
 \def\thetah{\hat{\theta}}
\def\phih{\hat{\phi}}

\begin{document}

 \begin{titlepage}

\begin{flushright}
SPHT-T07/008\\
hep-th/0701216
\end{flushright}

\bigskip\bigskip
\bigskip
\centerline{\Large \bf Black Holes, Black Rings and their Microstates
}
\bigskip\bigskip
\centerline{{\bf Iosif Bena$^1$ and Nicholas P. Warner $^2$}}
\medskip
\centerline{$^1$ Service de Physique Th\'eorique, CEA/Saclay}
\centerline{F-91191 Gif-sur-Yvette Cedex, FRANCE }
\medskip
\centerline{$^2$ Department of Physics and Astronomy}
\centerline{University of Southern California}
\centerline{Los Angeles, CA 90089-0484, USA}
\bigskip
\centerline{{\rm iosif.bena@cea.fr,~~ warner@usc.edu} }
\bigskip \bigskip

\begin{abstract}

In this review article we describe some of the recent
progress towards the construction and analysis of three-charge
configurations in string theory and supergravity.  We begin by
describing the Born-Infeld construction of three-charge supertubes
with two dipole charges, and then discuss the general method of
constructing three-charge solutions in five dimensions.  We explain in
detail the use of these methods to construct black rings, black holes,
as well as smooth microstate geometries with black hole and black ring
charges, but with no horizon. We present arguments that many of these
microstate geometries are dual to boundary states that belong to the same sector
of the D1-D5-P CFT as the typical states. We end with an extended
discussion of the implications of this work for the physics of black
holes in string theory.

\end{abstract}

\end{titlepage}

\section{Introduction}
\label{sec:1}

Black holes are very interesting objects, whose physics brings quantum
mechanics and general relativity into sharp contrast.  Perhaps the
best known, and sharpest, example of such contrast is Hawking's
information paradox \cite{hawking}.  This has provided a very valuable
guide and testing ground in formulating a quantum theory of gravity.
Indeed, it is one of the relatively few issues that we know {\it
  must} be explained by a viable theory of quantum gravity.

String theory is a quantum theory of gravity, and has had several
astounding successes in describing properties of black holes. In
particular, Strominger and Vafa have shown \cite{stromvafa} that one
can count microscopic configurations of branes and strings at zero
gravitational coupling, and exactly match their statistical entropy to
the Bekenstein-Hawking entropy of the corresponding black hole at
large effective coupling.

Another way to understand the Strominger-Vafa entropy matching is via
the $AdS$-CFT correspondence\footnote{Historically the AdS-CFT
  correspondence was found later.} \cite{adscft}. One can make a black
hole in string theory by putting together D5 branes and D1 branes and
turning on momentum along the direction of the D1's. If one takes a
near horizon limit of this system, one finds a bulk that is asymptotic
to $AdS_3 \times S^3 \times T^4$, and which contains a
BPS black hole. The dual boundary theory is the two-dimensional
conformal field theory that lives on the intersection of the D1 branes
and the D5 branes and is known as the D1-D5-P CFT.
If one counts the states with momentum $N_p$ and
R-charge $J$ in this conformal field theory, one obtains the entropy
\be
 S = 2 \pi \sqrt{N_1 N_5 N_p - J^2}~,
 \label{BMPVent}
\ee
which precisely matches the entropy of the dual black hole \cite{arkady}
in the bulk.

A very important question, with deep implications for the physics of
black holes, is: {\it ``What is the fate of these microscopic brane
  configurations as the effective coupling becomes large?''}
Alternatively, the question can be rephrased in $AdS$-CFT language as:
{\it ``What is the gravity dual of individual microstates of the
  D1-D5-P CFT?''}  More physically, {\it ``What do the black-hole
  microstates look like in a background that a relativist would
  recognize as a black hole?''}

\subsection{Two-charge systems}
\label{sec:11}

These questions have been addressed for the simpler D1-D5
system\footnote{Throughout these lectures we will refer to the D1-D5
  system and its U-duals as the two-charge system, and to the D1-D5-P
  system and its U-duals as the three-charge system.} by Mathur,
Lunin, Maldacena, Maoz, and others,
\cite{LuninJY,lm2,lmm,other2ch,babel,Kanitscheider:2006zf}, see
\cite{arkady-2ch} for earlier work in this direction, and
\cite{mathur-rev} for a review of that work.  They found that the
states of that CFT can be mapped into two-charge supergravity
solutions that are asymptotically $AdS_3 \times S^3 \times T^4$, and
have no singularity. These supergravity solutions are determined by
specifying an arbitrary closed curve in the space transverse to the D1
and D5 branes, and have a dipole moment corresponding to a
Kaluza-Klein monopole (KKM) wrapped on that curve\footnote{A system
  that has a prescribed set of charges as measured from infinity often
  must have additional dipole charge distributions.  We will discuss
  this further in Section \ref{supersection}, but for the present, one
  should note the important distinction between asymptotic charges and
  dipole charges.}.  Counting these configurations
\cite{LuninJY,counting} has shown that the entropy of the CFT is
reproduced by the entropy coming from the arbitrariness of the shape
of the closed curve.

While the existence of such a large number of two-charge supergravity
solutions might look puzzling -- again, these BPS solutions are
specified by arbitrary {\it functions} -- there is a simple
string-theoretic reason for this.  By performing a series of S and T
dualities, one can dualize the D1-D5 configurations with KKM dipole
charge into configurations that have F1 and D0 charge, and D2-brane
dipole moment. Via an analysis of the Born-Infeld action of the D2
brane, these configurations were found by Mateos and Townsend to be
supersymmetric, and moreover to preserve the same supersymmetries as
the branes whose asymptotic charges they carry (F1 and D0 charge),
independent of the shape of the curve that the D2 brane wraps
\cite{supertube,supercurve,sugrasuper}. Hence, they were named ``supertubes.''
Alternatively, one can also dualize the D1-D5 (+ KKM dipole)
geometries into F1 string configurations carrying left-moving
momentum. Because the string only has transverse modes, the
configurations carrying momentum will have a non-trivial shape:
Putting the momentum into various harmonics causes the shape to change
accordingly. Upon dualizing, the shape of the momentum wave on the F1
string can be mapped into the shape of the supertube \cite{Lunin:2001fv}.

Thus, for two-charge system, we see that the existence of a large
number of supergravity solutions could have been anticipated from this
earlier work on the microscopic two-charge stringy configurations
obtained from supertubes and their duals.  In Section
\ref{supersection} we will consider three-charge supertubes and
discuss how this anticipated the discovery of some of the
corresponding supergravity solutions that are discussed in Section
\ref{BlackRings}.

\subsection{Implications for black-hole physics}
\label{sec:12}

An intense research programme has been unfolding over the past few
years to try to see whether the correspondence between D1-D5 CFT states and
smooth bulk solutions also extends to the D1-D5-P system.  The crucial
difference between the two-charge system and the three-charge system
(in five dimensions) is that the latter generically has a macroscopic
horizon, whereas the former only has an effective horizon at the
Planck or string scale.  Indeed, historically, the link between
microstate counting and Bekenstein-Hawking entropy (at vanishing
string coupling) was first investigated by Sen \cite{Sen:1995in} for
the two-charge system.  While this work was extremely interesting and
suggestive, the result became compelling only when the problem was
later solved for the three-charge system by Strominger and Vafa
\cite{stromvafa}.  Similarly, the work on the microstate geometries of
two-charge systems is extremely interesting and suggestive, but to be
absolutely compelling, it must be extended to the three-charge
problem.  This would amount to establishing that the boundary D1-D5-P
CFT microstates are dual to bulk microstates -- configurations that
have no horizons or singularities, and which look like a black hole
from a large distance, but start differing significantly from the
black hole solution at the location of the would-be horizon.

\begin{figure}
\centering
\includegraphics[height=4cm]{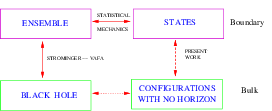}
\caption{An illustrative description of Mathur's conjecture. Most of
  the present research efforts go into improving the dictionary
  between bulk and boundary microstates (the dotted arrow), and into
  constructing more microstate geometries. }
\label{Mathur}
\end{figure}

String theory would then indicate that a black hole solution should
not be viewed as a fundamental object in quantum gravity, but rather
as an effective ``thermodynamic'' description of an ensemble of
horizonless configurations with the same macroscopic/asymptotic
properties. (See Fig. \ref{Mathur}.) The black hole horizon would be the place
where these configurations start differing from each other, and the classical
``thermodynamic'' description of the physics via the black hole
geometry stops making sense.

An analogy that is useful in understanding this proposal is to think
about the air in a room. One can use thermodynamics and fluid
mechanics to describe the air as a continuous fluid with a certain
equation of state. One can also describe the air using statistical
mechanics, by finding the typical configurations of molecules in the
ensemble, and noticing that the macroscopic features of these
configurations are the same as the ones found in the thermodynamic
description. For most practical purposes the thermodynamic description is the one to use; however, this description fails to capture the physics coming from the molecular structure of the air. To address problems like Brownian motion, one should not use the thermodynamic
approximation, but the statistical description. Similarly, to address
questions having to do with physics at the scale of the horizon (like
the information paradox) one should not use the thermodynamic
approximation, given by the black hole solution, but one should use the
statistical description, given by the microstate configurations.

This dramatic shift in the description of black holes, has been most
articulately proposed and strongly advocated by Mathur, and is thus
often referred to as ``Mathur's conjecture.''  In fact, one should be
careful and distinguish two variants of this conjecture. The weak
variant is that the black hole microstates are horizon-sized stringy
configurations that have unitary scattering, but cannot be described
accurately using the supergravity approximation. These configurations
are also sometimes called ``fuzzballs.''  If the weak Mathur
conjecture were true then the typical bulk microstates would be configurations where the curvature is Planck scale, and hence cannot be described in supergravity. The strong form
of Mathur's conjecture, which is better defined and easier to prove or
disprove, is that among the  {\it typical} black hole microstates
there are smooth solutions that can be described using supergravity.

Of course, the configurations that will be discussed and constructed in these notes are classical geometries with a moduli space. Classically, there is an infinite number of such configurations, that need to be quantized before one can call them {\it micro}states in the strictest sense of the word. In the analogy with the air in a room, these geometries correspond to classical configurations of molecules. Classically there is an infinite number of such configurations, but one can quantize them and count them to find the entropy of the system.

Whichever version of the conjecture is correct, we are looking for stringy
configurations that are very similar to the black hole from far away,
and start differing from each other at the location of the would-be
horizon.  Thus black hole microstates should have a size of the same
order as the horizon of the corresponding black hole.  From the
perspective of string theory, this is very a peculiar feature, since
most of the objects that one is familiar with become smaller, not larger,
as gravity becomes stronger.  We will see in these lectures how our
black hole microstates manage to achieve this feature.

If the strong form of this conjecture were true then it would not only
solve Hawking's information paradox (microstates have no horizon, and
scattering is unitary), but also would have important consequences for
quantum gravity. It also might allow one to derive 't Hooft's
holographic principle from string theory, and might even have
experimental consequences. A more detailed discussion about this can
be found in Section \ref{Conclusions}.

\subsection{Outline}
\label{sec:13}

As with the two-charge systems, the first step in finding three-charge
solutions that have no horizon and look like a black hole is to try to
construct large numbers of microscopic stringy three-charge
configurations.  This is the subject of Section \ref{supersection}, in
which we review the construction of three-charge supertubes -- string
theory objects that have the same charges and supersymmetries as the
three-charge black hole \cite{3chsuper}.

In Section \ref{BlackRings} we present the construction of
three-charge supergravity solutions corresponding to arbitrary
superpositions of black holes, black rings, and three-charge
supertubes of arbitrary shape. We construct explicitly a solution
corresponding to a black hole at the center of a black ring, and
analyze the properties of this solution.  This construction and the
material presented in subsequent sections can be read independently of
Section \ref{supersection}.

Section \ref{Geometry} is a geometric interlude, devoted to
Gibbons-Hawking metrics and the relationship between five-dimensional
black rings and four-dimensional black holes. Section \ref{GHsols}
contains the details of how to construct new microstate solutions
using an ``ambipolar'' Gibbons-Hawking space, whose signature
alternates from $(+,+,+,+)$ to $(-,-,-,-)$. Even though the sign of
the base-space metric can flip, the full eleven-dimensional solutions
are smooth.

In Section \ref{Bubbles} we discuss geometric transitions, and the way
to obtain smooth horizonless ``bubbling'' supergravity solutions that
have the same type of charges and angular momenta as three-charge
black holes and black rings. In Section \ref{Microstates} we construct
several such solutions, finding in particular microstates
corresponding to zero-entropy black holes and black rings.

In Section \ref{Mergers} we use mergers to construct and analyze
``deep microstates,'' which correspond to black holes with a
classically large horizon area. We find that the depth of these
microstates becomes infinite in the classical (large charge) limit,
and argue that they correspond to CFT states that have one long
component string. This is an essential (though not sufficient) feature
of the duals of typical black-hole microstates (for reviews of this,
see \cite{DavidWN,magoo}).  Thus the ``deep microstates'' are
either typical microstates themselves, or at least lie in the same
sector of the CFT as the typical microstates.

Finally, Section \ref{Conclusions} contains conclusions and an extensive
discussion of the implications of the work presented here on
for the physics of black holes in string theory.

Before beginning we should emphasize that the work that we present is
part of a larger effort to study black holes and their microstates in
string theory.  Many groups have worked at obtaining smooth microstate
solutions corresponding to five-dimensional and four-dimensional black
holes, a few of the relevant references include
\cite{Giusto:2004id,GiustoIP,GiustoKJ,GiustoZI,LuninUU,Bena:2005ay,
  Bena:2005va,Berglund:2005vb,Saxena:2005uk,Giusto:2006zi,BenaVA,
  Balasubramanian:2006gi,deep,Cheng:2006yq,Ford:2006yb}. Other groups
focus on improving the dictionary between bulk microstates and their
boundary counterparts, both in the two-charge and in the three-charge
systems \cite{other2ch,Bena:2004tk,Kanitscheider:2006zf}. Other groups
focus on small black holes\footnote{These black holes do not have a
  macroscopic horizon, but one can calculate their horizon area using
  higher order corrections \cite{Atish}. This area agrees with both
  the CFT calculation of the entropy, and also agrees (up to a
  numerical factor) with the counting of two-charge microstates.
  Hence, one could argue (with a caveat having to do with the fact
  that small black holes in IIA string theory on $T^4$ receive no
  corrections) that small black holes, which from the point of view of
  string theory are in the same category as the big black holes, are,
  in fact, superpositions of horizonless microstates.}  and study
their properties using the attractor mechanism \cite{attractor}, or
relating them to topological strings via the OSV conjecture
\cite{osv}.  Reviews of this can be found in \cite{small-review}, and
a limited sample of work that is related to the exploration presented
here can be found in \cite{small-others}.

\section{Three-charge microscopic configurations}
\label{supersection}

Our purpose here is to follow the historical path taken with the
two-charge system and try to construct three-charge brane
configurations using the  Born-Infeld (BI) action.  We are thus
considering the intrinsic action of a brane and we will not consider
the back-reaction of the brane on the geometry.  The complete
supergravity solutions will be considered later.

There are several ideas in the study of D-branes that will be
important here.  First, one of the easiest ways to create system with
multiple, different brane charges is to start with a
higher-dimensional brane and then turn on electromagnetic fields on
that brane so as to induce lower-dimensional branes that are
``dissolved'' in the original brane.  We will use this technique to
get systems with D0-D2-D4-D6 charges below.

In constructing multi-charge solutions, one should also remember that
the equations of motion are generically non-linear.  For example, in
supergravity the Maxwell action can involve Chern-Simons terms, or the
natural field strength may involve wedge products of lower degree
forms.  Similarly, in the BI action there is a highly non-trivial
interweaving of the Maxwell fields and hence of the brane charges.  In
practice, this often means that one cannot simply lay down independent
charges: Combinations of fields sourced by various charges may
themselves source other fields and thus create a distribution of new
charges.  In this process it is important to keep track of asymptotic
charges, which can be measured by the leading fall-off behaviour at
infinity, and ``dipole'' distributions that contribute no net charge
when measured at infinity.  When one discusses an $N$-charge system
one means a system with $N$ commuting asymptotic charges, as measured
at infinity.  For microstate configurations,  one often finds that the systems that
have certain charges will also have
fields sourced by other dipole charges. More precisely,   in
discussing the BI action of supertubes we will typically find that a
given pair of asymptotic charges, $A$ and $B$, comes naturally
with a third set of dipole charges, $C$.  We will therefore denote
this configuration by $A$-$B$ $\rightarrow$ $C$.

\subsection{Three-charge supertubes}

The original two-charge supertube \cite{supertube} carried two
independent asymptotic charges, D0 and F1, as well as a D2-brane
dipole moment; thus we denote it as a F1-D0 $\rightarrow$ D2
supertube.  It is perhaps most natural to try to generalize this object by
combining it with another set of branes to provide the third
charge\footnote{One might also have tried to generalize the F1-P dual
  of this system by adding a third type of charge.  Unfortunately,
  preserving the supersymmetry requires this third charge to be that
  of NS5 branes and, because of the dilaton throat of these objects,
  an analysis of the F1-P system similar to the two-charge one
  \cite{LuninJY} cannot be done.}.  Supersymmetry requires that this
new set be D4 branes.  To be more precise, supertubes have the same
sypersymmetries as the branes whose asymptotic charges they carry and
so one can naturally try to put together F1-D0 $\rightarrow$ D2
supertubes, F1-D4 $\rightarrow $ D6 supertubes, and D0-D4
$\rightarrow$ NS5 supertubes, and obtain a supersymmetric
configuration that has three asymptotic charges: D0, D4 and F1, and
three dipole distributions, coming from D6, NS5 and D2 branes wrapping
closed curves. Of course, the intuition coming from putting two-charge
supertubes together, though providing useful guidance, will not be
able to indicate anything about the size or other properties of the
resulting three-charge configuration.

\exbox{Show that the supertube with D2 dipole charge
and F1 and D0 charges  can be dualized into an F1-D4 $\rightarrow $ D6
supertube, and into a D0-D4 $\rightarrow$ NS5 supertube.}

To investigate objects with the foregoing charges and dipole charges
one has to use the theory on {\it one} of the sets of branes, and then
describe all the other branes as objects in this theory.  One route is
to consider tubular D6-branes\footnote{Tubular means it will only have
a dipole charge just like any loop of current in electromagnetism.},
and attempt to turn on world-volume fluxes to induce D4, D0 and F1
charges. As we will see, such a configuration also has a D2 dipole
moment. An alternative route is to use the D4 brane non-Abelian Born-Infeld action.
Both routes were pursued in \cite{3chsuper}, leading to
identical results.  Nevertheless, for simplicity we will only present
the first approach here.

One of the difficulties in describing three-charge supertubes in
this way is the fact that the Born-Infeld action and its non-Abelian
generalization cannot be used to describe NS5 brane dipole moments.
This is essentially because the NS5 brane is a non-perturbative object
from the perspective of the Born-Infeld  action \cite{pol-str}.
Thus, our analysis of three charge supertubes is limited to
supertubes that only have D2 and D6 dipole charge. Of course, one
can dualize these to supertubes with NS5 and D6 dipole charges, or
to supertubes with NS5 and D2. Nevertheless, using the action of a
single brane it is not possible to describe supertubes that have
three charges and three dipole charges. For that, we will have to
wait until Section \ref{BlackRings}, where we will construct the
full supergravity solution corresponding to these objects.

\subsection{The Born-Infeld construction}
\label{BIconstruction}

We start with a single tubular D6-brane, and attempt to turn on
worldvolume fluxes so that we describe a BPS configuration
carrying D4, D0 and F1 charges. We will see that this also necessarily
leads to the presence of D2-brane charges, but we will subsequently
introduce a second D6-brane to cancel this.

The D6-brane is described by the Born-Infeld action
\be
{ S= - T_6  \int\! d^7 \xi ~ \sqrt{ -\det(g_{ab} +\cF_{ab})},}
\label{za1}
\ee
where $g_{ab}$ is the induced worldvolume metric, $\cF_{ab} = 2\pi
F_{ab}$, $T_6$ is the D6-brane tension and we have set $\alpha'=1$.
The D6 brane also couples
to the background RR fields through the Chern-Simons action:
\be
{ S_{CS}=   T_6\,  \int  ~  \exp\big( \cF   + B\big) \, \wedge \,
\sum_q \, C^{(q)} \,.}
\label{zcs1}
\ee
By varying this with respect to the $C^{(q)}$ one obtains the
D4-brane, D2-brane  and D0-brane  charge densities:
\bea
Q_4& =& 2\pi \, T_6\,  \cF   \label{zb1}\\
Q_2& =& 2\pi \, T_6\,    \big( \coeff{1}{2}\, \cF \wedge \cF \label{zbxx1}\big)\\
Q_0 &=&  2\pi \, T_6\,   \big( \coeff{1}{3!}\, \cF \wedge \cF \wedge \cF \big) \,. \label{zc1}
\eea
To obtain the quantized Dp-brane charges, one takes the volume $p$-form on
any compact, $p$-dimensional spatial region, $R$,  and wedges this volume
form with $Q_p$ and integrates over the spatial section of the D6 brane.
The result is then the Dp-brane charge in the region $R$.

The F1 charge density can be obtained by varying the action with respect to the
time-space component of NS-NS two form potential, $B$. Since $B$ appears in  the combination
$\cF   + B$, one can differentiate with respect to the gauge field:
\be
{Q_1 ~=~  {\p {\cal L} \over \p  B_{0i}  } ~=~  {\p {\cal L} \over \p  F_{0i} }  ~=~
{\p {\cal L} \over \p \dot{\vec{A} }} ~=~  \vec{\pi}  \,,}
\label{zc2}
\ee
which is proportional to the canonical momentum conjugate to the vector potential,
$\vec{A}$.

Our construction will essentially follow that of the original D2-brane
supertube \cite{supertube}, except that we include four extra spatial
dimensions and corresponding fluxes.  We take our D6-brane to have the
geometry $\IR^{1,1} \times S^1 \times T^4$ and we choose coordinates
$(x^0,x^1)$ to span $\IR^{1,1}$ and $(x^6,x^7, x^8,x^9)$ to span the
$T^4$.  The $S^1$ will be a circle of of radius $r$ in the $(x^2,x^3)$
plane and we will let $\theta$ be the angular coordinate in this
plane.  We have also introduced factors of $2\pi$ in (\ref{zb1}),
(\ref{zbxx1}), (\ref{zc1}), (\ref{zc2}) to anticipate the fact that for
round tubes everything will be independent of $\theta$ and
so the integrals over $\theta$ will generate these factors of $2 \pi$.
Thus the D-brane charge densities above are really charge densities in
the remaining five dimensions, and the fundamental string charge is a
charge density per unit four-dimensional area. Note also that the
charges, $Q$, are the ones that appear in the Hamiltonian, and are
related to the number of strings or branes by the corresponding
tensions. These conventions will be convenient later on.

Since the $S^1$ is contractible and lies in the non-compact
space-time, any D-brane wrapping this circle will not give rise to
asymptotic charges and will only be dipolar.  In particular, the
configuration carries no asymptotic D6-brane charge due to its tubular
shape.  To induce D0-branes we turn on {\it constant} values of
$\cF_{1 \theta}$, $\cF_{67}, $ and $\cF_{89}$.  Turning on $\cF_{1
  \theta}$ induces a density of D4-branes in the $(x^6,x^7, x^8,x^9)$
plane, and since these D4 branes only wrap the $T^4$, their charge can
be measured asymptotically.  The fields $\cF_{67}, $ and $\cF_{89}$
similarly generate dipolar D4-brane charges.  To induce F1 charge in
the $x^1$ direction we turn on a constant value of $\cF_{01}$.  It is
also evident from (\ref{zbxx1}) that this configuration carries
asymptotic D2-brane charges in the $(x^6,x^7)$ and $(x^8,x^9)$ planes
and dipolar D2-brane charge in the $(x^1, \theta)$ direction.  The
asymptotic $D2$-brane charges will eventually be canceled by
introducing a second D6-brane.  This will also cancel the dipolar
D4-brane and D2-brane charges and we will then have a system with
asymptotic F1, D0 and D4 charges and dipolar D2 and D6 charges.

With these fluxes turned on we find
\be
S= - T_6  \int\! d^7 \xi ~
\sqrt{(1- \cF_{01}^2)r^2+\cF_{1\theta}^2
}\sqrt{(1+\cF_{67}^2)(1+\cF_{89}^2)} \,, \label{zd1}
\ee
where we use polar coordinates in the $(x^2,x^3)$  plane, and
the factors of $r^2$ come from $g_{\theta \theta}$.
By differentiating with respect to $\cF_{01}$ we find
\be
 Q_1 =2\pi T_6 {\cF_{01} r^2 \over  \sqrt{(1-
\cF_{01}^2)r^2+\cF_{1\theta}^2
}}\sqrt{(1+\cF_{67}^2)(1+\cF_{89}^2)}.\label{ze1}
\ee
The key point to observe now is that if we choose
\be
 \cF_{01}=1
\ee then $r^2$ drops out of the action (\ref{zd1}).  We will also
choose
\be
 \cF_{67} = \cF_{89}\,. \label{zg1}
\ee
We can then obtain the energy from the canonical Hamiltonian:
\bea
 H &=&   \int \! Q_1 \cF_{01}-L  \label{Ham}\\
&=&  \int \! \Big[ Q_1 +2\pi T_6
|\cF_{1\theta}| + 2\pi T_6 |\cF_{1\theta} \cF_{67} \cF_{89}|\Big]
\label{Ham1} \\
 & =& \int\! \left[ Q_1 + Q_4 +Q_0 \right].
\label{zh1}
\eea
The last two integrals are taken over the coordinates $(x^1,x^6,x^7, x^8,x^9)$
of the D6-brane. The radius of the system is determined by inverting
(\ref{ze1}):
\be
 r^2  = {Q_1 \over 2\pi T_6} {\cF_{1\theta}\over 1+
 \cF_{67} \cF_{89}}  = {1 \over (2\pi T_6)^2} {Q_1 Q_4^2 \over Q_0 +
 Q_4}.
 \label{zi1}
 \ee
 If we set $Q_0=0$ then (\ref{zi1}) reduces (with the obvious
 relabeling) to the radius formula found for the original D2-brane
 supertube \cite{supertube}. $\!\!$ From (\ref{zh1}) we see that we
 have saturated the BPS bound, and so our configuration must solve the
 equations of motion, as can be verified directly.

\exbox{Minimize the Hamiltonian in (\ref{Ham}) by  varying the radius, $r$,
while  keeping the F1, D0 and D4 charges constant.
Verify that the configuration with the radius
given (\ref{zi1}) solves the equations of motion.}

Supersymmetry can also be verified precisely as for the original
D2-brane supertube \cite{supertube}.  The presence of the electric field,
$\cF_{01}=1$, causes the D6-brane to drop out of the equations determining the
tension and the unbroken supersymmetry.  Indeed, just like the
two-charge system \cite{supercurve}, we can consider a D6-brane that
wraps an {\it arbitrary} closed curve in $\IR^4$; the only change in
(\ref{zd1}) and (\ref{ze1}) is that $r^2$ will be replace by the induced
metric on the D6 brane, $g_{\theta \theta} $. However, when
$\cF_{01}=1$ this does not affect equations (\ref{Ham1}) and (\ref{zh1}),
and therefore the configuration is still BPS. Moreover, if
$\cF_{1\theta} $ is not constant along the tube, or if $ \cF_{67}$ and
$ \cF_{89}$ remain equal but depend on $\theta$ the BPS bound is still
saturated.

Hence, classically, there exists an infinite number of three-charge
supertubes with two dipole charges, parameterized by several arbitrary
functions of one variable \cite{3chsuper}.  Four of these functions
come from the possible shapes of the supertube, and two functions
comes from the possibility of varying the D4 and D0 brane densities
inside the tube.  Anticipating the supergravity results, we expect
three-charge, three-dipole charge tubes to be given by {\it seven
  arbitrary functions}, four coming from the shape and three from the
possible brane densities inside the tube.  The procedure of
constructing supergravity solutions corresponding to these objects
\cite{Bena:2004de,var-charge} will be discussed in the next section,
and will make this ``functional freedom'' very clear.

As we have already noted, the foregoing configuration also carries
non-vanishing D2-brane charge associated with $\cF_{1\theta} \cF_{67}$
and $\cF_{1\theta} \cF_{89}$.  It also carries dipolar D4-brane charges
associated with $\cF_{67}$ and $\cF_{89}$.
To remedy this we can introduce one
more D6 brane with flipped signs of $\cF_{67}$ and $\cF_{89}$
\cite{wati}. This simply doubles the D4, D0, and F1 charges, while
canceling the asymptotic D2 charge and the dipolar D4-brane charges.
More generally, we can introduce $k$
coincident D6-branes, with fluxes described by diagonal $k\times k$
matrices. We again take the matrix-valued field strengths $\cF_{01}$
to be equal to the unit matrix, in order to obtain a BPS state.  We
also set $\cF_{67}=\cF_{89}$, and take $F_{1\theta}$ to have
non-negative diagonal entries to preclude the appearance of
$\overline{D4}$-branes.  The condition of vanishing D2-brane charge
 is then
\be
\Tr~ \cF_{1\theta}~ \cF_{67}  ~=~ \Tr~ \cF_{1\theta} \cF_{89} ~=~ 0.
\label{zj}
\ee
This configuration can also have D4-brane dipole charges, which we
may set to zero by choosing
\be
\Tr~ \cF_{67} ~=~\Tr~ \cF_{89} ~=~ 0.
\label{zj22}
\ee

Finally, the F1 charge is described by taking $Q_1$ to be an arbitrary
diagonal matrix with non-negative entries\footnote{Quantum
 mechanically, we should demand that $\Tr~ Q_1$ be an integer to
 ensure that the total number of F1 strings is integral.}. This
results in a BPS configuration of $k$ D6-branes wrapping curves of
arbitrary shape. If the curves are circular, the radius formula is now
given by (\ref{zi1}) but with the entries replaced by the corresponding
matrices.  Of course, for our purposes we are interested in situations
when we can use the Born-Infeld action of the D6 branes to describe
the dynamics of our objects. Since the BI action does not take into
account interactions between separated strands of branes, we will
henceforth restrict ourselves to the situations where these curves are coincident.
In analogy with the behaviour of other branes, if we take the $k$
D6-branes to sit on top of each other we expect that they can form a
marginally bound state.  In the classical description we should then
demand that the radius matrix (\ref{zi1}) be proportional to the unit
matrix.  Given a choice of magnetic fluxes, this determines the F1
charge matrix $Q_1$ up to an overall multiplicative constant that
parameterizes the radius of the combined system.

Since our matrices are all diagonal, the Born-Infeld action is
unchanged except for the inclusion of an overall trace. Similarly,
the energy is still given by $H =\int\! \Tr\left[ Q_1 + Q_4 +Q_0 \right]$.


Consider the example  in which all $k$ D6-branes are identical
modulo the sign of  $\cF_{67}$ and $\cF_{89}$, so that both
$\cF_{1\theta}$ and $\cF_{67} \cF_{89}$ are proportional to the unit
matrix\footnote{One could also take $\Tr \cF_{67} = \Tr
\cF_{89} = 0$ to cancel the D2 charge, but this does not affect the
radius formula. }. Then, in terms of the total charges, the radius
formula is
\be
 r^2   ~=~  {1 \over k^2(2\pi T_6)^2} {Q_1^{\rm tot}
(Q_4^{\rm tot})^2 \over Q_0^{\rm tot} + Q_4^{\rm tot}} \,.
\label{zia}
\ee
Observe that after fixing the conserved charges and imposing equal radii for
the component tubes, there is still freedom in the values of the
fluxes.   These can be partially parameterized in terms of various
non-conserved ``charges'', such as brane dipole moments. Due to the
tubular configuration, our solution carries non-zero D6, D4, and D2
dipole moments, proportional to
\bea
 Q^D_6 & =& T_6\, r\,   k \nonumber \\
 Q^D_4 & =& T_6\, r\,   \Tr\, \cF_{67} \nonumber \label{zk}\\
 Q^D_2 & =& T_6\, r\,   \Tr\, \cF_{67} \,\cF_{89} \equiv T_6\, r\,   k_2 .
\eea
When the $k$ D6-branes that form the tube are coincident, $k_2$
measures the local D2 brane dipole charge of the tube. It is also possible
to see that both for a single tube, and for $k$ tubes identical up
to the sign of $\cF_{67}$ and $\cF_{89}$, the dipole moments are
related via:
\be
 {Q_2^D \over Q_6^D} ~=~ {k_2 \over k} ~=~  {Q^{\rm tot}_0 \over
 Q^{\rm tot}_4} \,.
 \label{zl}
 \ee
 We will henceforth drop the superscripts on the $Q^{\rm tot}_p$ and
 denote them by  $Q_p$.
 One can also derive the microscopic relation,  (\ref{zl}), from the
 supergravity solutions that we construct in Section \ref{RoundRings}.
In the supergravity solution one has to set one of the three dipole
charges to zero to obtain the solution  with three asymptotic charges
 and two dipole charges.  One then finds that   (\ref{zl}) emerges
from are careful examination of the near-horizon limit and
the requirement that the solution be free of closed timelike curves
\cite{Bena:2004wv}.

If $\cF_{67}$ and $\cF_{89}$ are traceless, this tube has no D2 charge
and no D4 dipole moment. More general tubes will not satisfy
(\ref{zl}), and need not have vanishing D4 dipole moment when the D2
charge vanishes. We should also remark that the D2 dipole moment is an
essential ingredient in constructing a supersymmetric three-charge
tube of finite size. When this dipole moment goes to zero, the radius
of the tube also becomes zero.

In general, we can construct a tube of arbitrary shape, and this tube
will generically carry angular momentum in the $(x^2,x^3)$ and
$(x^4,x^5)$ planes. We can also consider a round tube, made of $k$
identical D6 branes wrapping an $S^1$ that lies for example in the
$(x^2,x^3)$ plane.  The microscopic angular momentum density of such a
configuration is given by the $(0,\theta)$ component of the
energy-momentum tensor:
\be
J_{23} ~=~  2\pi \, r \, T_{0\theta} ~=~ 2\pi T_6 \, k \, r^2\, \sqrt{(1+\cF_{67}^2)
(1+\cF_{89}^2)}\,.
\label{zm}
\ee
Now recall that supersymmetry requires $\cF_{67} =\cF_{89}$ and that
$\Tr(\cF_{67}  \cF_{89}) =  Q_0/ Q_4 $ and so this may be rewritten as:
\be
J_{23}  ~=~ 2\pi T_6 \, k \, r^2\,  \Big(1 ~+~ {Q_0\over Q_4} \Big) ~=~
{1 \over 2\pi T_6} \,   {Q_1\, Q_4 \over k}   \,,
\label{J23simp}
\ee
where we have used (\ref{zia}).  It is interesting to note that this
microscopic angular momentum density is not necessarily equal to the
angular momentum measured at infinity. As we will see in the next
section from the full supergravity solution, the angular momenta of
the three-charge supertube also have a piece coming from the
supergravity fluxes. This is similar to the non-zero angular momentum
coming from the Poynting vector, $ \vec E \times \vec B$, in the
static electromagnetic configuration consisting of an electron and a
magnetic monopole \cite{Jackson}.

Note also that when one adds D0 brane charge to a F1-D4 supertube, the
angular momentum does not change, even if the radius becomes
smaller. Hence, given charges of the same order, the angular momentum
that the ring carries is of order the square of the charge
(for a fixed number, $k$, of $D6$ branes).  For more general
three-charge supertubes, whose shape is an arbitrary curve inside
$\IR^4$, the angular momenta can be obtained rather straightforwardly
from this shape by integrating the appropriate components of
the BI energy-momentum tensor over the profile of the tube.

A T-duality along $x^1$ transforms our D0-D4-F1 tubes into the more
familiar D1-D5-P  configurations.   This T-duality is implemented
by the replacement $2\pi A^1 \rightarrow X^1$.  The non-zero value of
$\cF_{1\theta}$ is translated by the T-duality into a non-zero value of
$\partial_\theta X^1$.  This means that the resulting D5-brane
is in the shape of a helix whose axis is parallel to $x^1$.   This
is the same as the observation that the D2-brane supertube T-dualizes
into a helical D1-brane.
Since this helical shape is slightly less convenient to work with than
a tube, we have chosen to emphasize the F1-D4-D0 description instead.
Nevertheless, in the formulas that give the radius and angular momenta
of the three-charge supertubes we will use interchangingly the D1-D5-P and the
D0-D4-F1 quantities, related via U-duality $N_0 \rightarrow N_p$, $N_4
\rightarrow N_5$, and $N_1 \rightarrow N_1$, with similar replacements
for  the $Q$'s.

\exbox{Write the combination of S-duality and T-duality transformations
that corresponds to this identification of the D1-D5-P and F1-D4-D0
quantities.}

\subsection{Supertubes and black holes}

The spinning three-charge black hole (also known as
the BMPV black hole \cite{bmpv}) can only carry equal angular momenta,
bounded above by\footnote{In Section \ref{RoundRings} we will re-derive the
BMPV solution as part of a  more complex solution.  This bound can be seen
from   (\ref{BMPVent}) and follows from the  requirement that there are
no closed time-like curves outside the horizon.}:
\be
J_1^2  ~=~ J_2^2  ~\leq~  N_1 N_5 N_P \,.
\label{BHJbound}
\ee
For the three-charge supertubes, the angular momenta are not restricted to
be equal.  A supertube configuration can have arbitrary shape, and carry any
combination of the two angular momenta.  For example, we can choose a
closed curve such that the supertube cross-section lies in the $(x^2,x^3)$
plane, for which $J_{23} \neq 0 $ and $J_{45} =0$.
The bound on the  angular momentum can be obtained from (\ref{J23simp}):
\be
|J| ~=~  {1 \over 2\pi T_6} \,   {Q_1\, Q_4\over k}  ~ \leq~  {1 \over 2\pi T_6} \,
{Q_1\, Q_4}  ~=~ N_1 \, N_4 \,, \label{BRJbound}
\ee
where we have used $k \ge 1$ since it is the number of D6 branes.
The quantized charges\footnote{These charges are related to the charges that
appear in the Hamiltonian by the corresponding  tensions; more details about
this  can be found in \cite{3chsuper}.}  are given by $Q_1 ={1 \over 2\pi} N_1$,
$Q_4 = (2\pi)^2 T_6 N_4$.
We therefore see that a single D6 brane saturates the bound and
that by varying the number of D6 branes or by   appropriately changing the shape
and orientation of the tube cross section, we can span the entire range
of angular momenta between  $-N_1 N_4$ and $+N_1 N_4$.
Since (\ref{BRJbound}) is quadratic the charges,  one can easily exceed
the black hole angular momentum bound in (\ref{BHJbound}) by simply making
$Q_1$ and $Q_4$    sufficiently large.

One can also compare the size of the supertube with the size of the black hole.
Using (\ref{BRJbound}), one can rewrite (\ref{zia}) in terms of the angular momentum:
\be
r^2 ~=~  {J^2 \over  Q_1\, (Q_0 + Q_4) }  \,, \label{randJ}
\ee
Now recall that the tension of a D-brane varies as $g_s^{-1}$ and that
the charges, $Q_0$ and $Q_4$, appear in the Hamiltonian, (\ref{zh1}).
This means that  the quantization conditions on the D-brane charges
must have the form $Q_j \sim N_j/g_s$.   The energy of
the  fundamental string is independent of $g_s$ and so   $Q_1\sim N_1$,
with no factors of $g_s$.  If we take $N_0 \approx N_1\approx N_4 \approx N $
then we find:
\be
r_{\rm tube}^2 ~\sim~  g_s\,  {J^2 \over N^2} \,.\label{pii}
\ee

From the BMPV black hole metric \cite{bmpv,herdeiro} one can
compute the proper length of the circumference of the horizon
(as measured at one of the equator circles) to be
\be
{r_{\rm hole}^2  ~\sim~   g_s \, {N^3 - J^2 \over N^2}~.}
\label{pj}
\ee

The most important aspect of the equations (\ref{pii}) and (\ref{pj})
is that for comparable charges and angular momenta, the black hole and
the three-charge supertube have comparable sizes. Moreover, these
sizes grow with
$g_s$ in the same way.  This is a very counter-intuitive behavior.
Most of the objects we can think about tend to become smaller
when gravity is made stronger and this is consistent with our
intuition and the fact that gravity is an attractive force. The only
``familiar'' object that becomes larger with stronger gravity is a
black hole. Nevertheless, three-charge supertubes also become
larger as gravity becomes stronger!  The size of a tube is
determined by a balance between the angular momentum of the system and
the tension of the tubular brane. As the string coupling is increased,
the D-brane tension decreases, and thus the size of the tube grows, at
exactly the same rate as the Schwarzschild radius of the black
hole\footnote{Note that this is a feature only of three-charge
  supertubes; ordinary (two-charge) supertubes have a growth that is
  duality-frame dependent.}.

This is the distinguishing feature that makes the three-charge
supertubes (as well as the smooth geometries that we will obtain from
their geometric transitions) unlike any other configuration that one
counts in studying black hole entropy.

To be more precise, let us consider the counting of states that
leads to the black hole entropy ``\`a la
Strominger and Vafa.''   One counts microscopic brane/string configurations at
weak coupling where the system is of string scale in extent, and its
Schwarzschild radius even smaller. One then imagines increasing the
gravitational coupling; the Schwarzschild radius grows, becoming
comparable to the size of the brane configuration at the
``correspondence point'' \cite{HorowitzNW}, and larger thereafter.
When the Schwarzschild radius is much larger than the Planck scale,
the system can be described as a black hole.
There are thus two very different descriptions of the system: as a
microscopic string
theory object for small $g_s$, and as a black hole for large $g_s$.
One then compares the entropy in the two regimes and finds an
agreement, which is precise if supersymmetry forbids corrections
during the extrapolation.

Three-charge supertubes behave differently. Their size grows at the same
rate as the Schwarzschild radius, and thus they have no
``correspondence point.''  Their description is valid in the same
regime as the description of the black hole. If by counting such
configurations one could reproduce the entropy of the black hole, then
one should think about the supertubes as the large $g_s$ continuation
of the microstates counted at small $g_s$ in the string/brane picture,
and therefore as the microstates of the corresponding black hole.

It is interesting to note that if the supertubes did not grow with
{\it exactly} the same power of $g_s$ as the black hole horizon, they
would not be good candidates for being black hole microstates, and
Mathur's conjecture would have been in some trouble. The fact that
there exists a huge number of configurations that {\it do} have the
same growth with $g_s$ as the black hole is a non-trivial confirmation
that these configurations may well represent black-hole microstates
for the three-charge system.

We therefore expect that configurations constructed from three-charge
supertubes will give us a large number of three-charge BPS black hole
microstates.  Nevertheless, we have seen that three-charge supertubes
can have angular momenta larger than the BPS black hole, and
generically have $J_1 \neq J_2$.  Hence one can also  ask if there
exists a black object whose microstates those supertubes represent. In
\cite{3chsuper} it was conjectured that such an object should be a
three-charge BPS black ring, despite the belief at the time that there was
theorem that such BPS black rings could not exist.  After more evidence
for this conjecture came from the construction of the flat limit of
black rings \cite{Bena:2004wv}, a gap in the proof of the theorem was
found \cite{harvey-erratum}. Subsequently the BPS black ring with
equal charges and dipole charges was found in \cite{Elvang:2004rt},
followed by the rings with three arbitrary charges and three arbitrary
dipole charges \cite{Bena:2004de,blackring2,Gauntlett:2004qy}.
One of the morals of this story  is that whenever one encounters an
``established'' result that contradicts intuition one should really get to the
bottom of it and find out why the intuition is wrong or to expose
the cracks in established wisdom.

\section{Black rings and supertubes}
\label{BlackRings}

As we have seen in the D-brane analysis of the previous section,
three-charge supertubes of arbitrary shape preserve the same
supersymmetries as the three-charge black hole. Moreover, as we will
see, three-charge supertube solutions that have three dipole charges
can also have a horizon at large effective coupling, and thus become
black rings.  Therefore, one expects the existence of BPS
configurations with an arbitrary distribution of black holes, black
rings and supertubes of arbitrary shape. Finding the complete
supergravity solution for such configurations appears quite daunting.
We now show that this is nevertheless possible and that the entire problem
can be reduced to solving a linear system of equations in
four-dimensional, Euclidean electromagnetism.

\subsection{Supersymmetric configurations}
\label{SusyConfigs}

We begin by considering brane configurations that preserve the same
supersymmetries as the three-charge black hole.  In M-theory, the
latter can be constructed by compactifying on a six-torus, $T^6$, and
wrapping three sets of M2 branes on three orthogonal two-tori (see
the first three rows of Table \ref{braneconfig}).  Amazingly enough,
one can add a further three sets of M5 branes  while preserving the
same supersymmetries: Each set of M5 branes can be thought of as
magnetically dual to a set of M2 branes in that the M5 branes wrap
the four-torus, $T^4$, orthogonal to the $T^2$ wrapped by the M2
branes. The remaining spatial direction of the M5 branes follows a
simple, closed curve, $y^\mu(\sigma)$, in the  spatial section of
the five-dimensional space-time. Since we wish to make a single,
three-charge ring we take this curve to be the same for all three
sets of M5 branes.  This configuration is summarized in Table
\ref{braneconfig}.  In \cite{Bena:2004de}  it was argued that this
was the most general three-charge brane
configuration\footnote{Obviously one can choose add multiple curves
and black hole sources.} consistent with the supersymmetries of the
three-charge black-hole.

%
\begin{table}
\centering
%
\smallskip
\begin{tabular}{|c|ccccccccccc|}
\hline
~Brane~& ~0~ &  ~1~ &  ~2~ &  ~3~ &  ~4~ &  ~5~ &  ~6~  &  ~7~  &  ~8~  &  ~9~  &  ~10~   \\
\hline
M2 & $\updownarrow$ & $\star$  & $\star$ & $\star$ & $\star$ &$\updownarrow$ & $\updownarrow$ & $\leftrightarrow$ &$\leftrightarrow$ & $\leftrightarrow$ & $\leftrightarrow$ \\
M2 & $\updownarrow$ & $\star$  & $\star$ & $\star$ & $\star$ &$\leftrightarrow$ & $\leftrightarrow$ & $\updownarrow$ &$\updownarrow$ & $\leftrightarrow$ & $\leftrightarrow$ \\
M2 & $\updownarrow$ & $\star$  & $\star$ & $\star$ & $\star$ &$\leftrightarrow$ & $\leftrightarrow$ & $\leftrightarrow$ &$\leftrightarrow$ & $\updownarrow$ & $\updownarrow$ \\
M5 & $\updownarrow$ & \multicolumn{4}{c}{ $y^\mu(\sigma)$ } &$\leftrightarrow$ & $\leftrightarrow$ & $\updownarrow$  & $\updownarrow$ & $\updownarrow$ & $\updownarrow$ \\
M5 & $\updownarrow$ & \multicolumn{4}{c}{ $y^\mu(\sigma)$ } & $\updownarrow$ & $\updownarrow$ & $\leftrightarrow$ &$\leftrightarrow$ & $\updownarrow$ & $\updownarrow$ \\
M5 & $\updownarrow$ & \multicolumn{4}{c}{ $y^\mu(\sigma)$ } &$\updownarrow$ & $\updownarrow$ & $\updownarrow$ &$\updownarrow$ & $\leftrightarrow$ & $\leftrightarrow$ \\
\hline
\end{tabular}
\smallskip
\caption{Layout of the branes that give the supertubes and black rings
in an M-theory duality frame. Vertical arrows
$\updownarrow$, indicate the directions along which the branes are extended, and
horizontal arrows, $\leftrightarrow$, indicate the smearing directions. The functions, $y^\mu(\sigma)$,
indicate that the brane wraps a simple closed curve in $\Bbb{R}^4$ that defines
the black-ring or supertube profile. A star, $\star$, indicates that a brane is smeared along the
supertube profile, and pointlike on the other three directions.}
\label{braneconfig}
\end{table}

The metric corresponding to this brane configuration can be written as
\begin{eqnarray}
 ds_{11}^2  = ds_5^2 & + &    \left(Z_2 Z_3  Z_1^{-2}  \right)^{1\over 3}
 (dx_5^2+dx_6^2) \nonumber \\
 & + & \left( Z_1 Z_3  Z_2^{-2} \right)^{1\over 3} (dx_7^2+dx_8^2)    +
  \left(Z_1 Z_2  Z_3^{-2} \right)^{1\over 3} (dx_9^2+dx_{10}^2) \,,
\label{elevenmetric}
\end{eqnarray}
where the five-dimensional space-time metric has the form:
\begin{equation}
ds_5^2 ~\equiv~ - \left( Z_1 Z_2  Z_3 \right)^{-{2\over 3}}  (dt+k)^2 +
\left( Z_1 Z_2 Z_3\right)^{1\over 3} \, h_{\mu \nu}dy^\mu dy^\nu \,,
\label{fivemetric}
\end{equation}
for some one-form field, $k$, defined upon the spatial section of this
metric.  Since we want the metric to be asymptotic to flat $\Bbb{R}^{4,1} \times T^6$,
we require
\begin{equation}
ds_4^2 ~\equiv~ h_{\mu \nu}dy^\mu dy^\nu \,,
\label{fourmetric}
\end{equation}
 to limit to
the flat, Euclidean metric on $\Bbb{R}^4$ at spatial infinity and we require
the warp factors, $Z_I$, to limit to constants at infinity.  To fix the normalization of
the corresponding Kaluza-Klein $U(1)$ gauge fields, we will
take $Z_I  \to 1$ at infinity.

The supersymmetry, $\epsilon$, consistent with the brane configurations in
Table \ref{braneconfig} must satisfy:
\begin{equation}
\big(\oneone ~-~  \Gamma^{056}) \, \epsilon ~=~  \big(\oneone ~-~
\Gamma^{078}) \, \epsilon ~=~  \big(\oneone ~-~  \Gamma^{09\,10}) \, \epsilon ~=~  0  \,.
\label{susyproj}
\end{equation}
Since the product of all the gamma-matrices is the identity matrix, this implies
\begin{equation}
\big(\oneone ~-~  \Gamma^{1234}) \, \epsilon ~=~   0  \,,
\label{fourhelicity}
\end{equation}
which means that one of the four-dimensional helicity components
of the four dimensional supersymmetry must vanish identically.  The holonomy
of the metric, (\ref{fourmetric}), acting on the spinors is determined by
\begin{equation}
[\,\nabla_\mu \, , \, \nabla_\nu\,] \, \epsilon ~=~ \coeff{1}{4}\, R^{(4)}_{\mu \nu cd} \,
\Gamma^{cd}  \, \epsilon    \,,
\label{holonomy}
\end{equation}
where $R^{(4)}_{\mu \nu cd}$ is the Riemann tensor of (\ref{fourmetric}).  Observe that
(\ref{holonomy}) vanishes identically as a consequence of (\ref{fourhelicity}) if the Riemann
tensor is self-dual:
\begin{equation}
R^{(4)}_{abcd}  ~=~ \coeff{1}{2}\, {\varepsilon_{cd}}^{ef}\, R^{(4)}_{abef}    \,.
\label{halfflat}
\end{equation}
Such four-metrics are called ``half-flat.''  Equivalently, note that
the holonomy of a general  Euclidean four-metric is $SU(2) \times
SU(2)$ and that (\ref{halfflat})  implies that the holonomy lies
only in one of these $SU(2)$ factors and that the metric is flat in
the other factor.  The condition (\ref{fourhelicity}) means that all
the components of the supersymmetry upon which the non-trivial
holonomy would act actually vanish.  The other helicity components
feel no holonomy and so the supersymmetry can be defined globally.
One should also note that $SU(2)$ holonomy in four-dimensions is
equivalent to requiring that the metric be hyper-K\"ahler.

Thus we can preserve the supersymmetry if and only if we take the
four-metric to be hyper-K\"ahler.  However, there is a theorem that
states that any metric that is (i) Riemannian (signature $+4$) and
regular, (ii) hyper-K\"ahler and (iii) asymptotic to the flat metric
on $\Bbb{R}^4$, {\it must be globally} the flat metric on $\Bbb{R}^4$.
The obvious conclusion, which we will follow in this section, is that
we simply take (\ref{fourmetric}) to be the flat metric on
$\Bbb{R}^4$.  However, there are very important exceptions.  First, we
require the four-metric to be asymptotic to flat $\Bbb{R}^4$ because
we want to interpret the object in asymptotically flat,
five-dimensional space-time.  If we want something that can be
interpreted in terms of asymptotically flat, {\it four}-dimensional
space-time then we want the four-metric to be asymptotic to the flat
metric on $\Bbb{R}^3 \times S^1$.  This allows for a lot more
possibilities, and includes the multi-Taub-NUT metrics
\cite{Hawking:1976jb}.  Using such Taub-NUT metrics provides a
straightforward technique for reducing the five-dimensional solutions
to four dimensions \cite{Bena:2005ay,andy4d5d,Gaiotto:2005xt,Elvang:2005sa,Bena:2005ni}.

The other exception will be the subject of subsequent sections of this
review: The requirement that the four-metric be globally Riemannian is
too stringent.  As we will see, the metric can be allowed to change the
overall sign since this can be compensated by a sign change in the
warp factors of (\ref{fivemetric}).  In this section, however, we will
suppose that the four-metric is simply that of flat $\Bbb{R}^4$.

\subsection{The BPS equations}
\label{BPSeqns}

The Maxwell three-form potential is given by
\begin{equation}
C^{(3)}  = A^{(1)} \wedge dx_5 \wedge dx_6 ~+~  A^{(2)}   \wedge
dx_7 \wedge dx_8 ~+~ A^{(3)}  \wedge dx_9 \wedge dx_{10}  \,,
\label{Cfield:ring}
\end{equation}
where the six coordinates, $ x_A$, parameterize the compactification
torus, $T^6$, and $A^{(I)}$, $I=1,2, 3$, are one-form Maxwell
potentials in the five-dimensional space-time and depend only upon
the coordinates, $y^\mu$, that parameterize the spatial directions.
It is convenient to introduce the Maxwell
 ``dipole field strengths,''    $\Theta^{(I)}$,  obtained by removing the contributions
 of the electrostatic potentials
\begin{equation}
\Theta^{(I)} ~\equiv~  d A^{(I)} + d\big(  Z_I^{-1} \, (dt +k) \big)  \,,
\label{Thetadefn}
\end{equation}

The most general supersymmetric configuration is then obtained by solving the
{\it BPS equations}:
\begin{eqnarray}
 \Theta^{(I)}  &~=~&  \star_4 \, \Theta^{(I)} \label{BPSeqn:1} \,, \\
 \nabla^2  Z_I &~=~&  {1 \over 2  }  C_{IJK} \star_4 (\Theta^{(J)} \wedge
\Theta^{(K)})  \label{BPSeqn:2} \,, \\
 dk ~+~  \star_4 dk &~=~&  Z_I \,  \Theta^{(I)}\,,
\label{BPSeqn:3}
\end{eqnarray}
where $\star_4$ is the Hodge dual taken with respect to the
four-dimensional metric $h_{\mu \nu}$, and structure
constants\footnote{If the $T^6$ compactification manifold is replaced
  by a more general Calabi-Yau manifold, the $C_{IJK}$ change
  accordingly.}  are given by $C_{IJK} ~\equiv~ |\epsilon_{IJK}|$.  It
is important to note that if these equations are solved in the order
presented above, then one is solving a linear system.

\begin{figure}
\centering
\includegraphics[width=5.6cm]{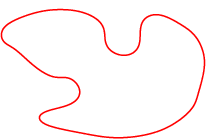}~~~~~~~~~
~\includegraphics[width=5.6cm]{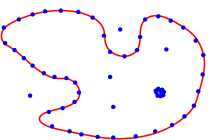}
\caption{The first two steps of the procedure to construct solutions. One first
chooses an arbitrary M5 brane profile, and then sprinkles the various types
of M2 branes, either on the M5 brane profile, or away from it. This gives a
solution for an arbitrary superposition of black rings, supertubes and black holes.}
\label{2steps}
\end{figure}

At each step in the solution-generating process one has the freedom to
add homogeneous solutions of the equations.  Since we are requiring
that the fields fall off at infinity, this means that these
homogeneous solutions must have sources in the base space and since
there is no topology in the $\IR^4$ base, these sources must be
singular.  One begins by choosing the profiles, in $\IR^4$, of the
three types of M5 brane that source the $\Theta^{(I)}$.  These fluxes
then give rise to the explicit sources on the right-hand side of
(\ref{BPSeqn:2}), but one also has the freedom to choose singular
sources for (\ref{BPSeqn:2}) corresponding to the densities,
$\rho_I(\sigma)$, of the three types of M2 branes.  The M2 branes can
be distributed at the same location as the M5 profile, and can also be
distributed away from this profile.  (See Fig. \ref{2steps}.)  The
functions, $Z_I$, then appear in the final solution as warp factors
and as the electrostatic potentials.  There are thus two contributions
to the total electric charge of the solution: The localized M2 brane
sources described by $\rho_I(\sigma)$ and the induced charge from the
fields, $\Theta^{(I)}$, generated by the M5 branes.  It is in this
sense that the solution contains electric charges that are dissolved
in the fluxes generated by M5 branes, much like in the
Klebanov-Strassler or Klebanov-Tseytlin solutions
\cite{Klebanov:2000nc,Klebanov:2000hb}.

The final step is to solve the last BPS equation, (\ref{BPSeqn:3}),
which is sourced by a cross term between the magnetic and electric
fields.  Again there are homogeneous solutions that may need to be
added and this time; however they need to be adjusted so as to ensure that
(\ref{fivemetric}) has no closed time-like curves (CTC's).  Roughly
one must make sure that the angular momentum at each point does not
exceed what can be supported by local energy density.

\subsection{Asymptotic charges}
\label{AsympCharges}

Even though a generic black ring is made from six sets of branes, there
are only three conserved electric charges that can be measured from infinity.
These are obtained from the three vector potentials, $A^{(I)}$, defined
in (\ref{Cfield:ring}), by integrating $\star_5 d A^{(I)}$ over the
three-sphere at spatial infinity.   Since the M5 branes run in a closed loop,
they do not directly contribute to the electric charges. The electric charges are
determined by electric fields at infinity, and hence by the functions
$Z_I$  (\ref{Thetadefn}). Indeed,  one has:
\begin{equation}
Z_I ~\sim~ 1 ~+~ c_1\, {Q_I \over \rho^2}~+~ \dots  \,, \qquad  \rho \to \infty \,,
\label{ZIinfty}
\end{equation}
where $c_1$ is a normalization constant (discussed below),
$\rho$ is the standard, Euclidean radial coordinate in $\IR^4$ and the $Q_I$
are the electric charges.
Note that while the M5 branes do not {\it directly}
contribute to the electric charges, they do contribute indirectly via
``charges dissolved in fluxes,''  that is, through the source terms on the
right-hand side  of (\ref{BPSeqn:2}).

To compute the angular momentum it is convenient to write the spatial
$\Bbb{R}^4$ as $\Bbb{R}^2 \times \Bbb{R}^2$ and pass to two sets of
polar coordinates, $(u,\theta_1)$ and $(v,\theta_2)$ in which the flat
metric on $\Bbb{R}^4$ is:
\begin{equation}
ds_4^2 ~=~ (du^2 ~+~ u^2 \, d \theta_1^2) ~+~
(dv^2 ~+~ v^2 \, d \theta_2^2)  \,.
\label{Rtwosqmet}
\end{equation}
There are two commuting angular momenta, $J_1$ and $J_2$, corresponding to
the components of rotation in these two planes.   One can then read off the
angular momentum by making an expansion at infinity of the angular momentum
vector, $k$, in (\ref{fivemetric}):
\begin{equation}
k  ~\sim~  c_2\, \bigg(J_1 \, {u^2  \over (u^2 +v^2 )^2 } ~+~ J_2 \, {v^2  \over
(u^2 +v^2 )^2 }  \bigg)~+~ \dots  ~\,, \qquad   u, v \to \infty \,,
\label{kinfty}
\end{equation}
where $c_2$ is a normalization constant.
The charges, $Q_I$, and the angular  momenta, $J_1,  J_2$, need to be correctly
normalized in order to express them  in terms of the quantized charges.
The normalization depends upon the eleven-dimensional
Planck length, $\ell_p$, and the volume of the compactifying torus, $T^6$.
The correct normalization can be  found \cite{Bena:2004de}, and has been
computed in many references. (For a good review, see \cite{Peet:2000hn}.)
Here we simply state that if  $L$ denotes the radius of the circles that make up the
$T^6$ (so that the compactification volume is $V_6 = (2\pi L)^6$), then one
obtains the canonically normalized quantities by using
\begin{equation}
c_1 ~=~ {\ell_p^6 \over L^4 }\,,  \qquad c_2 ~=~ {\ell_p^9 \over  L^6 }\,.
\label{normc}
\end{equation}
For simplicity, in most of the rest of this review we will take as system of units
in which $\ell_p =1$ and we will fix the torus volume so that $L=1$.  Thus
one has $c_1 = c_2 =1$.

\subsection{An example:  A three-charge black ring with a
black hole in the middle}
\label{RoundRings}

By solving the BPS equations, (\ref{BPSeqn:1})--(\ref{BPSeqn:3}),
one can, in principle,  find the supergravity solution for an
arbitrary distribution of black rings and black holes.
The metric for a general distribution of these objects will be extremely
complicated, and so to illustrate the technique we will
concentrate on a simpler system: A  BMPV black
hole at the center of a three-charge BPS black ring.
An extensive review of black rings, both BPS and non-BPS can be found in
\cite{Emparan:2006mm}.   Other interesting papers related to non-BPS black rings
include \cite{nonBPSring}.

Since the ring sits in an $\Bbb{R}^2$ inside $\Bbb{R}^4$, it is it is
natural to pass to the two sets of polar coordinates, $(u,\theta_1)$
and $(v,\theta_2)$ in which the base-space metric takes the form
(\ref{Rtwosqmet}) We then locate the ring at $u=R$ and $v=0$
and the black hole at $u=v=0$.

The best coordinate system for actually solving the black ring
equations is the one that has become relatively standard in the
black-ring literature (see, for example, \cite{Elvang:2004rt}).
The change of variables is:
\begin{eqnarray}
x &~=~& -{ u^2+v^2 - R^2 \over  \sqrt{((u-R)^2 + v^2)( (u+R)^2 + v^2 )}} \,, \\
y &~=~& -{ u^2+v^2 + R^2 \over  \sqrt{((u-R)^2 + v^2)( (u+R)^2 + v^2 )}} \,,
\end{eqnarray}
where $ -1 \le x  \le 1$, $-\infty < y \le -1$, and the ring is
located at $y = -\infty$.    This system has several advantages: it
makes the electric and magnetic two-form field strengths sourced by
the ring have a very simple form (see (\ref{Field-str})), and it
makes the ring look like a single point while maintaining
separability of the Laplace equation. In these coordinates the flat $\IR^4$
metric has the form:
\begin{equation}
ds_4^2 ~=~ {R^2 \over (x-y)^2}\left( {dy^2 \over y^2 -1}
+ (y^2-1) d \theta_1^2
+{dx^2 \over 1-x^2} + (1-x^2) d \theta_2^2  \right).
\end{equation}
The  self-dual\footnote{Our  orientation is $\epsilon^{yx\theta_1\theta_2} = +1$.}
field strengths that are sourced by the ring are then:
\begin{equation}
\Theta^{(I)} ~=~  2\, q_i\, (d x \wedge d \theta_2 ~-~  d y \wedge d
\theta_1) \,. \label{Field-str}
\end{equation}
The warp factors then have the form
\begin{equation}
Z_I ~=~ 1 ~+~ {\Qb_I \over R} (x-y) ~-~{2 \, C_{IJK}\,  q^J q^K
\over  R^2}\, (x^2 - y^2) ~-~ {Y_I \over R^2}\, {x-y \over x+y } \,,
\label{warpbrbh}
\end{equation}
and the angular momentum components are given by:
\begin{eqnarray}
k_{\psi} &=&   (y^2-1)\, g(x,y)  -  \, A \,(y+1)  \,, \qquad
k_{\phi}  =  (x^2-1)\,g(x,y)  \,; \\
g(x,y) &\equiv &  \left({C \over 3}\, (x+y)~+~
{B \over 2} - {D \over R^2(x+y)} + {K \over R^2 (x+y)^2 } \right)
\end{eqnarray}
where  $K$ represents the angular momentum of the BMPV black hole and
\begin{eqnarray}
A & ~\equiv~&  2 \big(\sum q^I\big) \,, \qquad \qquad
B ~\equiv~ {2\over R} \, (Q_I q^I )\,, \qquad \\
C & ~\equiv~&  -  {8\,  C_{IJK} \, q^I q^J q^K \over R^2}\,, \qquad
D  ~\equiv~  { 2 } \, Y_I q^I \,.
\end{eqnarray}
The homogeneous solutions of (\ref{BPSeqn:3}) have already been
chosen so as to remove any closed timelike curves (CTC).

The relation between the quantized ring and black-hole charges and the
parameters appearing in the solution are:
\begin{equation}
\Qb_I = {\Nb_I \, \ell_p^6 \over 2 L^4 R}\,, \qquad q^I={n^I \,
\ell_p^3 \over 4 L^2}\,, \qquad Y_I = {N^{\rm BH}_I \, \ell_p^6
\over L^4}\,, \qquad  K = { J^{\rm BMPV} \, \ell_p^9 \over L^6}\,,
\label{translation}
\end{equation}
where $L$ is the radius of the circles that make up the $T^6$ (so that
$V_6 = (2\pi L)^6$) and $\ell_p$ is the eleven-dimensional Planck length.

As we indicated earlier, the asymptotic charges, $N_I$, of the solution are the
sum of the microscopic charges on the black ring, $\Nb_I$, the charges of the
black hole, $N^{\rm BH}_I$, and the charges dissolved in fluxes:
\begin{equation}
N_I ~=~  \Nb_I ~+~ N^{\rm BH}_I ~+~ \coeff {1}{2} \, C_{IJK}\,n^J
n^K \,.
\label{chargeinf}
\end{equation}
%

\exbox{Derive this expression for the
charge from the asymptotic expansion of the $Z_I$ in (\ref{warpbrbh}). Derive the
relation between the parameters $q^I$ and the quantized M5 charges
$n^I$ in (\ref{translation}), by integrating the magnetic M-theory
four-form field strength around the ring profile.  (See, for example, \cite{Peet:2000hn}
in order to get the charge normalizations precisely correct.)}

The angular momenta of this solution are:
\begin{eqnarray}
J_1 &=& J_{\Delta}  ~+~ \Big( \coeff{1}{6}\, C_{IJK} \, n^I n^J n^K  ~+~
\coeff{1}{2}\,\Nb_I n^I ~+~ N^{\rm BH}_I n^I  ~+~  J^{\rm BMPV} \Big)\,, \\
J_2 &=& - \Big( \coeff{1}{6}\,C_{IJK} \, n^I n^J n^K  ~+~
\coeff{1}{2}\,\Nb_I n^I ~+~ N^{\rm BH}_I n^I ~+~  J^{\rm BMPV} \Big) \,,
\end{eqnarray}
where
\begin{equation}
J_{\Delta} ~\equiv~  {R^2 L^4 \over l_p^6}\, \Big(\sum n^I \Big)\,.
\end{equation}
The entropy of the ring is:
\begin{equation}
S ~= ~ {2 \pi A \over \kappa_{11}^2} ~=~  \pi\, \sqrt{\cal M}
\label{BRent}
\end{equation}
where
\begin{eqnarray}
{\cal M} &~ \equiv~ &  2\, n^1 n^2 \Nb_1 \Nb_2 + 2 \,n^1 n^3 \Nb_1 \Nb_3
+  2\, n^2 n^3 \Nb_2 \Nb_3 -  (n^1 \Nb_1)^2 \nonumber \\
& & \quad -  (n^2 \Nb_2)^2  -  (n^3 \Nb_3)^2
 -  4\, n^1 n^2 n^3\, J_T  \,.\label{mdef}
\end{eqnarray}
and
\begin{equation}
J_T ~\equiv~  J_{\Delta} + n^I N^{\rm BH}_I ~=~{R^2 L^4 \over l_p^6}\,
\Big(\sum n^I \Big) ~+~  n^I N^{\rm BH}_I \,.
\end{equation}
As we will explain in more detail in section \ref{4d5d}, black rings
can be related to four-dimensional black holes, and (\ref{mdef}) is
the square root of the $E_{7(7)}$ quartic invariant of the microscopic
charges of the ring \cite{Bena:2004tk}; these microscopic charges are
the $n^I$, the $\Nb_I $ and the angular momentum $J_T$. More
generally, in configurations with multiple black rings and black
holes, the quantity multiplying $n^1 n^2 n^3$ in ${\cal M} $ should be
identified with the microscopic angular momentum of the ring.  There
are several ways to confirm that this identification is correct.
First, one should note that $J_T$ is the quantity that appears in the
near-horizon limit of the metric and, in particular, determines the
horizon area and hence entropy of the ring as in (\ref{BRent}).  This
means that $J_T$ is an intrinsic property of the ring. In the next
section we will discuss the process of lowering a black hole into the
center of a ring and we will see, once again, that it is $J_T$ that
represents the intrinsic angular momentum of the ring.

The angular momenta of the solution may be re-written in terms
of fundamental charges as:
\begin{eqnarray}
J_1 &~=~ & J_{T}  ~+~ \Big( \coeff{1}{6}\,C_{IJK} \, n^I n^J n^K  ~+~  \coeff{1}{2}\,
\Nb_I n^I  ~+~  J^{\rm BMPV}\Big)\, \cr
J_2 &~=~ & - \Big( \coeff{1}{6}\, C_{IJK}\,  n^I n^J n^K  ~+~
\coeff{1}{2}\, \Nb_I n^I ~+~ N^{\rm BH}_I n^I ~+~  J^{\rm BMPV} \Big) \,.
\end{eqnarray}
Notice that in this form, $J_1$ contains no contribution coming from
the combined effect of the electric field of the black hole and the
magnetic field of the black ring.  Such a contribution only appears in
$J_2$.

\subsection{Merging black holes and black rings}
\label{OffCenter}

One can also use the methods above to study processes in which black holes
and black rings are brought together and ultimately merge.   Such processes
are interesting in their own right, but we will also see later that they can be
very useful in the study of microstate geometries.

\begin{figure}
\centering
\includegraphics[height=4cm]{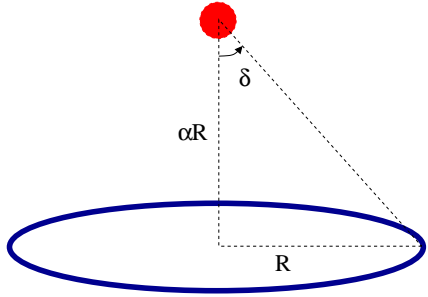}
\caption{The configuration  black ring with an off-set black hole on its axis.
The parameter,  $\alpha$, is related to the angle of approach, $\delta$, by
$\alpha \equiv \cot {\rm \delta}$.}
\label{BHring}
\end{figure}

It is fairly straightforward to generalize the solution of Section \ref{RoundRings}
to one that  describes a black ring with a black hole on the axis of the ring, but offset
above the ring by a distance, $a = \alpha  R$, where  $R$ is radius of
the ring. (Both $a$ and $R$ are measured in the $\IR^4$ base.)  This is depicted in
Fig. \ref{BHring}. The details of the exact solution may be found in \cite{Bena:2005zy}
and we will only  summarize the  main results here.

The total charge of the combined system is independent of $\alpha$ and
is given by (\ref{chargeinf}).  Similarly,  the entropy of the black ring is still given
by (\ref{BRent})  and (\ref{mdef}), but now with  $J_T$ defined by:
\begin{equation}
J_T ~=~  J_{\Delta} ~+~ {n^I N^{\rm BH}_I \over {1+\alpha^2}}
~\equiv~  {R^2 L^4 \over l_p^6}\, \Big(\sum n^I \Big) ~+~ {n^I N^{\rm BH}_I
\over {1+\alpha^2}}\,.
\label{JTnew}
\end{equation}
The horizon area of the black hole is unmodified by the presence of
the black ring and, in particular, its dependence on $\alpha$ only comes via
$J_T$.    Thus, for an adiabatic process, the quantity, ${\cal M}$, in  (\ref{mdef}) must
remain fixed, and therefore $J_T$ must remain fixed.  This is consistent
with identifying $J_T$ as the intrinsic angular momentum of the ring.

The two angular momenta of the system are:
\begin{eqnarray}
J_1 &~=~&  J_{T} ~+~ \Big(  \coeff{1}{6} \, C_{IJK} n^I n^J n^K  ~+~
\coeff{1}{2} \,  \Nb_I n^I  ~+~  J^{\rm BMPV} \Big)\, \label{Jonemerge} \,, \\
J_2 &~=~& - \Big(  \coeff{1}{6} \, C_{IJK} n^I n^J n^K  ~+~  \coeff{1}{2} \,
\Nb_I n^I ~+~ {N^{\rm BH}_I n^I  \over {1+\alpha^2}} ~+~  J^{\rm BMPV} \Big)\,.
 \label{Jtwomerge}
\end{eqnarray}
If we change the separation of the black hole and black ring while preserving the axial
symmetry, that is, if we vary $\alpha$, then  the symmetry requires   $J_1$ to be
conserved.  Once again we  see that this means that $J_T$ must remain fixed.

The constancy of $J_T$ along with (\ref{JTnew}) imply that as the black hole
is brought near the black ring, the embedding radius of the latter,
$R$, must change according to:
\begin{equation}
R^2 ~=~{ l_p^6  \over L^4 }\,
\,\big(\sum n^I \big )^{-1} \,  \Big(J_T - {n^I N^{\rm BH}_I \over
{1+\alpha^2}}\Big) \,.
\label{RJT}
\end{equation}
For fixed microscopic charges this formula gives the radius of the
ring as a function of the parameter $\alpha$.
The black hole will merge with the black ring if and only if $R$ vanishes
for some value of $\alpha$.   That is, if and only if
\begin{equation}
J_T  ~\le~   n^I N^{\rm BH}_I  \,.
\end{equation}
The vanishing of $R$ suggests that the ring is
pinching off, however, in the physical metric, (\ref{fivemetric}), the ring
generically has finite size as it settles onto the horizon of the black hole.
Indeed, the value of $ \alpha=\tan \zeta$ at the merger determines the latitude, $\zeta$,
at which the ring settles on the black hole.  If it occurs at $\alpha =0$ then the
ring merges by grazing the black hole at the equator.

At merger ($R=0$) one can see that $J_1 = J_2$ and so the resulting
object will have $J_{1} =J_{2 }$ given by (\ref{Jonemerge}).  This
will be a BMPV black hole and its electric charges are simply given by
(\ref{chargeinf}).  We can therefore use (\ref{BMPVent}) to determine
the final entropy after the merger.  Note that the process we are
considering is adiabatic up to the point where the ring touches the
horizon of the black hole.  The process of swallowing the ring is not
necessarily adiabatic, but we assume that the black hole does indeed
swallow the black ring and we can then compute the entropy from the
charges and angular momentum of the resulting BMPV black hole.

In general, the merger of a black hole and a black ring is
irreversible, that is, the total horizon area increases in the
process.  However, there is precisely one situation in which the
merger is reversible, and that requires {\it all} of the following to
be true:
\begin{enumerate}
\item The ring must have zero horizon area (with a slight abuse of terminology
we will also refer to such rings as {\it supertubes}).
\item The black hole that one begins with must have zero horizon area, {\it i.e.} it must be
{\it maximally spinning.}
\item The ring must meet the black hole by grazing it at the equator.
\item There are two integers, $\bar{P}$ and $P^{BH}$ such that
\begin{equation}
\Nb_I  ~=~ {\overline{P} \over n^I}    \qquad {\rm and} \qquad
 N^{BH}_I~=~   {P^{BH} \over n^I}  \,, \qquad I=1,2,3\,.
\end{equation}
\end{enumerate}
If all of these conditions are met then the end result is also a maximally spinning
BMPV black hole and hence also has zero horizon area.

Note that the last condition implies that
\begin{equation}
N_I ~\equiv~ \Nb_I + \coeff{1}{2}\, C_{IJK} \,n^J\,n^K
~=~ {(\Pb +  n^1\,n^2 \,n^3) \over n^I}  \,,
\end{equation}
and therefore the electric charges of black ring {\it and} its charges dissolved
in fluxes (${1 \over 2}\, C_{IJK} \,n^J\,n^K$) must both be aligned exactly
parallel to the electric charges of the black hole.
Conversely, if conditions 1--3 are satisfied, but the charge vectors
of the black hole and black ring are {\it not parallel} then the merger
will be irreversible.  This observation will be important in Section  \ref{Mergers}.

\section{Geometric interlude:  Four-dimensional black holes
 and five-dimensional foam}
\label{Geometry}

In Section \ref{SusyConfigs} we observed that supersymmetry allows us
to take the base-space metric to be any hyper-K\"ahler metric.  There
are certainly quite a number of interesting four-dimensional
hyper-K\"ahler metrics and in particular, there are the multi-centered
Gibbons-Hawking metrics.  These provide examples of asymptotically
locally Euclidean (ALE) and asymptotically locally flat (ALF) spaces,
which are asymptotic to $\IR^4/\ZZ_n$ and $\IR^3 \times S^1$
respectively.  Using ALF metrics provides a smooth way to transition
between a five-dimensional and a four-dimensional interpretation of a
certain configurations. Indeed, the size of the $S^1$ is usually a
modulus of a solution, and thus is freely adjustable. When this size
is large compared to the size of the source configuration, this
configuration is essentially five-dimensional; if the $S^1$ is small,
then the configuration has a four-dimensional description.

We noted earlier that a regular, Riemannian, hyper-K\"ahler metric
that is asymptotic to flat $\IR^4$ is necessarily flat $\IR^4$ {\it
globally.}  The non-trivial ALE metrics get around this by having a
discrete identification at infinity but, as a result, do not have an
asymptotic structure that lends itself to a space-time
interpretation. However, there is an unwarranted assumption here:
One should remember that the goal is for the five-metric
(\ref{fivemetric}) to be regular and Lorentzian and this might be
achievable if singularities of the four-dimensional base space were
canceled by the warp factors.  More specifically, we are going to
consider base-space metrics (\ref{fourmetric}) whose overall sign is
allowed to change in interior regions.  That is, we are going to
allow the signature to flip from $+4$ to $-4$.  We will call such
metrics {\it ambipolar}.

The potentially singular regions could actually be regular if the
warp factors, $Z_I$, all flip sign whenever the four-metric
signature flips.  Indeed, we suspect that the desired property may
follow quite generally from the BPS equations through the
four-dimensional dualization on the right-hand side of
(\ref{BPSeqn:2}).  Obviously, there are quite a number of details to
be checked before complete regularity is proven, but we will see
below that this can be done for ambipolar Gibbons-Hawking metrics.

Because of these two important applications, we now give a review of
Gibbons-Hawking geometries \cite{Hawking:1976jb, Gibbons:1979zt} and
their elementary ambipolar generalization.  These metrics have the
virtue of being simple enough for very explicit computation and yet
capture some extremely interesting physics.

\subsection{Gibbons-Hawking metrics}
\label{GHmetrics}

Gibbons-Hawking metrics have the form of a $U(1)$ fibration over a
flat $\IR^3$ base:
\begin{equation}
h_{\mu\nu}dx^\mu dx^\nu ~=~ V^{-1} \, \big( d\psi + \vec{A} \cdot d\vec{y}\big)^2  ~+~
 V\, (dx^2 +  dy ^2 +   dz^2) \,,
\label{GHmetric}
\end{equation}
where we write $\vec y =(x,y,z)$.  The function, $V$, is harmonic on
the flat $\IR^3$ while the connection, $A = \vec A \cdot d\vec{y} $,
is related to $V$ via
\begin{equation}
\vec \nabla \times \vec A ~=~ \vec \nabla V\,.
\label{AVreln}
\end{equation}
This family of metrics is the unique  set of  hyper-K\"ahler metrics with a
tri-holomorphic $U(1)$ isometry\footnote{Tri-holomorphic means that the
$U(1)$ preserves all three complex structures of the hyper-K\"ahler metric.}.
 Moreover,  four-dimensional hyper-K\"ahler manifolds with $U(1)\times
U(1)$ symmetry must, at least locally, be Gibbons-Hawking metrics with an extra $U(1)$
symmetry around an axis in the $\IR^3$ \cite{Gibbons:1987sp}.

In the standard form of the Gibbons-Hawking metrics one takes $V$ to have a
finite set  of isolated sources.   That is, let  $\vec{y}^{(j)}$ be  the positions of the
source points in the $\IR^3$  and let  $r_j \equiv |\vec{y}-\vec{y}^{(j)}|$.
Then one takes:
\begin{equation}
 V = \varepsilon_0 ~+~ \sum_{j=1}^N \,  {q_j  \over r_j} \,,
\label{Vform}
\end{equation}
where one usually takes $q_j \ge 0$ to ensure that the metric is
Riemannian (positive definite).   We will later relax this
restriction. There appear to be singularities in the metric at $r_j
=0$, however, if one changes to polar coordinates centered at $r_j
=0$ with radial coordinate to $\rho = 2 \sqrt{
|\vec{y}-\vec{y}^{(j)}|}$, then the metric is locally of the form:
\begin{equation}
ds_4^2 ~\sim~ d \rho^2 ~+~ \rho^2 \, d \Omega_3^2 \,,
\label{asympmet}
\end{equation}
where $d \Omega_3^2$ is the standard metric on $S^3/\ZZ_{|q_j|}$.
In particular, this means that one must have $q_j \in \ZZ$ and if $|q_j| = 1$
then the space looks locally like $\IR^4$.  If $|q_j| \ne 1$ then there
is an orbifold singularity, but since this is benign
in string theory, we will view such backgrounds as regular.

If $\varepsilon_0 \ne 0$, then $V \to \varepsilon_0$ at infinity and
so the metric (\ref{GHmetric}) is
asymptotic to flat $\IR^3 \times S^1$, that is, the base is
asymptotically locally flat (ALF). The five-dimensional space-time is
thus asymptotically compactified to a four-dimensional space-time.
This a standard
Kaluza-Klein reduction and the gauge field, $\vec  A$, yields a
non-trivial, four-dimensional Maxwell field whose sources, from the
ten-dimensional perspective, are simply D6 branes.  In Section
\ref{4d5d} we will  make extensive use of of the fact that introducing
a constant term into $V$ yields a further compactification and through this
we can relate five-dimensional physics to four-dimensional physics.

Now suppose that one has $\varepsilon_0 = 0$.  At infinity in $\IR^3$  one has
$V \sim q_0/r$, where $r \equiv |\vec y|$ and
\begin{equation}
q_0 ~\equiv~ \sum_{j=1}^N \, q_j \,.
\label{qzerodefn}
\end{equation}
Hence spatial infinity in the Gibbons-Hawking metric also has the form (\ref{asympmet}),
where
\begin{equation}
r ~=~ \coeff{1}{4}\, \rho^2 \,,
\label{rhorreln}
\end{equation}
and $d \Omega_3^2$ is the standard metric on $S^3/\ZZ_{|q_0|}$.
For the Gibbons-Hawking metric to be asymptotic to the positive definite, flat
metric on $\IR^4$ one must have $q_0 =1$.  Note that for the Gibbons-Hawking
metrics to be globally positive definite one would also have to take $q_j \ge 0$
and thus the only such metric would have to have $V \equiv {1 \over r}$.
The metric  (\ref{GHmetric})  is then the flat metric on $\IR^4$ globally, as can be
seen by using the change of variables  (\ref{rhorreln}).
The only way to get non-trivial metrics that are asymptotic to flat
$\IR^4$ is by taking some of the $q_j \in \ZZ$ to be negative.

\subsection{Homology and cohomology}
\label{homology}

\begin{figure}
\centering
\includegraphics[height=3cm]{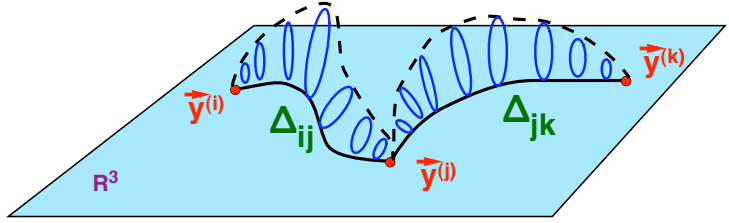}
\caption{This figure depicts some non-trivial cycles of the Gibbons-Hawking geometry.
The behaviour of the $U(1)$ fiber is shown along curves between the sources of
the potential, $V$. Here the fibers sweep out a pair of intersecting homology spheres.}
\label{GHcycles}
\end{figure}

The multi-center Gibbons-Hawking (GH) metrics also contain ${1 \over
  2} N(N-1)$ topologically non-trivial two-cycles, $\Delta_{ij}$, that
run between the GH centers.  These two-cycles can be defined by taking
any curve, $\gamma_{ij}$, between $\vec{y}^{(i)}$ and $\vec{y}^{(j)}$
and considering the $U(1)$ fiber of (\ref{GHmetric}) along the curve.
This fiber collapses to zero at the GH centers, and so the curve and
the fiber sweep out a $2$-sphere (up to $\ZZ_{|q_j|}$ orbifolds).  See
Fig. \ref{GHcycles}.  These spheres intersect one another at the
common points $\vec{y}^{(j)}$.  There are $(N-1)$ linearly independent
homology two-spheres, and the set $\Delta_{i\, (i+1)}$ represents a
basis\footnote{The integer homology corresponds to the root lattice of
  $SU(N)$ with an intersection matrix given by the inner product of
  the roots.}.

It is also convenient to introduce a set of frames
\begin{equation}
\hat e^1~=~ V^{-{1\over 2}}\, (d\psi ~+~ A) \,,
\qquad \hat e^{a+1} ~=~ V^{1\over 2}\, dy^a \,, \quad a=1,2,3 \,.
\label{GHframes}
\end{equation}
and two associated sets of two-forms:
\begin{equation}
\Omega_\pm^{(a)} ~\equiv~ \hat e^1  \wedge \hat
e^{a+1} ~\pm~ \coeff{1}{2}\, \epsilon_{abc}\,\hat e^{b+1}  \wedge
\hat e^{c+1} \,, \qquad a =1,2,3\,.\
\label{twoforms}
\end{equation}
The two-forms, $\Omega_-^{(a)}$, are anti-self-dual,  harmonic and
non-normalizable  and they define the
hyper-K\"ahler  structure on the base.  The forms, $\Omega_+^{(a)}$, are
self-dual and can be used to construct harmonic fluxes that are dual to the
two-cycles.  Consider  the self-dual two-form:
\begin{equation}
\Theta ~ \equiv~\sum_{a=1}^3 \, \big(\partial_a \big( V^{-1}\, H \big)\big) \,
\Omega_+^{(a)} \,.
\label{harmtwoform}
\end{equation}
Then $\Theta$ is closed (and hence co-closed and harmonic) if and
only if $H$ is harmonic in $\IR^3$, {\it i.e.}  $\nabla^2 H =0$. We
now have the choice of how to distribute sources of $H$ throughout the
$\IR^3$ base of the GH space; such a distribution may correspond to
having multiple black rings and black holes in this space.
Nevertheless, if we want to obtain a geometry that has no
singularities and no horizons, $\Theta$ has to be regular, and this
happens if and only if $H/V$ is regular; this occurs if and only if
$H$ has the form:
\begin{equation}
H ~ =~ h_0 ~+~ \sum_{j=1}^N \,  {h_j  \over r_j} \,.
\label{Hform}
\end{equation}
Also note that the ``gauge transformation:''
\begin{equation}
H ~ \to~ H ~+~ c\, V \,,
\label{gaugetrf}
\end{equation}
for some constant, $c$, leaves $\Theta$ unchanged, and so there
are only $N$ independent parameters in $H$.  In addition, if
$\varepsilon =0$ then one must take $h_0 =0$ for $\Theta$ to remain
finite at infinity.  The remaining $(N-1)$  parameters then describe
harmonic forms that are dual to the non-trivial two-cycles.
If $\varepsilon \ne 0$ then  the extra parameter is that of a Maxwell field whose
gauge potential gives the Wilson line around  the $S^1$ at infinity.

\exbox{ Show that the two-form, $\Theta$, defined by (\ref{harmtwoform}) and
(\ref{Hform}) is normalizable on standard GH spaces (with  $V>0$ everywhere).
That is, show that $\Theta$  square integrable:
\begin{equation}
\int  \Theta \wedge \Theta ~<~ \infty \,,
\end{equation}
where the integral is taken of the whole GH base space.}

It is straightforward to find a {\it local} potential  such that $\Theta = dB$:
\begin{equation}
B ~\equiv~  V^{-1}\,   H  \, (d\psi ~+~ A) ~+~  \vec{\xi} \cdot d \vec y \,,
\label{Bpot}
\end{equation}
where
\begin{equation}
\vec  \nabla \times \vec \xi  ~=~ - \vec \nabla H \,.
\label{xidefn}
\end{equation}
Hence, $\vec \xi$ is a vector potential for magnetic monopoles
located at the singular points of $H$.

To determine how these fluxes thread the two-cycles we
need the explicit forms for the vector potential, $B$,  and to find these we first need
the vector fields, $\vec v_i$, that satisfy:
\begin{equation}
\vec \nabla \times \vec v_{i} ~=~  \vec \nabla\, \bigg( {1\over r_i} \bigg) \,.
\label{vieqn}
\end{equation}
One then has:
\begin{equation}
\vec A ~=~  \sum_{j=1}^N \,  q_j \, \vec v_j \,, \qquad
\vec \xi  ~=~  \sum_{j=1}^N \, h_j \, \vec v_j \,.
\label{Asoln}
\end{equation}
If we choose coordinates so that $\vec y^{(i)} ~=~ (0,0,a)$  and let $\phi$ denote
the polar angle in the  $(x,y)$-plane, then:
\begin{equation}
\vec v_{i} \cdot d \vec y~=~ \Big( {(z -a) \over r_i} ~+~ c_i \Big) \, d \phi \,,
\label{vzdefns}
\end{equation}
where $c_i$ is a constant.  The  vector field, $\vec v_i$, is regular away
from the $z$-axis, but has a Dirac string along the $z$-axis.  By choosing
$c_i$ we can cancel the string along the positive or negative
$z$-axis, and by moving the axis we can arrange these strings  to run in
any  direction we choose, but they must start or finish at some $\vec y^{(i)}$,
or run out to infinity.

Now consider what happens to $B$ in the neighborhood of
$\vec y^{(i)}$.  Since the circles swept out by $\psi$ and $\phi$
are shrinking to zero size,  the string singularities near $\vec y^{(i)}$ are
of the form:
\begin{equation}
B~\sim~ { h_i  \over q_i} \, \Big (d \psi  +
q_i \,\Big( {(z -a) \over r_i}  +  c_i \Big) \, d \phi \Big)  ~-~
h_i \, \Big( { (z -a) \over r_i}  + c_i \Big) \, d \phi   ~\sim~
 {h_i  \over q_i} \, d \psi \,.
 \label{Basymp}
\end{equation}
This shows that the vector, $\vec \xi $, in (\ref{Bpot}) cancels the
string  singularities in the $\IR^3$.   The singular components of $B$ thus point
along the  $U(1)$ fiber of the GH metric.

Choose any curve, $\gamma_{ij}$,  between $\vec{y}^{(i)}$  and
$\vec{y}^{(j)}$ and define the two-cycle, $\Delta_{ij}$, as in Fig. \ref{GHcycles}.  If one has $V>0$ then
the vector field, $B$, is regular over the  whole of $\Delta_{ij}$ except at the end-points,
$\vec{y}^{(i)}$   and  $\vec{y}^{(j)}$.  Let $\widehat \Delta_{ij}$ be the
cycle $\Delta_{ij}$ with the poles excised.
Since   $\Theta$ is regular at the poles, then the expression for
the flux, $\flux_{ij}$, through $\Delta_{ij}$ can be obtained as follows:
\begin{eqnarray}
\flux_{ij} &~\equiv~ & {1 \over 4\, \pi}\,
\int_{\Delta_{ij}} \,
\Theta ~=~ {1 \over 4\, \pi}\, \int_{\widehat \Delta_{ij}} \,
\Theta ~=~
{1 \over 4\, \pi}\, \int_{\partial \widehat  \Delta_{ij}} \, B  \nonumber \\ &~=~&
{1 \over 4\, \pi}\,  \int_0^{4\pi} \,  d \psi \, \big( B |_{y^{(j)}} ~-~
 B |_{y^{(i)}}  \big) ~=~   \bigg( { h_j  \over q_j} ~-~
 { h_i \over q_i} \bigg) \,.
 \label{basicflux}
 \end{eqnarray}
We have normalized these periods for later convenience.

On an ambipolar GH space where the cycle runs between positive and
negative GH points, the flux, $\Theta$, and the potential $B$ are both
singular when $V=0$ and so this integral is a rather formal object.
However, we will see in Section \ref{BubbleEqnsSect} that when we
extend to the five-dimensional metric, the physical flux of the
complete Maxwell field combines $\Theta$ with another term so that the
result is completely regular.  Moreover, the physical flux through the
cycle is still given by (\ref{basicflux}).  We will therefore refer to
(\ref{basicflux}) as the magnetic flux even in ambipolar metrics and
we will see that such fluxes are directly responsible for holding up
the cycles

\section{Solutions on a Gibbons-Hawking base}
\label{GHsols}

\subsection{Solving the BPS equations}
\label{BPSequations}

Our task now is to solve the  BPS equations
(\ref{BPSeqn:1})--(\ref{BPSeqn:3}) but now with a Gibbons-Hawking
base metric.   Such solutions have been derived before for
positive-definite Gibbons-Hawking metrics \cite{Gauntlett:2002nw,
Gauntlett:2004qy}, and it is trivial to generalize to the ambipolar
form. For the present we will not impose any conditions on the
sources of the BPS equations.

In Section \ref{homology} we saw that there was a simple way to
obtain self-dual two-forms, $\Theta^{(I)}$, that satisfy
(\ref{BPSeqn:1}).  That is, we introduce three harmonic functions,
$K^I$,  on $\IR^3$ that satisfy ${\bf \nabla}^2  K^I = 0$, and define
$\Theta^{(I)}$ as in (\ref{harmtwoform}) by replacing $H$ with
$K^I$. We will not, as yet, assume any specific form for $K^I$.

\exbox{Substitute these two-forms into
(\ref{BPSeqn:2}) and show that the resulting equation has the
solution:
\begin{equation}
Z_I ~=~ \coeff{1}{2}  \, C_{IJK} \, V^{-1}\,K^J K^K  ~+~ L_I \,,
\label{ZIform}
\end{equation}
where the $L_I$ are three more independent harmonic functions.}

We now write the one-form, $k$, as:
\begin{equation}
k ~=~ \mu\, ( d\psi + A   ) ~+~ \omega
\label{kansatz}
\end{equation}
and then (\ref{BPSeqn:3}) becomes:
\begin{equation}
\vec \nabla \times \vec \omega ~=~  ( V \vec \nabla \mu ~-~
\mu \vec \nabla V ) ~-~ \, V\, \sum_{I=1}^3 \,
 Z_I \, \vec \nabla \bigg({K^I \over V}\bigg) \,.
\label{roteqn}
\end{equation}
Taking the divergence yields the following equation for $\mu$:
\begin{equation}
\nabla^2 \mu ~=~ \, V^{-1}\, \vec \nabla \cdot
\bigg( V \sum_{I=1}^3 \, Z_I ~\vec \nabla {K^I \over V} \bigg) \,,
\label{mueqn}
\end{equation}
which is solved by:
\begin{equation}
\mu ~=~ \coeff {1}{6} \, C_{IJK}\,  {K^I K^J K^K \over V^2} ~+~
{1 \over 2 \,V} \, K^I L_I ~+~  M\,,
\label{mures}
\end{equation}
where $M$ is yet another harmonic function on $\IR^3$.  Indeed, $M$
determines the anti-self-dual part of $dk$ that cancels out of
 (\ref{BPSeqn:3}). Substituting this result for $\mu$ into
 (\ref{roteqn})  we find that $\omega$ satisfies:
\begin{equation}
\vec \nabla \times \vec \omega ~=~  V \vec \nabla M ~-~
M \vec \nabla V ~+~   \coeff{1}{2}\, (K^I  \vec\nabla L_I - L_I \vec
\nabla K^I )\,.
\label{omegeqn}
\end{equation}
The integrability condition for this equation is simply the fact
that the divergence of both sides vanish, which is true because
$K^I, L_I, M$ and $V$ are harmonic.

\subsection{Some properties of the solution}

The solution is thus characterized by the harmonic functions
$K^I, L_I$, $V$ and $M$.  The gauge invariance, (\ref{gaugetrf}), extends
in a straightforward manner to the complete solution:
\begin{eqnarray}
K^I &~\rightarrow~&   K^I ~+~  c^I\, V \,, \nonumber \\
L_I &~\rightarrow~& L_I ~-~ C_{IJK}\,c^J \,K^K ~-~ \coeff{1}{2} \,
C_{IJK}\, c^J\, c^K\, V  \nonumber \,,\\
M &~\rightarrow~&M - \coeff{1}{2} \,c^I \, L_I +{1 \over 12}\, C_{IJK} \left( V \, c^I \, c^J
\, c^K +3\,  c^I\, c^J\, K^K\right) \,,
\label{fullgauge}
\end{eqnarray}
where the $c^I$ are three arbitrary constants\footnote{Note that
this gauge invariance exists for any $C_{ I J K}$, not only for those
coming from reducing M-theory on $T^6$.}.

The eight functions that give the solution may also be identified
with the eight independent parameters in the {\bf 56} of the $E_{7(7)}$ duality
group in four dimensions:
\begin{eqnarray}
x_{12} &~=~& L_1 \,, \qquad x_{34} ~=~ L_2 \,,
\qquad x_{56} ~=~ L_3  \,, \qquad x_{78} ~=~ - V  \,,  \nonumber \\
y_{12}  &~=~ & K^1 \,, \qquad y_{34} ~=~ K^2 \,,
\qquad y_{56} ~=~ K^3  \,, \qquad y_{78} ~=~ 2\, M \,.
\label{Esevenreln}
\end{eqnarray}
With these identifications, the right-hand side of
(\ref{omegeqn}) is the symplectic invariant of the {\bf 56} of
$E_{7(7)}$:
\begin{equation}
\vec \nabla \times \vec \omega ~=~   \coeff{1}{4}\,  \sum_{A,B = 1}^8 \,
(y_{AB}  \vec\nabla x_{AB} ~-~ x_{AB}  \vec\nabla y_{AB} )\,.
\label{newomega}
\end{equation}
We also note that the quartic invariant of  the {\bf 56} of $E_{7(7)}$
is determined by:
\begin{eqnarray}
J_4 &~=~& -{1 \over 4}(x_{12}y^{12} + x_{34}y^{34}
+x_{56}y^{56}+x_{78}y^{78})^2-  x_{12}x_{34}x_{56}x_{78} \nonumber \\
& & - y^{12}y^{34}y^{56}y^{78}   + x_{12}x_{34}y^{12}y^{34}
+ x_{12}x_{56}y^{12}y^{56} + x_{34}x_{56}y^{34}y^{56}  \nonumber \\
& & +x_{12}x_{78}y^{12}y^{78}+ x_{34}x_{78} y^{34}y^{78}
+ x_{56}x_{78}y^{56}y^{78} \,,
\label{quarticE}
\end{eqnarray}
and we will see that this plays a direct role in the  expression for
the scale of the $U(1)$ fibration. It also plays a central role in
the expression for the horizon area of a four-dimensional black hole
\cite{kallosh-kol}.

In principle we can choose the harmonic functions $K^I,L_I$ and $M$ to
have sources that are localized anywhere on the base. These solutions
then have localized brane sources, and include, for example,
supertubes and black rings in Taub-NUT \cite{Bena:2005ay,
  Gaiotto:2005xt, Elvang:2005sa, Bena:2005ni}, which we will review in
Section \ref{BlackTaub}.   Such solutions
also include  more general multi-center
black hole configurations in four dimensions, of the type considered
by Denef and collaborators \cite{denef}.

Nevertheless, our focus  for the moment is on obtaining smooth horizonless
solutions, which correspond to microstates of black holes and black
rings and we choose the harmonic functions so that there are no
brane charges anywhere, and all the charges come from the smooth
cohomological fluxes that thread the non-trivial cycles.

\subsection{Closed time-like curves}
\label{CTCsect}

To look for the presence of closed time-like curves in the metric one
considers  the space-space components of the metric given by (\ref{elevenmetric}),
(\ref{fivemetric}) and  (\ref{GHmetric}).  That is, one goes to the space-like slices
obtained by taking $t$ to be a constant. The $T^6$ directions immediately
yield the requirement that $Z_I Z_J > 0$ while the metric on the
four-dimensional base reduces to:
\begin{eqnarray}
ds^2_4   &= & - W^{-4}\, \big( \mu  (d \psi+ A ) + \omega  \big)^2 \nonumber \\
& & ~+~ {W^2 V^{-1}}\big( d\psi + A \big)^2 + W^2 V \big(dr^2 +
r^2 d\theta^2 + r^2 \sin^2 \theta \, d\phi^2\big)  \,,
 \label{dtzero}
\end{eqnarray}
where we have  chosen to write the metric on $\IR^3$ in terms of a generic
set of  spherical polar coordinates,  $(r,\theta, \phi)$ and where we have defined the
warp-factor, $W$, by:
\begin{equation}
W~\equiv~(Z_1\, Z_2\, Z_3)^{1/6} \,.
\label{Wdefn}
\end{equation}
There is some potentially singular behavior arising from the fact that
the $Z_I$, and hence $W$, diverge on the locus, $V=0$ (see
(\ref{ZIform})).  However, one  can show that if one expands
the metric (\ref{dtzero}) and uses the expression, (\ref{mures}), then
all the dangerous divergent terms cancel and the metric is regular.
We will discuss this further below and in Section \ref{CritSurfaces}.

Expanding (\ref{dtzero})  leads to:
\begin{eqnarray}
ds^2_4
& = &  W^{-4}\,  (W^6 V^{-1}-\mu^2 )\Big(  d\psi + A  - {\mu \,  \omega  \over
W^6 V^{-1}-\mu^2 }  \Big)^2  -  {W^2 \, V^{-1}   \over W^6 V^{-1}-\mu^2 } \,
\omega^2 \nonumber \\
& & ~+~  W^2 V \big(dr^2 + r^2 d\theta^2 + r^2 \sin^2 \theta\, d\phi^2\big)\nonumber \\
&= & {\cQ \over W^4 V^2} \Big( d\psi + A  - {\mu \, V^2 \over \cQ }\, \omega \Big)^2 ~+~
W^2 V \Big( r^2 \sin^2 \theta \, d \phi^2 -{\omega^2  \over \cQ} \Big) \nonumber \\
& & ~+~  W^2 V (dr^2 + r^2 d\theta^2) \,,
\label{dstil}
\end{eqnarray}
where we have introduced the quantity:
\begin{equation}
\cQ ~\equiv~  W^6\, V ~-~  \mu^2\,  V^2 ~=~  Z_1 Z_2 Z_3 V ~-~ \mu^2 \, V^2\,.
\label{Qdefn}
\end{equation}
Upon evaluating $\cQ$ as a function of the harmonic functions that determine
the solution one obtains a beautiful result:
\begin{eqnarray}
\cQ &~=~&  - M^2\,V^2   - \coeff{1}{3}\,M\,C_{IJK}{K^I}\,{K^J}\,{K^k} - M\,V\,{K^I}\,{L_I}
- \coeff{1}{4} \,(K^I L_I)^2 \nonumber \\
&& \quad +\coeff{1}{6} \, V C^{IJK}L_I L_J L_K +\coeff{1}{4} \,C^{IJK}
C_{IMN}L_J L_K K^M K^N
\label{QasEseven}
\end{eqnarray}
with $C^{IJK} \equiv C_{IJK}$.  We can straightforwardly see that when
we consider M-theory compactified on $T^6$, then $C^{IJK}=|\epsilon^{IJK}|$, and
$\cQ$   is nothing other than the $E_{7(7)}$ quartic invariant (\ref{quarticE}) where the
$x$'s and  $y$'s are identified as in (\ref{Esevenreln}). This is expected from
the fact that the solutions on a GH base have an extra $U(1)$ invariance, and
hence can be thought of as four-dimensional.  The four-dimensional supergravity
obtained by  compactifying M-theory on $T^7$ is $N=8$ supergravity, which has
an $E_{7(7)}$ symmetry group.  Of course, the analysis above and in
particular equation (\ref{QasEseven}) are valid for solutions of arbitrary five-dimensional
$U(1)^N$  ungauged supergravities on a GH base. More details on the explicit relation
for general theories can be found in \cite{Behrndt:2005he}.

\exbox{Check that $\cQ$ is invariant under the gauge
transformation (\ref{fullgauge}) }

Observe that (\ref{dstil}) only involves $V$ in the combinations $W^2 V$
and   $\cQ$ and both of these are regular as $V \to 0$.  Thus, at least the
spatial metric is  regular at $V=0$.  In Section \ref{CritSurfaces} we will show that
 the complete  solution is regular as one passes across the surface $V=0$.

From (\ref{dstil}) and (\ref{elevenmetric}) we see that to avoid CTC's, the following
inequalities must be true everywhere:
\begin{equation}
\cQ ~ \ge ~ 0\,, \quad W^2 \, V ~ \ge ~ 0\,, \quad
\big(Z_J \, Z_K \, Z_I^{-2}\big)^{1\over3} ~=~ W^2 Z_I^{-1}~ \ge ~ 0\,, \ \ I =1,2,3 \,.
\label{Qpos}
\end{equation}

The last two conditions can be subsumed into:
\begin{equation}
V\, Z_I ~=~  \coeff {1}{2}\, C_{IJK}\,  K^J \, K^K ~+~ L_I\,  V~ \ge ~ 0 \,,
\qquad  I = 1,2,3\,.
\label{VZpos}
\end{equation}
The obvious danger arises when $V$ is negative.   We will show in the next
sub-section that all these quantities remain finite and positive in a
neighborhood  of $V=0$, despite the fact that $W$ blows up. Nevertheless,
these quantities could possibly be negative away from the $V=0$ surface.
While we will, by no means, make a complete analysis of the positivity
of these quantities, we will discuss it further in Section \ref{CTCcomments}, and show
that (\ref{VZpos})  does not present a significant problem in a simple example.
One should also note that $\cQ \ge 0$ requires $\prod_I (V Z_I) \ge
\mu^2 V^4$, and so, given  (\ref{VZpos}),   the constraint  $\cQ \ge 0$
is still somewhat stronger.

Also note that there is a danger of CTC's arising from Dirac-Misner strings
in $\omega$.  That is, near $\theta=0, \pi$ the $-\omega^2$ term could be dominant
unless $\omega$ vanishes on the polar axis. We will analyze this issue completely
when we consider bubbled geometries in Section \ref{Bubbles}.

Finally, one can also try to argue \cite{Berglund:2005vb} that the complete
metric is stably causal and that the $t$ coordinate provides a global time function
\cite{HawkingEllis}.   In particular, $t$ will then be monotonic increasing on future-directed
non-space-like curves and hence there can be no CTC's.
The coordinate $t$ is a time function if and only if
\begin{equation}
- g^{\mu\nu} \partial_{\mu}t \, \partial_{\nu} t = - g^{tt} =  (W^2 V)^{-1}
(\cQ -  \omega^2) > 0\,,
\end{equation}
where $\omega$ is squared using the $\IR^3$ metric.
This is obviously a slightly stronger condition than $\cQ \ge 0$ in
(\ref{Qpos}).

\subsection{Regularity of the solution and critical surfaces}
\label{CritSurfaces}

As we have seen, the general solutions we will consider have
functions, $V$, that change sign on the $\IR^3$ base of the GH metric.
Our purpose here is to show that such solutions are completely
regular, with positive definite metrics, in the regions where $V$
changes sign.  As we will see the ``critical surfaces,'' where $V$
vanishes are simply a set of completely harmless, regular
hypersurfaces in the full five-dimensional geometry.

The most obvious issue is that if $V$ changes sign, then the overall
sign of the metric (\ref{GHmetric}) changes and there might be whole
regions of closed time-like curves when $V<0$.  However, we remarked
above that the warp factors, in the form of $W$, prevent this from
happening.  Specifically, the expanded form of the complete,
eleven-dimensional metric when projected onto the GH base yields
(\ref{dstil}). Moreover
\begin{equation}
W ^2 \, V ~=~ (Z_1\, Z_2 \, Z_3 \, V^3)^{1\over 3} ~\sim~
((K_1\, K_2 \, K_3)^2)^{1\over 3}
\label{WVsq}
\end{equation}
on the surface $V=0$. Hence $W ^2  V$ is regular and positive on
this surface, and therefore the space-space part (\ref{dstil}) of the full
eleven-dimensional metric is regular.

There is still the danger of singularities at $V=0$ for the other
background fields.  We first note that there is no danger of such
singularities being hidden implicitly in the $\vec \omega$ terms.
Even though (\ref{roteqn}) suggests that the source of $\vec \omega$
is singular at $V=0$, we see from (\ref{omegeqn}) that the source is
regular at $V=0$ and thus there is nothing hidden in $\vec \omega$. We
therefore need to focus on the explicit inverse powers of $V$ in the
solution.

The factors of $V$ cancel in the torus warp factors, which are
of the form $(Z_I Z_J Z_K^{-2})^{1 \over 3}$. The coefficient of $(dt
+ k)^2$ is $W^{-4}$, which vanishes as $V^2$. The singular part of the
cross term, $ dt \, k$, is $\mu \, dt \,(d \psi +A)$, which, from
(\ref{mures}), diverges as $V^{-2}$, and so the overall cross term,
$W^{-4} dt \, k$, remains finite at $V=0$.

So the metric is regular at critical surfaces.  The inverse metric is also regular
at $V=0$ because  the $dt \, d\psi$ part of the
metric remains finite and so the determinant is non-vanishing.

This surface is therefore not an event horizon even though the
time-like Killing vector defined by translations in $t$ becomes null
when $V = 0$. Indeed, when a metric is stationary but not static, the
fact that $g_{tt}$ vanishes on a surface does not make it an event horizon
(the best known example of this is the boundary of the ergosphere of
the Kerr metric). The necessary condition for a surface to be a
horizon is rather to have $g^{rr} = 0$, where $r$ is the coordinate
transverse to this surface. This is clearly not the case here.

Hence, the surface given by $V=0$ is like a boundary of an ergosphere,
except that the solution has no ergosphere\footnote{The non-supersymmetric
smooth three-charge solutions found in \cite{ross} do nevertheless
have ergospheres \cite{ross,rob}.} because this Killing vector is
time-like on both sides and does not change character across the  critical
surface.   In the Kerr metric the time-like Killing vector
becomes space-like and this enables energy extraction by the Penrose
process. Here there is no ergosphere and so energy extraction is not
possible, as is to be expected from a BPS geometry.

At first sight, it does appear that the Maxwell fields are singular on
the surface $V=0$.  Certainly the ``magnetic components,''
$\Theta^{(I)}$, (see (\ref{harmtwoform})) are singular when $V=0$.
However, one knows that the metric is non-singular and so one should
expect that the singularity in the $\Theta^{(I)}$ to be unphysical.
This intuition is correct: One must remember that the complete Maxwell
fields are the $A^{(I)}$, and these are indeed non-singular at $V=0$.
One finds that the singularities in the ``magnetic terms'' of
$A^{(I)}$ are canceled by singularities in the ``electric terms'' of
$A^{(I)}$, and this is possible at $V=0$ precisely because $g_{tt}$ goes to
zero, and so the magnetic and
electric terms can communicate.  Specifically, one has, from
(\ref{Thetadefn}) and (\ref{Bpot}):
\begin{equation}
d A^{(I)}~=~ d\, \bigg(B^{(I)} ~-~ {(dt + k) \over Z_I}\bigg) \,.
\label{dAI}
\end{equation}
Near $V=0$ the singular parts of this behave as:
\begin{eqnarray}
d A^{(I)} &~\sim~& d\, \bigg({ K^I \over V} ~-~ { \mu \over Z_I}\bigg) \,
(d \psi +A) \nonumber \\
& ~\sim~& d\, \bigg({ K^I \over V} ~-~ { K^1 \,  K^2 \,K^3
\over  \coeff{1}{2}\,V\, C_{IJK}\,   K^J \,K^K}\bigg) \, (d \psi +A)~\sim~0\,.
\label{dAIform}
\end{eqnarray}
The cancellations of the $V^{-1}$ terms here occur for much the same
reason that they do in the metric (\ref{dstil}).

Therefore, even if $V$ vanishes and changes sign and the base metric
becomes negative definite, the complete eleven-dimensional solution is
regular and well-behaved around the $V=0$ surfaces.
It is this fact that gets us around the uniqueness
theorems for asymptotically Euclidean self-dual (hyper-K\"ahler) metrics
in four dimensions, and as we will see, there are now a vast number of
candidates for the base metric.

\subsection{Black rings in Taub-NUT}
\label{BlackTaub}

Having analyzed the general form of solutions with a GH base, it is
interesting to  re-examine the black ring solution of
Section \ref{RoundRings} and rewrite it in the   form discussed in
Section \ref{BPSequations} with a trivial GH base (with  $V = {1 \over r}$).
We do this because it is then elementary to generalize the solution
to more complicated base spaces and most particularly to a Taub-NUT base.
This will then illustrate a very important technique that makes it
elementary to further compactify solutions to four-dimensional space-times
and establish the relationship between four-dimensional and five-dimensional
quantities.    For pedagogical reasons, we will focus on the
metric; details on the field strengths and the moduli can be found in
\cite{Bena:2005ni}.

\exbox{Show that the black ring warp factors and rotation vector, when written in
usual $\IR^4$ coordinates
\be
\label{R4coords}
ds^2= d\rt^2 + \rt^2(d\thetat^2 + \sin^2 \thetat
d\psit^2 +\cos^2 \thetat d\phit^2)
\ee
are given by:
\bea  \label{zaaaa}
Z_I & =& 1 + {\Qb_I
\over \Sigmat} + {1 \over 2} C_{IJK} q^J q^K {\rt^2 \over \Sigmat^2} \cr
k & =& -{\rt^2 \over 2 \Sigmat^2} \left(q^I \Qb_I + {2 q^1 q^2 q^3
\rt^2 \over \Sigmat}\right)(\cos^2 \thetat \, d\phit + \sin^2
\thetat \, d\psit) \cr
&-&  J_T \, {2 \rt^2 \sin^2 \thetat\over \Sigmat
(\rt^2 +\Rt^2 +\Sigmat)}\, d\psit \,,
\eea
where $C_{IJK}=1$ for $(IJK)=(123)$ and permutations thereof,
\be  \label{zbbbb}
{ \Sigmat \equiv \sqrt{(\rt^2-\Rt^2)^2 + 4 \Rt^2 \rt^2 \cos^2
\thetat }\,,}
\ee
and $J_T \equiv J_{\psit} -J_{\phit}~$. }

In the foregoing, we have written the solution in terms of the ring charges, $\Qb_I$ and,
as we have already noted,  for the five-dimensional  black ring these charges
differ from the charges measured at infinity
because of the charge ``dissolved'' in the M5-brane fluxes.  The charges measured at
infinity are   $Q_I = \Qb_I + {1 \over 2} C_{IJK}q^J q^K$.
We will also make a convenient choice of units in which
$G_5 = {\pi \over 4}$, and choose the three $T^2$'s  of the M-theory  metric
 to have equal size.

 \exbox{Show that in these units the charges $Q_I$, $\Qb_I$ and $q^I$
   that appear in the supergravity warp factors are the
   same as the corresponding quantized brane charges. \medskip \\
   Hint 1: Begin by relating $G_{11}$ and $G_5$ using the torus volumes.
\medskip\\
   Hint 2: You can cheat and use the relation between the charges in
   the supergravity formula and the integer quantized charges derived
   in \cite{Bena:2004de} and summarized in (\ref{translation}).  If
   you feel like doing honest character-building work, find the M2
   charges by integrating $F_7$ over the corresponding $S^3 \times
   T^2\times T^2$ at infinity; find the M5 charges by integrating
   $F_4$ over the corresponding $T^2 \times S^2$, where the $S^2$ goes
   around the ring. \medskip \\
   Hint 3: You can find the M5 dipole charges most easily if you use a coordinate
   system centered at $ \Sigma=0 $ described in equation (\ref{yb})
   below.}

From (\ref{RJT}), the radius of the ring, $\Rt$, and is related to $J_T$ by
\be  \label{jtt}
{J_T = (q^1 +q^2 +q^3) \Rt^2 .}
\ee

We now perform a change of coordinates, to bring the black ring to
a form that can easily be generalized to Taub-NUT.  Define
\be  \label{eb}
 \phi  = \phit-\psit ,\quad \psi =2\psit , \quad
\theta = 2\thetat, \quad \rho = {\rt^2 \over 4 }\,,
\ee
where the ranges of these coordinates are given by
\be
\label{ec}
\theta   \in(0,\pi), \quad (\psi,\phi) \cong (\psi +4\pi,\phi)
\cong( \psi, \phi+2\pi)~.
\ee
%

\exbox{Verify that when $V= {1\over \rho}$, the coordinate change (\ref{eb})
transforms the metric in the first line of (\ref{ze}) to that of flat $\IR^4$.}

In the new coordinates the black-ring metric is
\bea
\label{zd}
 ds^2 &=& -(Z_1 Z_2 Z_3)^{-2/3}(dt+k)^2+ (Z_1 Z_2
Z_3)^{1/3} h_{mn}dx^m dx^n  \,, \cr
Z_I & =& 1 + {\Qb_I \over
4\Sigma} + {1 \over 2} C_{IJK} q^J q^K {\rho \over 4\Sigma^2} \,, \cr
k&= & \mu\left(d\psi+(1+\cos \theta) d\phi \right)+\omega  \,, \cr
\mu & =& -{1 \over
16}{\rho \over \Sigma^2} \left(q^I\Qb_I + {2 q^1 q^2 q^3 \rho
\over \Sigma}\right) +
{J_T \over 16 R}
\left(1-{\rho\over \Sigma}-{R \over \Sigma}\right) \,, \cr
\omega & = &-
 {J_T \rho \over 4 \Sigma (\rho +R + \Sigma)} \sin^2\theta d\phi\,,
 \eea
with
\bea  \label{ze}
h_{mn}dx^m dx^n &=& V^{-1}\left(d\psi+ (1+\cos \theta)
d\phi\right)^2 +V(d\rho^2 + \rho^2(d\theta^2 +\sin^2 \theta d\phi^2))\,, \cr
 V &=&{1\over \rho}\,, \quad  \Sigma = \sqrt{\rho^2 +R^2 + 2R \rho\cos\theta} \,,
\quad R = {\Rt^2 \over 4}~.
\eea
%

\exbox{Check that the solution (\ref{zd}) has the
form  described in Section \ref{BPSequations} with the
eight harmonic functions:
\bea
 K^I &  =&  -{q^I \over 2 \Sigma} \,, \qquad
L_I  = 1+{\Qb_I \over 4  \Sigma}\,, \cr
M &=& {J_T \over 16}\left({1 \over R}- {1 \over \Sigma}\right)
\,, \qquad  V ={1 \over \rho} \,.
 \label{ej}
\eea
}

We should also note for completeness that the
conventions we use here for these harmonic functions are those of
\cite{Bena:2005ni}, and differ
from those of \cite{Gauntlett:2004qy} by various factors of two.
When $\IR^4 $ is written in Gibbons-Hawking form, the ring is
sitting at a distance $R$ along the negative $z$-axis of the
three-dimensional base. Adding more sources on the $z$ axis corresponds to
making concentric black rings \cite{concentric, Gauntlett:2004qy}.

\exbox{Show that adding sources on the same side of the origin in the
  $\IR^3$ base of (\ref{ze}), correspond to rings that sit in the same
  $\IR^2$ inside $\IR^4$. Show that rings that sit in orthogonal
  $\IR^2$'s inside $\IR^4$ correspond to sources sitting on opposite
  sides of the origin of the $\IR^3$ base of (\ref{ze}).}

To change the four-dimensional base metric into Taub-NUT one simply
needs to add a constant, $h$,  to the harmonic function $V$:
\be
  V ~=~ h+{1 \over \rho} \,.
 \label{zg}
\ee
Since the functions in the metric are harmonic, equations
(\ref{ZIform}), (\ref{kansatz}), (\ref{mures}), (\ref{omegeqn}) and (\ref{ej}),
still imply that we have a supersymmetric solution.  Actually, in order to avoid both Dirac
string singularities and closed time-like curves, the relation
(\ref{jtt}) between $J_T$ and the dipole charges must be modified to:
\be  \label{rtnew}
{J_T  \left(h+{1 \over R}\right) = 4 (q^1 +q^2 +q^3) \,.}
\ee
This is discussed in detail in
\cite{Bena:2005ni,Elvang:2005sa,Gaiotto:2005xt} and in later sections here,
but it follows because the absence of singularities in $\omega$ puts
constraints on the sources on the right hand side of (\ref{omegeqn}).

For small ring radius (or for small $h$), $R\ll h^{-1}$, this reduces to
the five-dimensional black  ring described earlier.  We now wish to
consider the opposite limit, $R\gg h^{-1}$. However,  to keep ``the same ring''
we must to keep all its quantized charges fixed, and so  (\ref{rtnew})
means $h+{1 \over R}$ must remain constant.  We can think of
this as keeping the physical radius of the ring fixed while
changing its position in Taub-NUT:  The ring slides to a point where the
physical ring radius is the same as the physical size of the compactification circle.
In the limit where $R$ is large, the black ring is far from the Taub-NUT
center and it is effectively wrapped around an infinite cylinder.
In other words, it has become a straight black string  wrapped on a circle
and, from the four-dimensional perspective, it is  point-like and is
nothing but a four-dimensional black hole.

To see this in more detail, we consider the geometry in the region
far from the tip, that is, for $\rho \gg 1$, where  we can take
$V=h$.  We also want to center the three-dimensional spherical
coordinates on the ring, and so we change to coordinates such that
$\Sigma$ is the radius away from the ring.  We then have:
\be
\label{yb}{ d\rho^2 + \rho^2(d\theta^2 +\sin^2 \theta  d\phi^2 ) = d\Sigma^2 +
\Sigma^2 (d\thetah^2 + \sin^2 \thetah d\phi^2)~,}
\ee
 and
\be  \label{ya}{ \rho = \sqrt{\Sigma^2 + R^2 -2R\Sigma \cos\thetah}, \quad
\cos \theta = {\Sigma \cos \thetah - R \over \rho}~.}
\ee

Taking $R \rightarrow \infty$, at fixed $(\Sigma,\thetah,\phih)$ and
$h+{1 \over R}$, we find that the metric is:
\begin{eqnarray}
 ds^2 &=& -(\tZ_1 \tZ_2 \tZ_3)^{-2/3}(d \tilde t+ \tilde \mu  d\psi)^2  \nonumber \\
&& \quad+~ (\tZ_1 \tZ_2 \tZ_3)^{1/3} \left(dr^2 + r^2(d\thetah^2 + \sin^2
\thetah d\phih^2) + d \psi^2 \right)\,, \label{yy}
\end{eqnarray}
where
\be  \label{coords}{  \tZ_I  \equiv {Z_I \over h} \,, \qquad \tilde \mu
\equiv {\mu \over h} \,, \qquad r    \equiv h \Sigma  \,, \qquad
\tilde t \equiv {t \over h} \,.}
\ee
Note that the spatial section of (\ref{yy}) is precisely $\IR^3 \times  S^1$.
When written in terms of the coordinate $r$ the  metric
functions become:
\be  \label{yz}{ \tZ_I   = {1 \over h} + {\Qb_I \over 4 r} +
{C_{IJK}q^Jq^K \over 8 r^2 } \,, \qquad \tilde \mu    =-{J_T \over
16 r} -{q^I \Qb_I \over 16 r^2} -{q^1 q^2 q^3 \over 8 r^3} \,, \qquad
\omega   =0\,.}
\ee
This is precisely the four-dimensional black hole  found by
wrapping the black string solution of \cite{Bena:2004wv}
on a circle.

As noted in \cite{Bena:2004tk}, the entropy of the five-dimensional
black ring takes a simple form in terms of the quartic invariant of $E_{7(7)}$:
\be  \label{ana}{ S = 2\pi \sqrt{J_4}~,}
\ee
where $J_4$ is given by  (\ref{quarticE}) with
\bea
x_{12}&=&\Qb_1, \quad x_{34}=\Qb_2, \quad
x_{56}=\Qb_3,\ \quad x_{78}=0, \cr  y^{12} &=& q^1, \quad
y^{34}=q^2, \quad y^{56} = q^3, \quad y^{78}= J_T = J_{\psit}
-J_{\phit}\,.
 \label{app}
\eea
Hence, the ``tube angular momentum,'' $J_T$, plays the role of
another charge in the four-dimensional black hole picture.  From the
five-dimensional perspective, $J_T$  is the difference of the two independent
angular momenta and is given by (\ref{jtt}).  Upon compactification on
the Taub-NUT circle, $J_T$ represents the momentum around that
circle and, as is very familiar in Kaluza-Klein (KK)  reduction,  a
KK momentum becomes a conserved charge in the lower dimension.

It has long been known that maximal supergravity in four dimensions
has $E_{7(7)}$ duality group and that the general entropy for the
corresponding class of four-dimensional black holes can be expressed
in terms of the quartic invariant \cite{kallosh-kol}.  The observation
in \cite{Bena:2004tk} thus provided the first clue as to the
relationship between five-dimensional black rings and four-dimensional
black holes.  We now examine this relationship in more detail.

\subsection{Parameters, charges and the ``4D-5D'' connection}
\label{4d5d}

As we have seen, the ability to introduce a constant, $h$, into $V$ as
in (\ref{zg}) enables us to interpolate between configurations in
five-dimensional space-time and configurations in four-dimensional
space-time.  For small $h$, the Taub-NUT circle is very large and the
configuration behaves as if it were in a five-dimensional space-time
while, for large $h$, the Taub-NUT circle is small and the
configuration is effectively compactified.  The first connection
between a five-dimensional configuration and such a four-dimensional
solution was made in \cite{Bena:2005ay}, where the simple two-charge
supertube \cite{supertube} was put in Taub-NUT, and was related to a two-centered,
four-dimensional configuration of the type previously analyzed in
\cite{denef}.  One can also consider a four-dimensional black hole
that has a non-trivial KKM charge, and that sits at the center of
Taub-NUT. When the KKM charge is one, this black hole also has two
interpretations, both as a four-dimensional and as a five-dimensional
black hole \cite{andy4d5d}.  Since one can interpolate between the
five-dimensional and the four-dimensional regimes by changing the
moduli of the solution, one can give microscopic descriptions of black
rings and black holes both from a four-dimensional perspective and
from a five-dimensional perspective. This is called the ``4D-5D''
connection.  This connection enables us to relate the
 parameters
and charges appearing in the five-dimensional description of a system to
those appearing in the four-dimensional description. We now examine
this more closely and we will encounter some important subtleties.  To
appreciate these, we need to recall some of the background behind the
BPS black ring solutions.

One of the reasons that makes the BPS black ring solution so interesting
is that it shows the failure of black-hole uniqueness in five
dimensions.  To be more specific, for the round ($U(1) \times U(1)$
invariant) BPS black ring solution there are only five conserved quantities:
The two angular momenta, $J_1$, $J_2$, and the three electric
charges, $Q_I$, {\it as measured from infinity.}  However, these rings are
determined by seven parameters: $\Qb_I$, $q^I$ and $J_T$.
We have seen how these
parameters are related to details of the constituent branes and we
have stressed, in particular, that the $q^I$ are dipole charges that,
{\it a priori}, are not conserved charges and so cannot be measured from
infinity in five dimensions.  As discussed in Section \ref{RoundRings}
the true conserved charges in five dimensions are non-trivial
combinations of the fundamental ``brane parameters,'' $\Qb_I$, $q^I$
and $J_T$.

\noindent{\bf $\bullet$~  5D dipole charges and 4D charges}.

In discussing the conserved charges of a system there is a very
significant assumption about the structure of infinity.  To determine
the charges one  integrates various field strengths and their duals on
certain Gaussian surfaces.  If one changes the structure of
infinity, one can promote dipoles to conserved charges or lose
conserved charges.  One sees this very explicitly in the case of Taub-NUT
space (\ref{ze}): by turning one a constant piece in harmonic function $V$,
one replaces the $S^3$ at infinity of $\IR^4$  by an $S^2 \times S^1$.
In particular, the ``dipole''
charges, $q^I$, of the five-dimensional black-ring become conserved
magnetic charges in the Taub-NUT space.  This is evident from the
identification in (\ref{app}) in which the $x^{AB}$ and $y^{AB}$
respectively represent conserved electric and magnetic charges
measured on the Gaussian two-spheres at infinity in $\IR^3$.  More
generally, from (\ref{Esevenreln}) we see that the leading behaviour
of each of the eight harmonic functions, $K^I$, $L_I$, $M$, and $V$
yields a conserved charge in the Taub-NUT compactification.

In terms of the thermodynamics of black holes and black rings, the
conserved charges measured at infinity are thermodynamic state
functions of the system and the set of state functions depends upon
the asymptotic geometry of the ``box'' in which we place the system.
If a solution has free parameters that cannot be measured by the
thermodynamic state functions then these parameters should be thought
of as special properties of a particular microstate, or set of
microstates, of the system. Thus, in a space-time that is asymptotic
to flat $\IR^{4,1}$, one cannot identify the microstates of a
particular round black-ring solution by simply looking at the charges
and angular momenta at infinity. Moreover, given a generic microstate
with certain charges it is not possible to straightforwardly identify
the black ring (or
rings) to which this microstate corresponds. The only situation in which
one can do this is when there exists a box in which one can place both
the ring and the microstate, and one uses the box to define extra
state functions that the two objects must share. Putting these objects in
Taub-NUT and changing the moduli such that both the ring and the microstate
have a four-dimensional interpretation, allows one to define a box that
can be used to measure the ``specialized
microstate structure'' (i.e. the dipoles), as charges at
infinity in four dimensions.

A good analogy is the thermodynamics and the kinetic
theory of gases.  The conserved charges correspond to the state
functions while the internal, constituent brane parameters correspond
to details of the motions of molecules in particular microstates.  The
state functions are non-trivial combinations of parameters of
microstate, but do not capture all the individual microstate
parameters.  If the box is a simple cube then there is no state
function to capture vorticity, but there is such a state function for
a toroidal box.

Thus solutions come with two classes of parameters: Those that are
conserved and can be measured from infinity and those that represent
particular, internal configurations of the thermodynamic system. There
are two ways in which one can hope to give a microscopic
interpretation of black rings. One is to take a near-horizon limit in
which the black ring solution becomes asymptotically $AdS_3 \times S^3
\times T^4$ \cite{blackring2,Bena:2004tk}, and to describe the ring in
the D1-D5-P CFT dual to this system. The other
is to focus on the near-ring geometry (or to put the ring in Taub-NUT)
and describe it as a four-dimensional BPS black hole
\cite{Bena:2004tk,CyrierHJ,Gaiotto:2005xt,Bena:2005ni}, using the
microscopic description of 4D black holes constructed in given in
\cite{MaldacenaDE,BertoliniYA}.

If one wants to describe black rings in the D1-D5-P CFT, it is, {\it a priori},
unclear how the dipole charges, which are not conserved charges (state
functions) appear in this CFT. A phenomenological proposal for this
has been put forth in \cite{Bena:2004tk}, but clearly more work
remains to be done. Moreover, the obvious partition functions that one
can define and compute
in this CFT \cite{farey}, which only depend on the charges and angular
momenta, cannot be compared to the bulk entropy of a particular black ring.
One rather needs to find the ring (or rings) with the largest entropy
for a given set of charges, and match their entropy to that computed in the CFT.

Moreover, if one wants to describe the ring using a CFT corresponding to a
four-dimensional black hole, it is essential to identify the correct M2 charges
of the ring. The beauty of the brane
description (or any other stringy description) of supertubes
and black rings is that it naturally points out what these charges are.

\noindent{\bf $\bullet$~ 5D electric charges and their 4D interpretation.}

There has been some discussion in the literature about the correct
identification of the charges of the black-ring system.  In
particular, there was the issue of whether the $Q_I$ or the $\Qb_I$
are the ``correct'' charges of the black-ring.  There is no dispute
about the charge measured at infinity, the only issue was the physical
meaning, if any, to the $\Qb_I$.  In \cite{Horowitz:2004je} it was
argued that the only meaningful charge was the ``Page charge'' that
measures $Q_I$ and not $\Qb_I$ even when the Gaussian surface is small
surface surrounding the black ring.  This is an interesting,
mathematically self-consistent view but it neglects a lot of the
important underlying physics.  It also generates some confusion as to
the proper identification of the microscopic charges of the underlying
system.  The competing view \cite{Bena:2004de} is the one we have
presented here: The $\Qb_I$ represent the number of constituent M2
branes and the $Q_I$ get two contributions, one from the $\Qb_I$ and
another from the ``charges dissolved in fluxes'' arising from the M5
branes.  It is certainly true that the $\Qb_I$ and the $q^I$ are not
conserved individually, but they do represent critically important
physical parameters.

This is easily understood in analogy with a heavy nucleus. The energy
of the nucleus has two contributions, one coming from the rest mass of
the neutrons and protons, and the other coming from the interactions
between them.  In trying to find the ``microscopic'' features of the
nucleus, like the number of nucleons, one obtains an incorrect result
if one simply divides the total energy by the mass of a nucleon. To
find the correct answer one should first subtract the energy coming
from interactions, and then divide the remainder by the mass of the
nucleon.

One of the nice features of Taub-NUT compactification and the
``4D-5D'' connection is that it provides a very simple resolution of
the foregoing issue in the identification of constituent microscopic
charges. If one simply compactifies M-theory on $T^6 \times S^1$ from
the outset, wrapping $q^I$ M5 branes on the $S^1$ and each of the tori
as shown in Table \ref{braneconfig}, then the $q^I$ simply emerge as
magnetic charges in four dimensions as in (\ref{app}).  Similarly, the
$\Qb_I$ are, unambiguously, the conserved electric charges of the
system.  This is also true of the Taub-NUT compactification of the
black ring and the fact that we can adiabatically vary $h$ in
(\ref{zg}) means we can bring the ring from a region that looks like
M-theory on $T^6 \times S^1 \times \IR^{3,1} $ into a region that
looks like M-theory on $T^6 \times \IR^{4,1} $, and still have
confidence that the identification is correct because M2 and M5 brane
charges are quantized and cannot jump in an adiabatic process.  This
establishes that the microscopic charges of the black ring are not the
same as the charges measured at infinity in the five-dimensional black
ring solution.

There are, of course, many situations where the rings cannot be put in
Taub-NUT, and one cannot obtain the microscopic charges using the
4D-5D connection. The simplest example is the black ring with a black
hole offset from is center \cite{Bena:2005zy} that we reviewed in
Section \ref{OffCenter}. However, based upon our experience with the
single black ring, we expect that the values of $\Qb_I$ in the
near-ring geometry will yield the number of M2 brane constituents of
each individual ring.

There has also been a proposal to understand the  entropy of
BPS black rings in terms of microscopic charges, in which $Q_I$ are
interpreted as  the M2 brane charges.  This is  based on a
four-dimensional black hole CFT with charges $Q_I$ rather than
$\Qb_I$, and with momentum $J_{\psi}$ rather than $J_T$
\cite{CyrierHJ}. In order to recover the entropy formula
(\ref{BRent},\ref{mdef})
an important role in that description was played by a non-extensive
zero point energy shift of $L_0$. In light of our analysis,
it is rather mysterious why this gives the right
entropy, since we have shown explicitly that the relevant
four-dimensional black hole CFT is the one with charges $\Qb_I$,
momentum $J_T$, and no zero point shift of $L_0$. We should also note
that the approach of \cite{CyrierHJ} seems to run into problems when
describing concentric black rings   because the total charge $Q_{A}$ is
not simply a sum of the individual $Q_{A,i}$, but gets contributions
from cross terms of the form $C_{ABC}~q_i^B~ q_j^C$.  The
approach of \cite{CyrierHJ} also appears not to correctly
incorporate some of the  higher
order corrections to the black ring entropy \cite{corrections,per-cor}.

One of the other benefits of the 4D-5D connection is that it also
unites what have been two parallel threads in research.  Prior to this
there had been extensive, and largely independent bodies of research
on four-dimensional objects and upon on five-dimensional objects.  It
is now evident that the four-dimensional two-center solution
corresponding to black rings and supertubes in Taub-NUT
\cite{Bena:2005ay, Gaiotto:2005xt, Elvang:2005sa, Bena:2005ni} is part
of the family of multi-center solutions that have been explored by
Denef and collaborators \cite{denef}. In fact, one can also imagine
putting in Taub-NUT multiple concentric black rings of the type
studied by Gauntlett and Gutowski in
\cite{concentric,Gauntlett:2004qy}. These descend in four dimensions
to a multi-black hole configuration, in which the center of the rings
becomes a center of KKM charge one, the black rings in one plane
become black holes on the right of the KKM center, and the black rings
in the other plane becomes black holes on the left of this center.

More generally, we expect that the 4D-5D connection will lead to a
valuable symbiosis.  For example, the work on attractor flows in
Calabi-Yau manifolds and the branching of these flows could have
important consequences for the bubbled geometries that we will discuss
in the next section.

\section{Bubbled geometries}
\label{Bubbles}

\subsection{The geometric transition}
\label{geomtrans}

The main purpose of our investigation thus far has been to construct
smooth horizonless geometries starting from three-charge supertubes.
We have seen that if one considers a process in which one takes a
three-charge, three-dipole charge supertube to a regime where the
gravitational back-reaction becomes important, the resulting
supergravity solution is generically that of a BPS black ring.
Although black rings are very interesting in their own right, they
do have event horizons and therefore do not correspond to microstates
of the boundary theory.

Hence it is natural to try to obtain microstates by starting with
brane configurations that do not develop a horizon at large effective
coupling, or alternatively to consider a black ring solution in the
limit where its entropy decreases and becomes zero. However, the
geometry of a zero-entropy black ring is singular. This singularity is
not a curvature singularity, since the curvature is bounded above by
the inverse of the dipole charges. Rather, the singularity is caused
by the fact that the size of the $S^1$ of the horizon shrinks to zero
size and the result is a ``null orbifold.'' One can also think about
this singularity as caused by the gravitational back-reaction of the
branes that form the three-charge supertube, which causes the $S^1$
 wrapped by these branes to shrink to zero size.

\begin{figure}
\centering
\includegraphics[width=6cm]{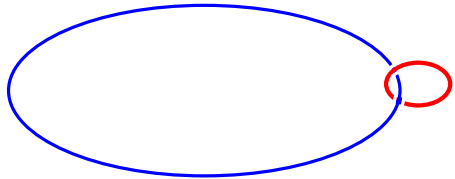}
\caption{ {\it Geometric transitions}: The branes wrap the large
(blue) cycle; the flux through the Gaussian (small, red) cycle
measures the brane charge. In the open-string picture the small (red) cycle
has non-zero size, and the large (blue) cycle is contractible. After the
geometric transition the size of the  large (blue) cycle becomes zero, while
the small (red) cycle becomes topologically non-trivial.}
\label{transition-i}
\end{figure}

 Fortunately, string theory is very good at solving this kind of
 singularities, and the mechanism by which it does is that of
 ``geometric transition.'' To understand what a geometric transition
 is, consider a collection of branes wrapped on a certain cycle. At
 weak effective coupling one can describe these branes by studying the
 open strings that live on them. One can also find the number of
 branes by integrating the corresponding flux over a ``Gaussian''
 cycle dual to that wrapped by the branes. However, when one increases
 the coupling, the branes back-react on the geometry, and shrink the
 cycle they wrap to zero size. At the same time, the ``Gaussian
 cycle'' becomes large and topologically non-trivial. (See Fig.
 \ref{transition-i}.)  The resulting geometry has a different
 topology, and {\it no brane sources}; the only information about the
 branes is now in the integral of the flux over the blown-up dual
 ``Gaussian cycle.'' Hence, even if in the open-string (weakly
 coupled) description we had a configuration of branes, in the
 closed-string (large effective coupling) description these branes
 have disappeared and have been replaced by a non-trivial topology
 with flux.

Geometric transitions appear in many systems \cite{Gopakumar:1998ki,
  Klebanov:2000hb, Vafa:2000wi, Lin:2004nb}.  A classic example of
such system are the brane models that break an $\cN=2$ superconformal
field theory down to an $\cN=1$ supersymmetric field theory \cite{
  Klebanov:2000hb, Cachazo:2001jy}.  Typically, the $\cN=2$
superconformal field theory is realized on a stack of D3 branes in
some Calabi-Yau compactification.  One can then break the
supersymmetry to $\cN=1$ by introducing extra D5 branes that wrap a
two-cycle. When one investigates the closed-string picture, the two
cycle collapses and the dual three-cycle blows up (this is also known
as a conifold transition).  The D5 branes disappear and are replaced
by non-trivial fluxes on the three-cycle.  The resulting geometry has
no more brane sources, and has a different topology than the one we
started with.

\begin{figure}
\centering
\includegraphics[width=5.6cm]{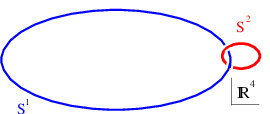}~~~~~~~~~
~\includegraphics[width=5.6cm]{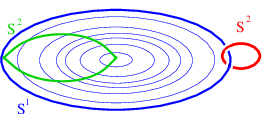}
\caption{{\it The geometric transition of the black ring:}  Before the
  transition the branes wrap the large (blue) $S^1$; the flux through the
Gaussian $S^2$ (small, red) cycle measures the brane charge. After the
transition the Gaussian $S^2$ (small, red) cycle is topologically non-trivial
and of finite size and a new (green) $S^2$ appears, coming from the fact that
the blue $S^1 $ shrinks to zero so that the disk spanning the $S^1$ becomes
an $S^2$. The
resulting geometry has two non-trivial $S^2$'s and no brane sources.}
\label{tran}
\end{figure}

Our purpose here is to see precisely how geometric transitions resolve
the singularity of the zero-entropy black ring ({\it supertube}) of
Section \ref{BlackRings}.   Here the ring wraps a curve
$y^\mu(\sigma)$, that is topologically an $S^1$ inside $\IR^4$.  (In
Fig. \ref{tran} this $S^1$ is depicted as a large, blue cycle.) The
Gaussian cycle for this $S^1$ is a two-sphere around the ring
(illustrated by the red small cycle in Fig. \ref{tran}). If one
integrates the field strengths $\Theta^{(I)}$ on the red Gaussian
two-cycle one obtains the M5 brane dipole charges of the ring, $n^I$.

After the geometric transition the large (blue) $S^1$ becomes of zero length,
and the red $S^2$ becomes topologically non-trivial. Moreover, because
the original topology is trivial, the curve $y^\mu(\sigma)$ was the
boundary of a disk. When after the transition this boundary curve
collapses, the disk becomes a (topologically non-trivial) two-sphere.
Alternatively, one can think about this two-sphere (shown in Fig.
\ref{tran} in green) as coming from having an $S^1$ that has zero size
both at the origin of the space $r=0$ and at the location of the ring.
Hence, before the transition we had a ring wrapping a curve
of arbitrary shape inside $\IR^4$, and after the transition we have a
manifold that is asymptotically $\IR^4$, and has two non-trivial
two-spheres, and no brane sources.

Can we determine the geometry of such a manifold? If the curve has an arbitrary
shape the only information about this manifold is that it is asymptotically
$\IR^4$ and that it is hyper-K\"ahler, as required by
supersymmetry\footnote{This might cause the faint-hearted to give up hope
because of the theorem that the only such manifold is flat $\IR^4$.
  This is the second instance when such theorems appear to preclude
  further progress in this research programme (the first is discussed
  at the end of Section \ref{supersection}). As in the previous
example, we will proceed guided by the string-theory intuition, and
  will find a way to avoid the theorem.}. If the curve wrapped by the
supertube has arbitrary shape, this is not enough to determine the
space that will come out after the geometric transition. However, if
one considers a circular supertube, the solution before the transition
has a $U(1) \times U(1) $ invariance, and so one naturally expects
the solution resulting from
the transition should also have this invariance.

With such a high level of symmetry  we do have enough information to determine
what the result of the geometric transition is: \\
$\bullet $ By a theorem of Gibbons and Ruback \cite{Gibbons:1987sp},
a hyper-K\"ahler manifold that has a $U(1) \times U(1) $ invariance must have
a translational $U(1)$ invariance and hence, must be Gibbons-Hawking. \\
$\bullet $ We also know that this manifold should have two non-trivial
two-cycles, and hence, as we have discussed in Section \ref{GHmetrics}
it should have three centers. \\
$\bullet $ Each of these centers must have integer GH charge.\\
$\bullet $ The sum of the three charges must be 1, in order for the
manifold to be asymptotically $\IR^4$.\\
$\bullet $ Moreover, we expect the geometric transition to be
something that happens locally near the ring, and so we expect the
region near the center of the ring (which is also the origin of our
coordinate system) to remain the same. Hence, the GH center at the
origin of the space must have charge + 1.

The conclusion of this argument is that the space that results from
the geometric transition of a $U(1) \times U(1)$ invariant supertube
must be a GH space with three centers, that have charges $1,~ +Q,~-Q$,
where $Q$ is any integer. As we have seen in Section  \ref{BlackTaub},
equation  (\ref{ej}), if we think about $\IR^4$ as a
trivial Gibbons-Hawking metric with $V = {1 \over r}$, the black ring solution
of Section \ref{RoundRings} has a GH center at the
origin, and the ring at a certain point on the $\IR^3$ base of the GH space.
In the ``transitioned'' solution, the singularity of the zero-entropy black ring is
resolved by the nucleation, or ``pair creation,'' of two equal and
oppositely charged GH points.

\begin{figure}
\centering
\includegraphics[width=11cm]{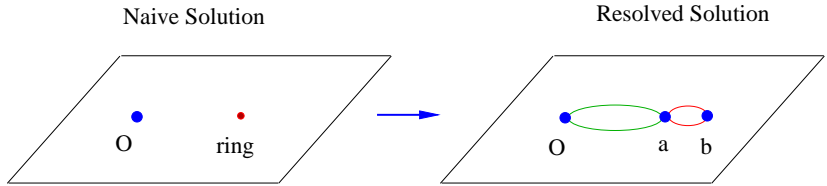}
\caption{{\it Geometric transition of supertube}: The first diagram
  shows the geometry before the transition. The second shows the
  resolved geometry, where a pair of GH charges has nucleated at positions
$a$ and $b$.
}
\label{resol}
\end{figure}

This process is depicted in Fig. \ref{resol}. The nucleation of a GH
pair of oppositely-charged centers blows up a pair of two-cycles. In
the resolved geometry there are no more brane sources, only fluxes
through the two-cycles. The charge of the solution does not come from
any brane sources, but from having non-trivial fluxes over intersecting
two-cycles (or ``bubbles'').

Similarly, if one considers the geometric transition of multiple
concentric black rings, one will nucleate one pair of GH points for
each ring, resulting in a geometry with no brane sources, and with a
very large number of positive and negative GH centers. As we will see,
these centers are not restricted to be on a line, but can have
arbitrary positions in the $\IR^3$ base of the GH space, as long as certain
algebraic equations (discussed in Section \ref{BubbleEqnsSect}) are satisfied.

There is one further piece of physical intuition that is extremely
useful in understanding these bubbled geometries.  As we have already
remarked, GH points can be interpreted, from a ten-dimensional
perspective, as D6 branes.  Since these branes are mutually BPS, there
should be no force between them.  On the other hand, D6 branes of
opposite charge attract one another, both gravitationally and
electromagnetically.  If one simply compactifies M-theory on an
ambipolar GH space, one can only hold in equilibrium GH points of
opposite charge at the cost of having large regions where the metric
has the wrong signature and CTC's.  To eliminate these singular
regions, one must hold the GH points apart by some other mechanism.
In the geometries we seek, this is done by having fluxes threading the
bubbles: Collapsing a bubble concentrates the energy density of the
flux and increases the energy in the flux sector.  Thus a flux tends
to blow up a cycle.  The regular, ambipolar BPS configurations that we
construct come about when these two competing effects - the tendency
of oppositely charged GH points to attract each other and the tendency
of the fluxes to make the bubbles large - are in balance.  We will see
precisely how this happens in Section \ref{BubbleEqnsSect}.

Before proceeding to construct these solutions, we should note that
there are two other completely different ways of arriving at the
conclusion that three-charge black hole microstates can have a base
given by a GH space with negative centers.

One direction, mostly followed by Mathur, Giusto and Saxena
\cite{GiustoIP,GiustoKJ,GiustoZI} is to construct microstates by
taking a novel extremal limit of the non-extremal five-dimensional
black hole \cite{cveticlarsen}. This limit produces a smooth
horizonless geometries that have a GH base with two centers, of charges
$N+1$ and $-N$. These geometries have a known CFT interpretation, and
form a subset of the class described above. A solution that is locally
identical (but differs by a global identification of charges) was also
found in \cite{LuninUU} by doing a spectral flow on a two-charge
solution and extending the resulting solution to an asymptotically
flat one.

The second direction, followed by Kraus and one of the present authors
is to consider the four-dimensional black hole with D1-D5-KKM-P
charges, when the momentum is taken to zero. The resulting naive
solution for the zero-entropy four-dimensional black hole is singular,
and is resolved by an intriguing mechanism: The branes that form the
black hole split into two stacks, giving a non-singular solution
\cite{Bena:2005ay}. One can then relate the black ring to a
four-dimensional black hole by putting it in a Taub-NUT background, as
discussed in Section \ref{4d5d} and in
\cite{Gaiotto:2005xt,Elvang:2005sa,Bena:2005ni}, and then the
nucleation of a pair of oppositely-charged GH centers corresponds, from
a four-dimensional point of view, to the splitting of the zero-entropy
four-dimensional black hole into two stacks of branes, giving a smooth
resulting solution.

Hence, we have three completely independent routes for obtaining
three-charge microstates and resolving the singularity of the
zero-entropy black ring, and all three routes support the same
conclusion: {\it The singularity of the zero-entropy black ring is
resolved by the nucleation of GH centers of opposite charge. The
solutions that result, as well as other three-charge microstate
solutions, are topologically non-trivial, have no brane sources, and
are smooth despite the fact that they are constructed using an ambipolar
GH metric (with regions where the metric is negative-definite). }

\subsection{The bubbled solutions}
\label{bubbledsols}

We now proceed to construct the general form of bubbling solutions
constructed using an ambipolar Gibbons-Hawking base
\cite{Bena:2005va,Berglund:2005vb,Saxena:2005uk}. In Section
\ref{homology} we saw that the two-forms, $\Theta^{(I)}$, will be {\it
  regular}, self-dual, harmonic two-forms, and thus representatives of
the cohomology dual to the two-cycles, provided that the $K^I$ have
the form:
\begin{equation}
K^I ~=~ k^I_0 ~+~  \sum_{j=1}^N \, {k_j^I \over r_j} \,.
\label{KIdefn}
\end{equation}
Moreover, from (\ref{basicflux}), the flux  of the two-form, $\Theta^{(I)}$, through
the two-cycle $\Delta_{ij}$ is given by
\begin{equation}
\flux_{ij}^{(I)}   ~=~       \bigg( {k^I_j   \over q_j} ~-~
 {k^I_i \over q_i} \bigg) \,, \qquad  1 \le i,j \le N \,.
 \label{fluxes}
 \end{equation}

The functions, $L_I$ and $M$, must  similarly be chosen to ensure that the
warp factors, $Z_I$, and the function, $\mu$, are  regular as $r_j \to 0$.
This means that we must take:
\begin{equation}
L^I ~=~ \ell^I_0 ~+~  \sum_{j=1}^N \, {\ell_j^I \over r_j} \,, \quad
M ~=~ m_0 ~+~  \sum_{j=1}^N \, {m_j \over r_j} \,,
\label{LMexp}
\end{equation}
with
\begin{eqnarray}
\ell_j^I  &~=~& -  \coeff{1}{2}\,  C_{IJK} \,
{ k_j^J \, k_j^K  \over q_j} \,,  \qquad j=1,\dots, N \,;\\
m_j &~=~&   \coeff {1}{12}\,C_{IJK} {k_j^I \, k_j^J \, k_j^K \over q_j^2}  ~=~
\coeff{1}{2}\,  {k_j^1 \, k_j^2 \, k_j^3 \over q_j^2} \,,  \qquad j=1,\dots, N \,.
\label{lmchoice}
\end{eqnarray}

Since we have now fixed the eight harmonic functions,  all that remains is
to solve for $\omega$ in equation (\ref{omegeqn}).
The right-hand  side of (\ref{omegeqn}) has two kinds of terms:
\begin{equation}
{1\over r_i}\, \vec \nabla\, {1\over r_j} ~-~ {1\over r_j}\, \vec \nabla\,
{1\over r_i}   \qquad {\rm and } \qquad \vec \nabla \, {1\over r_i} \,.
\label{rhsterms}
\end{equation}
Hence  $\omega$  will be built from the vectors $\vec v_{i}$ of
(\ref{vieqn}) and some new vectors, $\vec w_{ij}$, defined by:
\begin{equation}
 \vec \nabla \times \vec w_{ij} ~=~  {1\over r_i}\, \vec
\nabla\, {1\over r_j} ~-~ {1\over r_j}\, \vec \nabla\, {1\over r_i} \,.
\label{wijeqn}
\end{equation}

To find a simple expression for $ \vec w_{ij}$ it is convenient to
use the coordinates outlined earlier with the $z$-axis running
through $\vec y^{(i)}$ and  $\vec y^{(j)}$.   Indeed,
choose coordinates so that $\vec y^{(i)} ~=~ (0,0,a)$ and $\vec
y^{(j)} ~=~ (0,0,b)$ and one may take $a > b$.   Then the explicit solutions may be
written very simply:
\begin{equation}
w_{ij} ~=~ - {(x^2 +  y^2 + (z-a)(z-b)) \over (a-b)  \, r_i  \, r_j } \, d \phi \,.
\label{wijres}
\end{equation}
This is then easy to convert  to a more general system of coordinates.  One
can then add up all the contributions to $\omega$ from all the
pairs of points.

There is, however, a more convenient basis of vector fields that
may be used instead of the $w_{ij}$.  Define:
\begin{equation}
\omega_{ij} ~\equiv~ w_{ij} +  \coeff{1}{(a-b)} \, \big(  v_{i} - v_{j}  + d\phi \big) ~=~ -
{(x^2 +  y^2 + (z-a+ r_i)(z-b - r_j)) \over (a-b)  \, r_i  \, r_j } \, d \phi  \,,
\label{omegaijdefn}
\end{equation}
These vector fields then satisfy:
\begin{equation}
  \vec \nabla \times \vec \omega_{ij} ~=~  {1\over r_i}\, \vec
\nabla\, {1\over r_j} ~-~ {1\over r_j}\, \vec \nabla\, {1\over r_i}~+~
 {1 \over r_{ij} } \, \bigg(\vec \nabla\,  {1\over r_i} ~-~ \vec \nabla\,
 {1\over r_j} \bigg) \,,
\label{omijeqn}
\end{equation}
where
\begin{equation}
r_{ij} ~\equiv~ |\vec y^{(i)} ~-~ \vec y^{(j)}  |
\label{dijdefn}
\end{equation}
is the distance between the $i^{\rm th}$ and $j^{\rm th}$ center
in the Gibbons-Hawking metric.

We then see that the general solution for $\vec \omega$ may be written
as:
\begin{equation}
\vec \omega ~=~  \sum_{i,j}^ N \, a_{ij} \, \vec \omega_{ij} ~+~
 \sum_{i}^ N \, b_{i} \, \vec v_{i}\,,
\label{omsol}
\end{equation}
for some constants $a_{ij}$, $b_i$.

The important point about the $\omega_{ij}$ is that they have {\it no
string singularities whatsoever.}  They can be used to solve
(\ref{omegeqn}) with the first set of source terms in (\ref{rhsterms}), without
introducing Dirac-Misner strings, but at the cost of adding new source
terms of the form of the second term in (\ref{rhsterms}).  If there are $N$
source points, $\vec{y}^{(j)}$, then using the $w_{ij}$ suggests that
there are ${1 \over 2} N(N-1)$ possible string singularities
associated with the axes between every pair of points $\vec{y}^{(i)}$
and $\vec{y}^{(j)}$.  However, using the $\omega_{ij}$ makes it far
more transparent that all the string singularities can be reduced to
those associated with the second set of terms in (\ref{rhsterms}) and so
there are at most $N$ possible string singularities and these can be
arranged to run in any direction from each of the points
$\vec{y}^{(j)}$.

Finally, we note that the constant terms in (\ref{Vform}), (\ref{KIdefn}) and
(\ref{LMexp})  determine the behavior of the solution at infinity.  If the asymptotic
geometry is Taub-NUT, all these term can be non-zero, and they correspond to
combinations of the moduli.   However, in order to obtain solutions that
are asymptotic to five-dimensional Minkowski space, $\IR^{4,1}$, one
must take $\varepsilon_0 = 0$ in (\ref{Vform}), and $k_0^I =0$ in (\ref{KIdefn}).
Moreover, $\mu$ must vanish at infinity, and this fixes $m_0$.  For simplicity
we also fix the asymptotic values of the moduli that give the size of
the three $T^2$'s, and take $Z_I \to 1$ as $r \to \infty$. Hence, the
solutions that are asymptotic to five-dimensional Minkowski space
have:
\begin{equation}
\varepsilon_0 = 0 \,, \qquad  k_0^I =0\,, \qquad   l_0^I =1\,, \qquad
 m_0  = -\coeff{1}{2}\, q_0^{-1} \, \sum_{j=1}^N\, \sum_{I=1}^3 k_j^I \,.
\label{fiveDsol}
\end{equation}
It is straightforward to generalize these results to solutions with
different asymptotics, and in particular to Taub-NUT.

\subsection{The bubble equations}
\label{BubbleEqnsSect}

In Section \ref{CTCsect} we examined the conditions for the
absence of CTC's and in general   the following must be true globally:
\begin{equation}
\cQ ~ \ge ~ 0\,, \qquad
V\, Z_I ~=~  \coeff {1}{2}\, C_{IJK}\,  K^J \, K^K ~+~ L_I\,  V~ \ge ~ 0 \,,
\qquad  I = 1,2,3\,.
\label{noCTCs}
\end{equation}
As yet, we do not know how to verify these conditions in general,
but one can learn a great deal by studying the limits in which one approaches
a Gibbons-Hawking point, {\it i.e.} $r_j \to 0$.  From this one can derive
some simple, physical conditions (the {\it Bubble Equations}) that in some examples
ensure that (\ref{noCTCs}) are satisfied globally.

To study the limit in which $r_j \to 0$, it is simpler to use (\ref{dtzero}) than
(\ref{dstil}).  In particular, as $r_j \to 0$, the functions, $Z_I$, $\mu$ and $W$
limit to finite values while $V^{-1}$ vanishes.   This means that the circle defined
by $\psi$ will be a CTC unless we impose the additional condition:
\begin{equation}
 \mu(\vec y = \vec y^{(j)} ) ~=~ 0\,, \quad j=1,\dots,N \,.
\label{mucond}
\end{equation}
There is also potentially another problem: The small circles in $\phi$
near $\theta =0$ or $\theta = \pi$ will be CTC's if $\omega$ has a
finite $ d\phi$ component near $\theta =0$ or $\theta = \pi$.  Such a
finite $ d\phi$ component corresponds precisely to a Dirac-Misner
string in the solution and so we must make sure that $\omega$ has no
such string singularities.

It turns out that these two sets of constraints are exactly the same.
One can check this explicitly, but it is also rather easy to see from
(\ref{roteqn}).  The string singularities in $\vec \omega$ potentially
arise from the $\vec \nabla (r_j^{-1})$ terms on the right-hand side
of (\ref{roteqn}).  We have already arranged that the $Z_I$ and $\mu$
go to finite limits at $r_j =0$, and the same is automatically true of
$K^I V^{-1}$.  This means that the only term on the right hand side of
(\ref{roteqn}) that could, and indeed will, source a string is the
$\mu \vec \nabla V$ term.  Thus removing the string singularities is
equivalent to (\ref{mucond}).

One should note that by arranging that $\mu$, $\omega$ and $Z_I$ are
regular one has also guaranteed that the physical Maxwell fields,
$dA^{(I)}$, in (\ref{dAI}) are regular.  Furthermore, by removing the
Dirac strings in $\omega$ and by requiring $\mu$ to vanish at GH
points one has guaranteed that the physical flux of $dA^{(I)}$ through
the cycle $\Delta_{ij}$ is still given by (\ref{basicflux}) (and
(\ref{fluxes})).  This is because the extra terms, $d(Z_I^{-1} k)$, in
(\ref{dAI}), while canceling the singular behaviour when $V=0$, as in
(\ref{dAIform}), give no further contribution in (\ref{basicflux}).
Thus the fluxes, $\flux_{ij}^{(I)}$, are well-defined and represent
the true physical, magnetic flux in the five-dimensional extension of
the ambipolar GH metrics.

Performing the expansion of $\mu$ using (\ref{mures}),
(\ref{KIdefn}), (\ref{LMexp}) and (\ref{lmchoice}) around each
Gibbons-Hawking point one finds that (\ref{mucond}) becomes the {\it
  Bubble Equations}:
\begin{equation}
 \sum_{{\scriptstyle j=1} \atop {\scriptstyle j \ne i}}^N \,
\,  \flux^{(1)}_{ij} \,   \flux^{(2)}_{ij} \,  \flux^{(3)}_{ij} \   {q_i \, q_j  \over r_{ij} } ~=~
-2\, \Big(m_0 \, q_i ~+~  \coeff{1}{2} \sum_{I=1}^3  k^I_i \Big) \,,
\label{BubbleEqns}
\end{equation}
where $r_{ij} \equiv |\vec y^{(i)} -\vec y^{(j)}|$.  One obtains the
same set of equations if one collects all the Dirac string
contributions to $\omega$ and sets them to zero by imposing $b_i =0$
in (\ref{omsol}).  If one adds together all of the bubble equations,
then the left-hand side vanishes identically, and one obtains the
condition on $m_0$ in (\ref{fiveDsol}).  This is simply the condition
$\mu \to 0$ as $r \to \infty$ and means that there is no Dirac-Misner
string running out to infinity.  Thus there are only $(N-1)$
independent bubble equations.

We refer to (\ref{BubbleEqns}) as the bubble equations because they
relate the flux through each bubble to the physical size of the
bubble, represented by $r_{ij}$.  Note that for a generic
configuration, a bubble size can only be non-zero if and only if {\it
  all three} of the fluxes are non-zero.  Thus the bubbling transition
will only be generically possible for the three-charge system. We
should also note that from a four-dimensional perspective these
equations describe a collection of BPS stacks of branes, and are thus
particular case of a collection of BPS black holes. Such
configurations have been constructed in \cite{denef}, and the
equations that must be satisfied by the positions of the black holes
are called ``integrability equations'' and reduce to the equations
(\ref{BubbleEqns}) when the charges are such that the five-dimensional
solution is smooth.

\subsection{The asymptotic charges}
\label{AsympChGH}

As in Section \ref{AsympCharges}, one can obtain the electric charges
and angular momenta of bubbled geometries by expanding $Z_I$ and
$k$ at infinity.   It is, however, more convenient to translate the asymptotics
into the standard coordinates of the Gibbons-Hawking spaces.
Thus, remembering that $r = {1 \over 4} \rho^2$, one has
\begin{equation}
Z_I ~\sim~ 1 ~+~ \, {Q_I \over 4\, r }~+~ \dots  \,, \qquad  \rho \to \infty \,,
\label{ZIexpGH}
\end{equation}
and from (\ref{ZIform}) one easily obtains
\begin{equation}
Q_I ~=~ -2 \, C_{IJK} \, \sum_{j=1}^N \, q_j^{-1} \,
\tilde  k^J_j \, \tilde  k^K_j\,,
\label{QIchg}
\end{equation}
where
\begin{equation}
\tilde  k^I_j ~\equiv~ k^I_j ~-~    q_j\, N  \,  k_0^I  \,,
\qquad {\rm and} \qquad k_0^I ~\equiv~{1 \over N} \, \sum_{j=1}^N k_j^I\,.
\label{ktilde}
\end{equation}
Note that $\tilde  k^I_j$ is gauge invariant under (\ref{gaugetrf}).

One can read off the angular momenta using an expansion like
that of  (\ref{kinfty}).  However, it is easiest to re-cast this in terms
of the Gibbons-Hawking coordinates.  The flat GH metric (near infinity) has
$V= {1 \over r}$ and making the change of variable $r = {1 \over 4} \rho^2$,
one obtains the metric in spherical polar coordinates:
\begin{equation}
ds_4^2 ~=~ d\rho^2 ~+~ \coeff{1}{4} \,\rho^2 \,(d\theta^2 + \sin^2 \theta \,
d \phi^2 ~+~ (d \psi + \cos \theta\, d \phi)^2) \,.
\end{equation}
This can be mapped to the form of (\ref{Rtwosqmet}) via the change
of variable:
\begin{equation}
u\, e^{i \theta_1} ~=~  \rho \cos(\coeff{1}{2}\theta)\,e^{{i \over 2}(\psi + \phi)} \,,
\qquad
v\, e^{i \theta_2} ~=~  \rho \sin(\coeff{1}{2}\theta)\,e^{{i \over 2}(\psi - \phi)} \,.
\label{varchg}
\end{equation}
Using this in (\ref{kinfty}) one finds that
\begin{equation}
k ~\sim~ {1 \over 4 \,\rho^2} \, \big((J_1+J_2) ~+~ (J_1-J_2) \, \cos \theta   \big) \,
d\psi ~+~ \dots \,.
\label{angmomform}
\end{equation}
Thus, one can get the angular momenta from the asymptotic expansion of
$g_{t \psi} $, which is given by the coefficient of $d \psi$ in the
expansion of $k$, which is proportional to
$\mu$.  There are two types of such terms, the simple ${1 \over r}$
terms and the dipole terms arising from the expansion of $V^{-1} K^I
$.  Following \cite{Berglund:2005vb}, define the dipoles
\begin{equation}
\vec D_j ~\equiv~  \, \sum_I  \, \tilde k_j^I \, \vec y^{(j)} \,,
\qquad \vec D ~\equiv~ \sum_{j=1}^N \, \vec D_j \,.
\label{dipoles}
\end{equation}
and then the expansion of $k$ takes the form (\ref{angmomform})
if one takes $\vec D$ to define the polar axis from which $\theta$ is measured.
One then arrives at
\begin{equation}
J_R ~\equiv~ J_1 + J_2 ~=~ \coeff{4}{3}\, \, C_{IJK} \, \sum_{j=1}^N q_j^{-2} \,
\tilde  k^I_j \, \tilde  k^J_j \,  \tilde  k^K_j  \,,
\label{Jright}
\end{equation}
\begin{equation}
 J_L ~\equiv~ J_1 - J_2 ~=~ 8 \,\big| \vec D\big|  \,.
\label{Jleft}
\end{equation}
While we have put modulus signs around $\vec D$ in (\ref{Jleft}), one
should note that it does have a meaningful orientation, and so we
will sometimes consider $\vec J_L = 8 \vec D$.

One can use the bubble equations to obtain another, rather more
intuitive expression for $J_1 -J_2$.  One should first note  that the right-hand side
of the bubble equation, (\ref{BubbleEqns}), may be written as $-  \sum_I  \tilde k_i^I$.
Multiplying this by $\vec y^{(i)}$ and summing over $i$ yields:
\begin{eqnarray}
\vec J_L &~\equiv~&  8\, \vec D  ~=~  -  \coeff{4}{3}\, C_{IJK} \,
 \sum_{{\scriptstyle i, j=1} \atop {\scriptstyle j \ne i}}^N \,  \,  \Pi^{(I)}_{ij} \,
  \Pi^{(J)}_{ij} \,  \Pi^{(K)}_{ij} \ {q_i \, q_j \,  \vec y^{(i)}  \over r_{ij} } \nonumber \\
 & ~=~ & -  \coeff{2}{3}\, C_{IJK} \, \sum_{{\scriptstyle i, j=1} \atop
 {\scriptstyle j \ne i}}^N \,  q_i \, q_j \,  \Pi^{(I)}_{ij} \,   \Pi^{(J)}_{ij} \,  \Pi^{(K)}_{ij} \
{(\vec y^{(i)} - \vec y^{(j)}) \over \big|\vec y^{(i)} - \vec y^{(j)}\big| }\,,
\label{JLnice}
\end{eqnarray}
where we have used the skew symmetry $\Pi_{ij} = - \Pi_{ji}$ to obtain
the second identity.  This result suggests that one should define an angular
momentum flux vector  associated with the $ij^{\rm th}$ bubble:
\begin{equation}
\vec J_{L\, ij} ~\equiv ~ -  \coeff{4}{3}\,q_i \, q_j \, C_{IJK} \,
\Pi^{(I)}_{ij} \,   \Pi^{(J)}_{ij} \,  \Pi^{(K)}_{ij} \, \hat y_{ij} \,,
\label{angmomflux}
\end{equation}
where $\hat y_{ij}$ are {\it unit} vectors,
\begin{equation}
\hat y_{ij} ~\equiv ~  {(\vec y^{(i)} - \vec y^{(j)}) \over
 \big|\vec y^{(i)} - \vec y^{(j)}\big| } \,.
 \label{unitvecs}
\end{equation}
This means that the flux terms on the left-hand side of the bubble equation
actually have a natural spatial direction, and once this is incorporated, it yields the
contribution of the bubble to $J_L$.

\subsection{Comments on closed time-like curves and the bubble equations}
\label{CTCcomments}

While the bubble equations are necessary to avoid CTC's near the
Gibbons-Hawking points, they are not sufficient to guarantee the
absence of CTC's globally.  Indeed, there are non-trivial examples
that satisfy the bubble equations and still have CTC's.  On the other
hand, there are quite a number of important physical examples in which
the bubble equations {\it do} guarantee the absence of CTC's globally.
For example, the simplest bubbled supertube will be discussed in
Section \ref{Simplest} and it has been shown numerically in some
examples that the bubble equations do indeed ensure the global absence
of CTC's.  Some more complex examples that are free of CTC's are
described in Section \ref{Numerics}.  It is an open question as to how
and when a bubbled configuration that satisfies (\ref{BubbleEqns}) is
globally free of CTC's. In this section we will make some simple
observations that suggest approaches to solving this problem.

First we need to dispel a ``myth'' or, more precisely, give a correct
statement of an often mis-stated theorem that in a BPS solution
of extremal black holes, all the black holes must have electric
charges of the same sign.  The physical intuition is simple: If two
BPS black holes have opposite charges, then they necessarily attract
both gravitationally and electromagnetically and cannot be stationary
and this time dependence breaks the supersymmetry of the original BPS
solutions.  While this is physically correct, there is an implicit
assumption that we are not going to allow physical solutions to have
CTC's, changes in metric signature, or regions with complex metrics.
A simple example
is to make a solution with BMPV black holes given by the harmonic function:
\begin{equation}
Z_I ~=~ Z ~\equiv~ 1 ~+~ {Q \over r_1} ~-~ {Q \over r_2} \,.
\label{sillyZ}
\end{equation}
We are not advocating that solutions like this, or ones with CTC's in
general, should be taken as physically sensible. Nevertheless, this
solution does satisfy all the equations of motion. The point we wish
to make is that if one takes a completely standard, multi-centered BPS
solution one can get lots of CTC's or imaginary metric coefficients if
one is sloppy about the relative signs of the distributed charges.
For this reason, one cannot expect to take a bubbled geometry and
randomly assign some flux parameters and expect to avoid CTC's even if
one has satisfied the bubble equations. Indeed, a bubbled analogue of
the BMPV configuration (\ref{sillyZ}) could easily satisfy the bubble
equations thereby avoiding CTC's near the Gibbons-Hawking points, only
to have all sorts of pathology in between the two bubbled black holes.

There must therefore be some  kind of positivity condition on the flux
parameters.  One suggestion might be  to look at every subset, $\cS $,
of the Gibbons-Hawking  points.  To such a subset  one can associate a
contribution,  $Q^{(\cS)}_I$,  to  the  electric  charges  by  summing
(\ref{QIchg})  only over the  subset, $\cS$.   One could  then require
that the  $Q^{(\cS)}_I$ have the same  sign for all  choices of $\cS$.
This  would exclude bubbled  analogs of  (\ref{sillyZ}), but  it might
also  be too  stringent.  It  may be  that one  can tolerate  a ``mild
failure'' of the  conditions on relative signs of  electric charges if
the Gibbons-Hawking points are all  clustered; the  danger might only
occur in ``classical limits'' when some fluxes are very large so that
the  solution  decomposes  into  two  widely  separated  ``blobs''  of
opposite charge.

Another natural, and possibly related condition is to remember that
given $N$ Gibbons-Hawking centers, the cycles are related to the root
lattice of $SU(N)$ and the dual fluxes can be labeled by the weight
lattice.  In this language, the obvious positivity condition is to
insist that the fluxes all lie in the positive Weyl chamber of the
lattice.  Moreover, when there are $N_1$ positive and $N_2$ negative
GH points, it may be more appropriate to think in terms of the weight
lattice of a super Lie algebra, $SU(N_1|N_2)$.  In this context, one
can rewrite $Z_I V$ in a rather more suggestive manner:
\begin{equation}
Z_I\,V ~=~ V~-~   \coeff{1}{4} \, C_{IJK} \, \sum_{i, j=1}^N \,
\Pi^{(J)}_{ij} \,   \Pi^{(K)}_{ij} \,    { q_i \, q_j \over  r_i\, r_j} \,.
\label{ZIVexp}
\end{equation}
With suitable positivity conditions on the fluxes one can arrange all
the terms with $q_i  q_j< 0$ to be positive.  It is not clear why, in general,
 these terms dominate the terms with $q_i    q_j > 0$, but one can certainly
 verify it in examples like the one below.

Consider the situation where all the flux parameters corresponding to a given
$U(1)$  are
equal and positive:
\begin{equation}
k_j^1 ~=~  a \,, \qquad k_j^2 ~=~  b \,, \qquad k_j^3 ~=~  c \,, \qquad
a,b,c >0\,,\quad  j=1,\dots, N  \,.
\label{constk}
\end{equation}
Also suppose that $q_j = \pm 1$ and let $P_\pm$ be the subsets of integers, $j$,
for  which $\pm q_j >0$.  Define
\begin{equation}
V_\pm  ~\equiv~   \sum_{j \in P_\pm}  {1 \over r_j} \,.
  \label{Vpmdefn}
\end{equation}
Then one has
\begin{equation}
Z_I \,V ~=~ V~+~  4\, h_I \, V_+ \, V_- \,,
\label{ZIsimp}
\end{equation}
where $h_1 \equiv bc$, $h_2 \equiv ac$, $h_3 \equiv ab$.

For this flux distribution the bubble equations reduce to:
\begin{eqnarray}
8 \, abc\,V_-(\vec y = \vec y^{(i)}) &=&  (N-1)\,(a+b+c)  \qquad {\rm for \ all } \ i \in P_+ \,,
\label{simpBubbs1} \\
8 \, abc\, V_+ (\vec y = \vec y^{(j)}) &=& (N+1) \,(a+b+c)  \qquad  {\rm for \ all } \ j \in P_-\,.
\label{simpBubbs2}
\end{eqnarray}
Multiply the first of these equations by $r_i^{-1}$ and sum, and multiply
the second equation by $r_j^{-1}$ and sum, and one obtains:
\begin{eqnarray}
 V_+  &=&  {8 \, abc \over (N-1)\,(a+b+c)}  \,\sum_{i \in P_+} \sum_{j \in P_-}
 {1 \over r_{ij}} \, {1 \over r_i} \,,
\label{Vplusexp} \\
 V_-  &=&  {8 \, abc \over  (N+1)\,(a+b+c)}  \,\sum_{i \in P_+} \sum_{j \in P_-}
 {1 \over r_{ij}} \, {1 \over r_j} \,.
\label{Vminusexp}
\end{eqnarray}
Now note that $V = V_+ - V_-$ and use the foregoing expressions in
(\ref{ZIsimp}) to obtain:
\begin{eqnarray}
Z_I \,V   &=&   \sum_{i \in P_+} \sum_{j \in P_-}  { 1\over r_i\, r_j \,r_{ij}} \,
\bigg[  {8 \, abc \over (a+b+c)}  \Big({r_j \over (N-1)}  ~-~
{r_i \over  (N+1)}\,  \Big)\nonumber \\
&&  \qquad \qquad \qquad \qquad\qquad~+~    4\, h_I \,  r_{ij} \bigg]  \,.
\label{ZIVexpsimp}
\end{eqnarray}
Since $a,b,c>0$ and $N > 1$, one has
\begin{equation}
{8 \, abc \over (N+1) \,(a+b+c)} ~<~   4\, h_I\,, \qquad I=1,2,3\,,
\label{obviousbound}
\end{equation}
and thus the positivity of $Z_I  V$ follows trivially from the triangle
inequality:
\begin{equation}
 r_j ~+~  r_{ij}  ~\ge~ r_i \,.
\label{triangle}
\end{equation}

Note that it was relatively easy to prove positivity under the
forgoing assumptions and that there was a lot of ``wiggle room'' in
establishing the inequality.  More formally, one can show that
(\ref{ZIVexpsimp}) is uniformly bounded below in a large compact
region and so one can allow some variation of the flux parameters with
$j$ and still preserve positivity.  It would, of course, be very nice
to know what the possible ranges of flux parameters are.

\section{Microstates for black holes and black rings}
\label{Microstates}

Having explored the general way to construct smooth three-charge
bubbling solutions that have charges and angular momenta of the same
type as three-charge black holes and black rings, we now turn to
exploring such solutions in greater generality. We will begin by
describing several simple examples, like the simplest bubbled black
ring, or a bubbled black hole made of several bubbles. We will find
that when the number of bubbles is large, and the fluxes on them are
generic, these solutions have the same relation between charges and
angular momenta as the maximally-spinning (zero-entropy) three-charge
BPS black hole ($J^2=N_1 N_5 N_p$). Moreover, when all the GH centers
except one are in the same blob, and one GH center sits away from the
blob, the solutions have the same charges, dipole charges and angular
momenta as a zero-entropy, three-charge BPS black ring.  Thus, to any
zero-entropy black hole or any round three-charge supertube there corresponds
a very large number of bubbled counterparts.

It is interesting to recall how the upper bound on the angular
momentum is obtained for the BMPV black hole: One takes a solution
with $J^2 < N_1 N_2 N_3$, and imagines spinning it faster. As this
happens, the closed timelike curves (CTC's) inside the horizon get
closer and closer to the horizon. When $J^2$ becomes larger than $N_1
N_2 N_3$, these CTC's sit outside of the horizon and the solution has
to be discarded as unphysical. A similar story happens with the black
ring. What is remarkable is that this relation between the charges and
angular momentum, which came from studying the solution near the
horizon of the black hole and black ring, also comes out from
investigating horizonless solutions with a large number of bubbles and
generic fluxes. The fact that this coincidence happens both for black
holes, and for black rings (as well as for BPS black holes and rings
in any $U(1)^N$ ${\cal N}=2 $, five-dimensional,  ungauged supergravity),
is indicative of a stronger connection between black holes and their bubbling
counterparts.

Nevertheless, the fact that generic bubbling solutions correspond to
zero-entropy black holes or to zero-entropy black rings means that
we have only found a special corner of the  microstate geometries.
One might suspect, for example, that this feature comes from using a
GH base space, and that obtaining microstates of positive-entropy black holes
might be impossible unless one considers a more general base space.
As we will see, this is not the case:  We will be able to obtain microstates of black
holes with $J^2 <N_1 N_2 N_3 $ by merging together zero-entropy black hole
 microstates and zero-entropy black ring  microstates\footnote{Obviously,
 the term ``zero-entropy''   applies to the
black hole and black ring whose microstate geometries we discuss,
and {\it not} to the horizonless microstate geometries themselves.
Such horizonless microstate geometries trivially have zero entropy.}.

As we have seen in  Section \ref{OffCenter}, unlike the merger of
two BPS black holes, which is always irreversible, the merger of a BPS
black hole and a BPS black ring can be reversible or irreversible,
depending on the charges of the two objects. We therefore expect the
merger of microstates to result in an zero-entropy microstate or a positive-entropy
black-hole microstate, depending on the charges of the merging microstates.
Moreover, since the merger can be achieved  in a Gibbons-Hawking base,
we will obtain positive-entropy black-hole microstates that have a
Gibbons-Hawking base. However, as we will see in the following  sections,
the merger process will be full of surprises.

We will find there is a huge qualitative difference
between the behaviour of the internal structure of microstates in
``reversible'' and ``irreversible'' mergers\footnote{With an obvious abuse
of terminology, we will refer to such solutions as ``reversible'' and
``irreversible'' mergers of microstates with the understanding that
the notion of reversibility refers to the classical black-hole and black-ring
solutions to which the microstates correspond.}. A ``reversible''
merger of an zero-entropy black-hole microstate and  an zero-entropy black-ring
microstate produces another  zero-entropy black-hole microstate.
For reversible mergers we find  the bubbles
corresponding to the ring simply join the bubbles corresponding to the
black hole, and form a bigger bubbled structure.

In an ``irreversible'' merger, as the ring bubbles and the black
hole bubbles get closer and closer, we find that the distances between the GH
points that form the black hole foam and the black ring foam also
decrease. As one approaches the merger point, all the sizes in the
GH base scale down to zero while preserving their relative proportions.
In the limit in which the merger occurs, the solutions have $J_1=J_2 <
\sqrt{Q_1 Q_2 Q_3}$, and all the points have scaled down to zero
size on the base.   Therefore, it naively looks like the
configuration is singular; however, the full physical size of the
bubbles also depends on the warp factors, and taking these into
account one can show that the physical size of all the bubbles
that form the black hole and the black ring remains the same. The
fact that the GH points get closer and closer together implies
that the throat of the solution becomes deeper and deeper, and
more and more similar to the throat of a BPS black hole (which is
infinite).

\subsection{The simplest bubbled supertube}
\label{Simplest}

As we have discussed in  Section \ref{geomtrans}, we expect the
solution resulting from the geometric transition of a zero-entropy
black ring to contain three GH centers, of charges $q_1=1$, $q_2=-Q$
and $q_3=+Q$. The integral of the flux on the Gaussian two-cycle bubbled at the
position of the ring gives the dipole charges of the latter, $d^I$. It
is useful to define another physical variables $f^I$, measuring the
fluxes through the other two-cycle:
\be
 d^I ~\equiv~ 2 \,(k_2^I ~+~ k_3^I) \,, \qquad
 f_I  ~\equiv~ 2\,  k_1^I ~+~ \big(1+  \coeff{1}{Q} \big)\, k_2^I ~+~
\big (1-  \coeff{1}{Q} \big)\,k_3^I    \,.\label{dfring}
\ee
Note that $d^I$ and $f^I$ are invariant under (\ref{fullgauge}).

The electric charges of the bubbled tube are:
\be
{Q_I ~=~ C_{IJK} \, d^J \, f^K\,,}\label{simpringchg}
\ee
and the angular momenta are:
\be
{J_1 ~=~  - {  (Q   - 1) \over 12\, Q} \, C_{IJK}\,
{d^I}\,{d^J}\,{d^K}~+~ {1\over2} \, C_{IJK}\, {d^I}\,{d^J}\,{f^K} \,,}
\label{jonering}
\ee
\be
{ J_2 ~=~  {{ ( Q -1 ) }^2 \over 24 \, Q^2} \, C_{IJK}\,
{d^I}\,{d^J}\,{d^K} ~+~ {1\over2} \,C_{IJK}\, {f^I}\,{f^J} \,{d^K} \,.}
\label{jtworing}
\ee
In particular, the angular momentum of the tube is:
\begin{eqnarray}
 J_T ~=~  J_2-J_1~=~ && {1\over2} \,  C_{IJK}\, ( {f^I}\,{f^J}\,{d^K} -
{d^I}\,{d^J}\,{f^K} )\nonumber \\
&&\quad ~+~   \Big( {3 \,Q^2 - 4\,Q +1 \over 24\, Q^2} \Big)
\, C_{IJK}\, {d^I}\,{d^J}\,{d^K} \,.
\label{Jsimptube}
\end{eqnarray}
When the size of the  2-3  bubble (between GH charges $q_2$ and $q_3$)
 is small, this configuration can be
thought of as the resolution of the singularity of the zero-entropy
supertube, and has the same charges, angular momenta and size as the
naive zero-entropy black ring solution. In the bubble equations, the
size of the  2-3  bubble comes from a balance between the attraction
of oppositely charged $GH $ points, and the fluxes that have a lot of
energy when the cycle they wrap becomes very small. Hence, both when $Q$ is large
and when $d $ is much smaller than $f$ the solution approaches the naive
zero-entropy black ring solution

\exbox{Verify that in the limit of large $Q$, as well as in the limit
  $d/f \rightarrow 0$ equations (\ref{jonering}) and (\ref{jtworing}) match
  exactly the charges and angular momenta of a three-charge black
  ring of zero entropy.}

One can also estimate, in this limit, the distance from the small  2-3 bubble
to the origin of space, and find that this distance asymptotes to the
radius, $R_T$, of the unbubbled black ring solution (as measured
in the $\IR^3$ metric of the GH base), given by
\be
{J_T ~=~  4 \, R_T\, (d^1 + d^2 + d^3) \,.}
\label{JTRreln}
\ee

\subsection{Microstates of many bubbles}
\label{ManyBub}

We now consider bubbled solutions that have a large number of
localized centers, and show that these solutions correspond to
maximally-spinning (zero-entropy) BMPV black holes, or to maximally
spinning BPS black rings \cite{BenaVA}.  The ring microstates have a
blob of GH centers of zero total charge with a single GH center away from the
blob while the black hole microstates have all the centers in one blob
of net GH charge one.  We will see that this apparently small
difference can very significantly influence the solution of the bubble
equations.

\subsubsection{A black-hole blob}

We first consider a configuration of $N$ GH centers that lie is a
single ``blob'' and take all these centers to have roughly the same flux
parameters, to leading order in $N$.  To argue that such a blob
corresponds to a BMPV black hole, we first need to show that $J_1
=J_2$.  If the overall configuration has three independent $\ZZ_2$
reflection symmetries then this is trivial because the $\vec D_j$ in
(\ref{Jleft}) will then come in equal and opposite pairs, and so one
has $J_L =0$.  More generally, for a ``random''
distribution\footnote{Such a distribution must, of course, satisfy the
  bubble equations, (\ref{BubbleEqns}), but this will still allow a
  sufficiently random distribution of GH points.}  the vectors $\hat
y_{ij}$ (defined in (\ref{unitvecs})) will point in ``random''
directions and so the $\vec J_{L \, ij}$ will generically cancel one
another at leading order in $N$.  The only way to generate a non-zero
value of $J_L$ is to bias the distribution such that there are more
positive centers in one region and more negative ones in another. This
is essentially what happens in the microstate solutions constructed
and analyzed by \cite{GiustoIP,GiustoKJ}.  However, a single generic
blob will have $J_1 - J_2$ small compared to $|J_1|$ and $|J_2|$.

To compute the other properties of such a blob, we will simplify
things by taking $N=2 M +1$ to be odd, and assume that $q_j =
(-1)^{j+1}$.  Using the gauge invariance, we can take all of $k^I_i$
to be positive numbers, and we will assume that they have small
variations about their mean value:
\def\cO{{\cal O}}
\be
\label{kvariations}{k^I_j  ~=~ k_0^I  \, (1 ~+~ \cO(1)) \,,}
\ee
where $k_0^I $ is defined in (\ref{ktilde}).   The charges are given by:
\bea
Q_I &=& - 2 \, C_{IJK }
\sum_j  q_j^{-1}\, (k^J_j- q_j N k_0^J )\, (k^K_j- q_j N k_0^K )  \nonumber \\
&=& - 2 \, C_{IJK } \bigg( \sum_j q_j^{-1} k^J_j k^K_j - N k_0^J
\sum_j k^K_j -  N k_0^K \sum_j k^J_j + N^2 k_0^J k_0^K \sum_j q_j \bigg)
\nonumber\\
&=& 2\, C_{IJK } \Big( N^2 k^J k^K  - \sum_j k^J_j k^K_j q_j^{-1}\,   \Big)
\nonumber \\
&\approx&  2\, C_{IJK }  \big( N^2  +  \cO(1)\big )\, k_0^J k_0^K
\label{charges}
\eea
where we used (\ref{kvariations}) and the fact that $|q_i| =1$ only in
the last step.  In the large $N$ limit, for M theory on $T^6$ we have
\be
\label{MtheoryQ}{Q_1 \approx  4 N ^2 k_0^2 k_0^3 + \cO(1)\,, \quad Q_2 \approx
4 N^2 k_0^1 k_0^3 + \cO(1)\,, \quad Q_3 \approx 4 N^2 _0k^1 k_0^3 + \cO(1)\,.}
\ee
We can make a similar computation for the angular momenta:
\bea
J_R &~=~& \coeff{4}{3}\,  C_{IJK} \sum_j  q_j^{-2}\,
(k^I_j- q_j N k_0^I)\, (k^J_j- q_j N k_0^J )\, (k^K_j- q_j N k_0^K ) \nonumber
\\
&~=~&  \coeff{4}{3}\,  C_{IJK} \bigg( \sum_j   q_j^{-2}\,k^I_j k^J_j k^K_j
- 3  N k_0^I  \sum_j  q_j^{-1}\, k^J_j  k^K_j  \nonumber \\
& &\qquad \qquad \qquad
+ 3   N^2 k_0^I k_0^J \sum_j  k^K_j  -  N^3 k_0^I k_0^J  k_0^K
\sum_j q_j  \bigg) \nonumber \\
& ~ \approx ~&  \coeff{4}{3}\,  C_{IJK}
\big( N   - 3  N   + 3  N^3   - N^3  + \cO(N) \big) \, k_0^I k_0^J k_0^K\,,
\label{Jfin}
\eea
where we used the fact that, for a ``well behaved'' distribution of
positive $k_i^I$ with $|q_j| =1$,  one has:
\be
\label{sums}{  \sum_i q_i^{-1}  k^J_i  k^K_i ~=~
\sum_i q_i   k^J_i  k^K_i ~\approx~  k_0^J  k_0^K \,, \qquad
\sum_i  k^I_i k^J_i k^K_i ~\approx~ N k_0^I  k_0^J k_0^K \,.}
\ee
Therefore we simply have:
\be
\label{Japprox}{J_R ~\approx~ 16 N^3 k_0^1 k_0^2 k_0^3 + O(N) ~.}
\ee

Since $J_L\approx 0$ for a generic blob at large $N$ we therefore have,
at leading order:
\be
\label{bmpvj}{J_1^2 ~\approx~ J_2^2 ~\approx~ \coeff{1}{4}\, J_R^2 ~\approx~
 Q_1 Q_2 Q_3 \,,}
\ee
and so, in the large-$N$ limit, these microstates always correspond to a
maximally spinning BMPV black hole.

\exbox{Show that at sub-leading order in $N$
\be
\label{deviation}{{J_R^2 \over 4  Q_1 Q_2 Q_3} ~-~ 1\sim
O\left({1 \over N^2}\right)\,.}
\ee
}

Interestingly enough, the value of $J_R $ is slightly bigger than
$\sqrt{4 Q_1 Q_2 Q_3}$. However, this is not a problem because in the
classical limit this correction vanishes. Moreover, it is possible to
argue that a classical black hole with zero horizon area will receive
higher-order curvature corrections, that usually increase the horizon
area; hence a zero-entropy configuration will have $J_R$ slightly
larger then the maximal classically allowed value, by an amount that
vanishes in the large $N$ (classical) limit.

\subsubsection{A supertube blob}

The next simplest configuration to consider is one in which one starts
with  the blob considered above and then moves a single GH point of charge
$+1$ out to a very large distance from the blob.  That is, one considers a blob
of total GH charge zero with a single very distant point of GH-charge $+1$.
Since one now has a strongly ``biased'' distribution of GH charges one
should now expect $J_1 - J_2 \ne 0$.

Again we will assume $N$ to be odd, and take the
a  GH charge distribution to be $q_j = (-1)^{j+1}$, with the distant
charge being the $N^{\rm th}$ GH charge.
The blob therefore has ${1 \over 2}(N-1)$ points of GH charge $+ 1$ and
 ${1 \over 2}(N-1)$ points of GH charge $- 1$. When seen from far away
one might expect this blob  to resemble the three-point
solution described above with $Q= {1 \over 2}(N-1)$.  We will show that
this is exactly what happens in the large-$N$ limit.

To have  the $N^{\rm th}$ GH charge far from the blob means that
 all the two-cycles, $\Delta_{j \, N}$ must support a very large flux
 compared to the fluxes on the $\Delta_{ij}$ for $i,j < N$.
 To achieve this we therefore take:
\be
\label{ringblobvars}{k^I_j  ~=~ a_0^I  \, (1 ~+~ \cO(1))  \,, \quad j=1,\dots,N -1 \,, \qquad
k^I_N  ~=~ - b_0^I \, N \,.}
\ee
where
\be
\label{aavg}{a_0^I ~\equiv~{1 \over (N-1)} \, \sum_{j=1}^{N-1} k_j^I\,.}
\ee
We also assume that $a_0^I$ and $b_0^I$ are of the same order.
The fluxes of this configuration are then:
\bea
\label{blobfluxes}\Pi^{(I)}_{ij} &~=~& \bigg({ k_j^I \over q_j} ~-~  { k_i^I \over q_i}
\bigg) \,, \qquad \\
 \Pi^{(I)}_{i\,N} &~=~&  - \Pi^{(I)}_{ N \, i} ~=~ -\bigg(
{ k_i^I  \over q_i} ~+~    N\, b_0^I \bigg)
 \,,  \quad i,j =1, \dots, N-1\,. \nonumber
\eea
For this configuration one has:
\bea
\label{abvariations1}
k_0^I  &~=~ & {(N-1)\over N} \, a_0^I ~-~b_0^I \,,
\qquad \tilde k_N^I  ~=~  - (N-1) \, a_0^I   \,, \\
 \tilde k^I_j  &~=~ & k_j^I ~+~ q_j \, (N\, b_0^I  - (N-1)\, a_0^I)\,, \quad
j=1,\dots, N-1 \,.
\label{abvariations2}
\eea
Motivated by the bubbling black ring of \cite{BenaVA} and Section \ref{Simplest},
define the physical parameters:
\be
\label{dandf}{d^I ~\equiv~  2\,(N-1)\, a_0^I \,, \qquad f^I ~\equiv~
(N-1)\, a_0^I - 2\,N\, b_0^I \,.}
\ee
Keeping only the terms of leading order in $N$
in (\ref{QIchg}) and (\ref{Jright}), one finds:
\bea
\label{rcharges}Q_I &~=~& C_{IJK} d^J f^K \,, \\
J_1 + J_2  &~=~& \coeff{1}{2} \, C_{IJK}  (d^I d^J f^K + f^I f^J d^K) ~-~
 \coeff{1}{24} \, C_{IJK}  d^I d^J d^K\,.\label{rcharges2}
\eea
Since the $N^{\rm th}$ point is far from the blob,
we can take $r_{i N} \approx r_0$ and then the $N^{\rm th}$ bubble
equation reduces to:
\be
\label{bubblerad}{\coeff{1}{6}\, C_{IJK} \,\sum_{j=1}^{N-1}\,
\bigg({k_j^I \over q_j} + N\, b_0^I\bigg)\, \bigg({k_j^J \over q_j} + N\, b_0^J\bigg)\,
\bigg({k_j^K \over q_j} + N\, b_0^K\bigg)\, {q_j  \over r_0} =(N-1)\, \sum_I \, a_0^I\,.}
\ee
To leading order in $N$ this means that the distance from the
blob to the $N^{\rm th}$ point, $r_0$, in the GH space is given by:
\bea
r_0 &~\approx~&   \coeff{1}{2} \, N^2\, \bigg[\sum_I \, a^I\bigg]^{-1}
C_{IJK} \,   a_0^I \, b_0^J \, b_0^K \nonumber \\
&~=~&  \coeff{1}{32} \,
\bigg[\sum_I \, d^I\bigg]^{-1}  C_{IJK} \,   d^I \, (2f^J - d^J) \, (2f^K - d^K) \,.
\label{ringrad}
\eea
Finally, considering  the dipoles (\ref{dipoles}), it is evident that, to leading
order in $N$, $\vec D$ is dominated by the contribution coming from the
$N^{\rm th}$ point and that:
\bea
 J_1 - J_2 &~=~ & 8\, | \vec D| ~=~  8\, N \, \bigg(
\sum_I \, a_0^I \bigg)\,  r_0 ~=~ 4 \, N^3 \, C_{IJK} \,
a_0^I \, b_0^J \, b_0^K \\
& ~=~ & \coeff{1}{8} \,    C_{IJK} \,   d^I \, (2f^J - d^J) \, (2f^K - d^K) \,.\label{ringJL}
\eea
%

\exbox{ Verify that the angular momenta and the radius of this
  bubbling supertube (equations (\ref{ringJL}), (\ref{ringrad}),
  \ref{rcharges}) and (\ref{rcharges2})) match those of the simplest
  bubbling supertube described in Section \ref{Simplest}, and therefore match
those of a zero-entropy black ring.}

Thus, the bubbling supertube of many centers also has {\it exactly}
the same size, angular momenta, charges and dipole charges as a
zero-entropy black ring, and should be thought of a microstate of the
later.

\section{Mergers and deep microstates}
\label{Mergers}

As we have seen in Section \ref{OffCenter}, a merger of a zero-entropy
black ring and a zero-entropy black hole can produce both a
zero-entropy black hole (reversible merger) or a non-zero-entropy one
(irreversible merger). We expect that in a similar fashion, the merger
of the microstates of zero-entropy black holes and zero-entropy black
rings should produce microstates of both zero-entropy and
positive-entropy black holes. Since we have already constructed
zero-entropy black-hole microstates, we will mainly focus on
irreversible mergers and their physics. One can learn more about
reversible mergers of microstates in Section 6 of \cite{deep}.

Even though the original black-ring plus black-hole solution that
describes the merger in Section \ref{OffCenter} and \cite{Bena:2005zy}
does not have a tri-holomorphic $U(1)$ symmetry (and thus the merger
of the corresponding microstates cannot be done using a GH base), one
can also study the merger of black rings and black holes by
considering a $U(1) \times U(1)$ invariant solution describing a black
ring with a black hole in the center.  As the ring is made smaller and
smaller by, for example, decreasing its angular momentum, it
eventually merges into the black hole.  At the point of merger, this
solution is identical to the merger described in Section
\ref{OffCenter} with the black ring grazing the black-hole horizon.
Hence this $U(1) \times U(1)$ invariant solution can be used to study
mergers where the black ring grazes the black hole horizon at the
point of merger. As we have seen in Section \ref{OffCenter}, all the
reversible mergers and some of the irreversible mergers belong to this
class.

In the previous section we have seen how to create bubbled solutions
corresponding to a zero-entropy black ring and maximally-spinning
black holes.  The generic bubbled solutions with GH base have a $U(1)$
symmetry corresponding to $J_R \equiv J_1 + J_2$ and if the GH points
all lie on an axis then the solution is $U(1) \times U(1)$ invariant.
We can therefore study the merger of bubbled microstates by
constructing $U(1) \times U(1)$ invariant bubbling solutions
describing a black ring with a black hole in the center. By changing
some of the flux parameters of the solution, one can decrease the radius of
the bubbling black ring and merge it into the bubbling black hole to
create a larger bubbling black hole.

In this section we consider a bubbling black hole with a very large
number of GH centers, sitting at the center of the simplest bubbled
supertube, generated by a pair of GH points\footnote{Of course it is
  straightforward to generalize our analysis to the situation where
  both the supertube and the black hole have a large number of GH
  centers. However, the analysis is simpler and the numerical
  stability is better for mergers in which the supertube is composed
  of only two points, and we have therefore focussed on this.}. We
expect two different classes of merger solution depending upon whether
the flux parameters on the bubbled black hole and bubbled black ring
are parallel or not.  These correspond to reversible and irreversible
mergers respectively.  The reversible mergers involve the GH points
approaching and joining the black-hole blob to make a similar,
slightly larger black-hole blob \cite{deep}.  The irreversible merger
is qualitatively very different and we will examine it in detail.
First, however, we will establish some general results about the
charges and angular momenta of the bubbled solutions that describe a
bubbled black ring of two GH centers with a bubbling black-hole at the
center.

\subsection{Some exact results}
\label{ExactRes}

We begin by seeing what may be deduced with no approximations
whatsoever.  Our purpose here is to separate all the algebraic
formulae for charges and angular momenta into those associated with
the black hole foam and those associated with the bubbled supertube.
We will consider a system of $N$ GH points in which the first $N-2$
points will be considered to be a blob and the last two points will
have $q_{N-1} = -Q$ and $q_{N} = Q$.  The latter two points can then
be used to define a bubbled black ring.

Let $\hat k_0^I$ denote the average of the flux parameters
over the first $(N-2)$ points:
\be \label{kIavg}{\hat k_0^I ~\equiv~{1 \over (N-2)} \, \sum_{j=1}^{N-2} \, k_j^I \,,}
\ee
and introduce $k$-charges that have a vanishing average  over the
first $(N-2)$ points:
\be \label{khat}{\hat  k^I_j ~\equiv~ k^I_j ~-~ (N-2) \,  q_j  \,  \hat k_0^I  \,,
\qquad j=1,\dots,N-2 \,.}
\ee
We also parameterize the last two $k^I$-charges in exactly the same
manner as for the bubbled supertube (see equation (\ref{dfring})):
\bea \label{ringchgs}
d^I &~\equiv~& 2 \, \big(  k_{N-1}^I + k_{N}^I \big) \,, \\
f^I &~\equiv~ &2 \,(N-2) \,   \hat k_0^I +  \big(1 + \coeff{1}{Q}\big)\, k_{N-1}^I  +
\big(1 - \coeff{1}{Q}\big)\, k_{N}^I  \,. \nonumber
\eea
One can easily show that the charge (\ref{QIchg}) decomposes into
\be \label{chgdecomp}{Q_I ~=~ \widehat Q_I  ~+~C_{IJK}\, d^J\, f^K \,,}
\ee
where
\be \label{QIhat}{\widehat Q_I ~\equiv~ - 2\, C_{IJK} \, \sum_{j=1}^{N-2}  q_j^{-1} \,
\hat  k^J_j \, \hat  k^K_j  \,.}
\ee
The $\widehat Q_I$ are simply the charges of the black-hole blob, made of the
first $(N-2)$ points.  The second term in (\ref{chgdecomp})
is exactly the expression, (\ref{simpringchg}), for the charges of
a bubbled supertube with GH centers of charges $+1$, $-Q$ and $Q$ and
$k$-charges $(N-2) \hat k_0^I$, $k_{N-1}^I$ and $k_{N}^I$, respectively.
Thus the charge of this configuration decomposes exactly as if
it were a black-hole blob of $(N-2)$ centers and a bubbled supertube.

There is a similar result for the angular momentum, $J_R$.  One
can easily show that:
\be \label{Rangmomdecomp}{J_R ~=~ \widehat J_R  ~+~  d^I \,
\widehat Q_I~+~j_R\,,}
\ee
where
\be \label{JRhat}{\hat J_R  ~\equiv~  {4 \over 3}\, \, C_{IJK} \,
\sum_{j=1}^{N-2} q_j^{-2} \,  \hat  k^I_j \, \hat  k^J_j \,  \hat  k^K_j  \,,}
\ee
and
\be \label{JringR}{j_R  ~\equiv~ \coeff{1}{2} \, C_{IJK}\, \big(f^I f^J d^K ~+~
f^I  d^J  d^K\big)  ~-~ \coeff{1}{24} \,(1-Q^{-2})\, C_{IJK}\,  d^I d^J d^K \,.}
\ee
The term, $\hat J_R$, is simply the right-handed angular momentum of
the black-hole blob made from $N-2$ points.  The ``ring'' contribution
to the angular momentum, $j_R$, agrees precisely with $J_1+ J_2$ given
by (\ref{jonering}) and (\ref{jtworing}) for an isolated bubbled
supertube.  The cross term, $d^I \widehat Q_I$ is represents the
interaction of the flux of the bubbled ring and the charge of the
black-hole blob.  This interaction term is exactly the same as that
found in Section \ref{OffCenter} and in
\cite{Bena:2004de,Gauntlett:2004qy,Bena:2005zy} for a concentric black
hole and black ring.

Thus, as far as the charges and $J_R$ are concerned, the complete
system is behaving as though it were a black-hole blob of $(N-2)$
points interacting with a bubbled black ring defined by the points
with GH charges $\pm Q$ and a single point with GH charge $+1$
replacing the black-hole blob. Note that no approximations were made
in the foregoing computations, and the results are true independent
of the locations of the GH centers.

To make further progress we need to make some assumptions about
the configuration of the points.    Suppose, for the moment, that all the GH
charges lie on the $z$-axis at points $z_i$ with $z_i < z_{i+1}$.
In particular, the GH charges, $-Q$ and $+ Q$,  are located at $z_{N-1}$ and
$z_{N}$ respectively.

With this ordering of the GH points, the expression for $\vec J_L$ collapses to:
\be \label{linJL}
 J_L ~=~  \coeff{4}{3}\, C_{IJK} \, \sum_{1 \le i < j \le N}
\,  q_i \, q_j \,  \Pi^{(I)}_{ij} \,   \Pi^{(J)}_{ij} \,  \Pi^{(K)}_{ij}   \,.
\ee
This expression can then be separated, just as we did for $J_R$, into a black-hole
blob component, a ring component, and interaction cross-terms.  To that
end, define the left-handed angular momentum of the blob to be:
\be \label{linJLh}{\widehat J_L ~=~  \coeff{4}{3}\, C_{IJK} \, \sum_{1 \le
i < j \le N-2} \,  q_i \, q_j \,  \Pi^{(I)}_{ij} \,   \Pi^{(J)}_{ij}
\,  \Pi^{(K)}_{ij}   \,.}
\ee
Note that
\be
\Pi^{(I)}_{ij} ~\equiv~  \bigg( {k_j^I \over q_j} ~-~
{k_i^I \over q_i} \bigg)  ~=~  \bigg( {\hat k_j^I \over q_j} ~-~
{\hat k_i^I \over q_i} \bigg)  \,, \qquad  1 \le i, j \le N-2 \,,
\ee
and so this only depends upon the fluxes in the blob.

The remaining terms  in (\ref{linJL}) may then be written in terms of $\hat k_j^I$,
$d^I$ and $f^I$ defined in (\ref{khat}) and (\ref{ringchgs}).  In particular, there are terms
that depend only upon $d^I$ and $f^I$, and then there are terms that are
linear, quadratic and cubic in $\hat k_j^I$ (and depend upon $d^I$ and $f^I$).
The linear terms vanish  because $\sum_{j=1}^{N-2}\hat k_j^I \equiv 0$, the quadratic terms
assemble into $\widehat Q_I$  of (\ref{QIhat}) and the cubic terms assemble
into $\hat J_R$ of  (\ref{JRhat}).    The terms proportional to (\ref{JRhat}) cancel between
the terms with $j=N-1$ and $j=N$, and one is left with:
\be \label{Langmomdecomp}{J_L ~=~ \widehat J_L  ~-~  d^I \,
\widehat Q_I~+~j_L\,,}
\ee
where $j_L$ is precisely the angular momentum, $J_T$, of the tube:
\be \label{JringL}
{j_L ~\equiv~ \coeff{1}{2} \, C_{IJK}\, \big(d^I f^J f^K ~-~
f^I d^J d^K \big)  ~+~ \Big( {3 \,Q^2 - 4\,Q +1 \over 24\, Q^2} \Big) \,
C_{IJK}\,  d^I d^J d^K \,.}
\ee
Observe that (\ref{JringR}) and (\ref{JringL})  are exactly the angular momenta
of a simple bubbled ring, (\ref{jonering}) and (\ref{jtworing}).  Again we see
the cross-term from the interaction of the ring dipoles and the
electric charge of the blob.  Indeed, combining (\ref{Rangmomdecomp})
and (\ref{Langmomdecomp}), we obtain:
\be \label{angmomcomp}{J_1 ~=~ \widehat J_1   ~+~j_1\,, \qquad
J_2 ~=~ \widehat J_2   ~+~ j_2 ~-~  d^I \,  \widehat Q_I  \,,}
\ee
which is exactly how the angular momenta of the classical ring-hole solution in
Section \ref{OffCenter} decomposed.   In particular, the term coming from the
 interaction of the ring dipole moment with the black hole charge only
 contributes to $J_2$.

\exbox{Check the decompositions (\ref{QIhat}), (\ref{JRhat}) and
(\ref{Langmomdecomp}).}

The results obtained above are independent of whether the blob of
$N-2$ points is a BMPV black-hole blob, or a more generic configuration.
However, to study mergers we now take the blob to be a black-hole
microstate, with $\widehat J_L ~=~ 0$.  The end result of the merger process is
also a BMPV black hole microstate, and so $J_L ~=~  0$. Therefore, the {\it exact}
merger condition is simply:
\bea
\Omega &\equiv& \coeff{1}{2} \, C_{IJK}\,
\big(d^I f^J f^K - f^I d^J d^K \big)  +  \Big( {3 \,Q^2 - 4\,Q +
1 \over 24\, Q^2} \Big) \,  C_{IJK}\,  d^I d^J d^K -  d^I \,  \widehat Q_I \nonumber \\
&=&0 \,. \label{exactMcond}
\eea
Using (\ref{Jsimptube}), this may be written:
\be
\label{ClassMerger}{ J_T ~-~  d^I \,  \widehat Q_I ~=~0 \,,}
\ee
which is precisely the condition obtained in  Section \ref{OffCenter} and \cite{Bena:2005zy} for a classical
black ring to merge with a black hole at its equator.

One should note that the argument that led
 to the expressions (\ref{Langmomdecomp}) and (\ref{JringL}) for $J_L$,  and to the exact merger  condition, (\ref{exactMcond}), apply far more generally.  In particular
we only needed the fact that the unit vectors, $\hat y_{ij}$, defined in
 (\ref{unitvecs}), are all parallel for $j = N-1$ and $j=N$.   This is
 approximately true in many contexts, and most particularly if the
 line between the  $(N-1)^{\rm th}$ and  $N^{\rm th}$ points runs
 through the blob and the width of the  blob is small compared
 to the distance to the two exceptional points.

One should also not be surprised by the generality of the result in equation (\ref{JringL}).
 The angular momentum, $J_T$, is an  intrinsic property of a black ring, and hence
for a zero-entropy black ring, $J_T$ can only depend on the $d$'s
and $f$'s, and cannot depend on the black hole charges (that is, the $\hat k_j^I$).
Therefore, we could have obtained (\ref{JringL}) by simply setting the black hole charge
to zero, and then reading off $J_T$ from the bubbling black ring solution of
Section \ref{Simplest}.   Hence, one should think
about the expression of  $J_T$ in (\ref{Jsimptube}) as a universal relation between
intrinsic properties of the bubbled ring: $J_T, d^I$ and $f^I$.

\subsection{Some simple approximations}
\label{Approximations}

We now return to a general distribution of GH points, but we will assume
that the two ``black ring points'' (the $(N-1)^{\rm th}$ and $N^{\rm th}$ points)
are close  together but very far from the black-hole blob of the remaining $(N-2)$
points.   Set up coordinates in the geometric center of the black-hole blob, {\it i.e.}
choose the origin so that
\be \label{geomcent}{\sum_{i=1}^{N-2} \, \vec r_i ~=~ 0 \,.}
\ee
Let $r_0  \equiv |\vec r_{N-1}|$ be the distance from the geometric center
of the blob to the first exceptional point, and let $\hat r_0$ be the the unit
vector in that direction.  The vector, $\vec \Delta \equiv \vec r_{N } - \vec r_{N-1}$,
defines the width of the ring.   We will assume  that the
magnitudes $\Delta \equiv  |\vec \Delta|$  and $ r_j \equiv  |\vec  r_j|$ are small
compared to $r_0$.  We will also need the first terms of the multipole expansions:
\bea
{1\over |\vec r_{N-1}-\vec r_j |}& ~=~ &
{1 \over r_0} ~+~  { \vec r_j \cdot \hat r_0  \over r_0^2}  ~+~ \dots  \\
{1\over | \vec r_N  -\vec r_j |}& ~=~ &{1 \over r_0} ~+~
{(\vec r_j - \vec \Delta)\cdot \hat r_0  \over r_0^2}  ~+~ \dots \,.
\label{dipoleexp}
\eea
For simplicity, we will also assume that the two black-ring points
(we will also call these points ``exceptional points'')
are co-linear with the origin so that
\be \label{rNval}{r_N ~\equiv~  | \vec r_{N}  | ~=~ r_0 + \Delta \,.}
\ee
The last two bubble equations are then:
\be \label{bubbleone}{ { \gamma \over  \Delta} ~-~  \sum_{j=1}^{N-2}\,
{q_j\, \alpha_j  \over  |\vec r_N - \vec r_j|}   ~=~  \sum_I \,
\big(N \,Q\, k^I_0 - k_{N}^I\big) \,,}
\ee
\be \label{bubbletwo}{ - { \gamma \over  \Delta} ~+~  \sum_{j=1}^{N-2}\,
{q_j\, \beta_j   \over |\vec r_{N-1} -  \vec r_j|}   ~=~ - \sum_I \,
\big(N \,Q\, k^I_0  + k_{N-1}^I\big)\, }
\ee
where $k^I_0$ is given in (\ref{ktilde}) and
\bea \label{aldefn}
\alpha_j &\equiv& \coeff{1}{6} \,Q\, C_{IJK}
\, \Pi_{j\,N}^{(I)} \,  \Pi_{j\,N}^{(J)}\,  \Pi_{j\,N}^{(K)} \\
&=&   \coeff{1}{6} \,Q\,  C_{IJK} \,
\bigg({k_{N}^I \over Q} - {k_j^I \over q_j} \bigg)\, \bigg({k_{N}^J \over Q} -
{k_j^J \over q_j} \bigg)\,\bigg({k_{N}^K \over Q} - {k_j^K \over q_j} \bigg) \,,
\eea
\bea \label{bedefn}
\beta_j & \equiv & \coeff{1}{6}  \,Q\, C_{IJK}
\, \Pi_{j\,(N-1)}^{(I)} \,  \Pi_{j\,(N-1)}^{(J)}\,  \Pi_{j\,(N-1)}^{(K)} \\
&= & - \coeff{1}{6} \,Q\,  C_{IJK} \,
\bigg({k_{N-1}^I \over Q} + {k_j^I \over q_j} \bigg)\, \bigg({k_{N-1}^J \over Q} +
{k_j^J \over q_j} \bigg)\,\bigg({k_{N-1}^K \over Q} + {k_j^K \over q_j} \bigg)  \,,
\eea
\be \label{gadefn}{\gamma ~\equiv~  \coeff{1}{6}\, Q^2 \, C_{IJK}\,
\Pi_{ (N-1)\,N}^{(I)} \,  \Pi_{ (N-1)\,N}^{(J)} \, \Pi_{ (N-1)\,N}^{(K)}
~=~  \coeff{1}{48}\,  Q^{-1} \,  C_{IJK} \, d^I\, d^J\,d^K  \,.}
\ee
It is also convenient to introduce
\be \label{ABzero}{\alpha_0 ~\equiv~ \sum_{j=1}^{N-2} \,q_j \, \alpha_j\,, \qquad
\beta_0 ~\equiv~ \sum_{j=1}^{N-2} \,q_j \, \beta_j\,.}
\ee
If one adds (\ref{bubbleone}) and (\ref{bubbletwo})  then the terms involving $\gamma$
cancel and using  (\ref{dipoleexp}) one then obtains:
\be \label{expbubr}{
\sum_{j=1}^{N-2}\,q_j\,  \bigg[\, \alpha_j \, \bigg({1 \over r_0}  +  {(\vec r_j - \vec \Delta)
\cdot \hat r_0 \over r_0^2}  \bigg)    ~-~ \beta_j  \, \bigg({1 \over r_0} +
 { \vec r_j \cdot \hat r_0  \over r_0^2} \bigg)\,\bigg] ~=~
 \coeff{1}{2} \,\sum_I \,  d^I \,.}
\ee
One now needs to perform the expansions with some care.   Introduce the flux
vector:
\be \label{Xdefn}{X^I ~\equiv~  2\,f^I - d^I - 4\,(N-2)\,\hat k_0^I \,, }
\ee
and note that the fluxes between the blob and ring points are given by:
\bea \label{ringholeflux}
\Pi^{(I)}_{j \,(N-1)} &~=~& -\coeff{1}{4}\, \big[\,
X^I + Q^{-1}\, d^I  + 4\, q_j^{-1} \, k_j^I \,\big]\,, \\
\Pi^{(I)}_{j \,N} &~=~& -\coeff{1}{4}\, \big[\, X^I - Q^{-1}\, d^I  +
4\, q_j^{-1} \, k_j^I \,\big] \,.
\eea
In particular, the difference of these fluxes is simply the flux through the two-cycle
running between the two ring points:
\be \label{fluxdiff}{\Pi^{(I)}_{j \,N}~-~ \Pi^{(I)}_{j \,(N-1)} ~=~   {d^I  \over 2\, Q}  ~=~
\Pi^{(I)}_{(N-1)\, N} \,.}
\ee
For the ring to  be far from the black hole, the fluxes $\Pi^{(I)}_{j \,(N-1)}$ and
$\Pi^{(I)}_{j \,N}$ must be large.  For the ring to be thin ($\Delta \ll  r_0$), these fluxes
must be of similar order, or $\Pi^{(I)}_{(N-1)\, N}$ should be small.  Hence
we are assuming that $ {d^I  \over 2 Q} $ is small compared to $X^I$.
We are also going to want the black hole and the black ring to have similar
charges and angular momenta, $J_R$,  and one of the ways of achieving this
is to make $f^I$, $d^I$ and  $N \hat k_0^I$ of roughly the same order.

Given this,  the leading order terms in (\ref{expbubr}) become:
\be \label{redexpbubr}{
\sum_{j=1}^{N-2}\, q_j\, \bigg[\, {(\alpha_j - \beta_j) \over r_0} ~-~ \alpha_j \,
{\Delta    \over r_0^2} \,\bigg] ~=~  \coeff{1}{2} \,\sum_I \,  d^I \,.}
\ee
One can then determine the ring width, $\Delta$, using (\ref{bubbleone}) or (\ref{bubbletwo}).
In particular, when the ring width is small while the ring radius is large, the left-hand
side of each of these equations is the difference of two very large numbers of similar
magnitude.  To leading order we may therefore neglect the right-hand sides and
use the leading monopole term to obtain:
\be \label{deltares}{\beta_0 \,{\Delta \over r_0} ~\approx~
\alpha_0 \,{\Delta \over r_0} ~=~  \bigg[\, \sum_{j=1}^{N-2}\, q_j\,
\alpha_j  \,\bigg] \,  {\Delta \over r_0}~\approx~\gamma    \,, }
\ee
and hence (\ref{expbubr}) becomes:
\be \label{simpexpbubr}{
- \gamma   ~+~ \sum_{j=1}^{N-2}\, q_j\, (\alpha_j - \beta_j)    ~\approx~
 \bigg[\,  \coeff{1}{2} \,\sum_I \,  d^I\, \bigg] \, r_0 \,.}
\ee
Using the explicit expressions for $\alpha_j, \beta_j$ and $\gamma$, one
then finds:
\bea
 r_0 ~\approx &~&  \bigg[4\, \sum_I \, d^I\bigg]^{-1}
 \bigg[ \, \coeff{1}{2} \, C_{IJK} \,      (d^I f^J  f^K  -   f^I d^J d^K) \nonumber \\
 & +& \bigg({3\,Q^2 -4\,Q +1 \over 24\, Q^2}\bigg) \,
C_{IJK} \, d^I d^J d^K  ~-~ d^I \widehat Q_I \, \bigg] \,.
 \label{newringrad}
\eea
This is exactly the same as the formula for the tube radius that one obtains
from (\ref{JTRreln}) and (\ref{Jsimptube}).  Note also that we have:
\be \label{rradangmom}
{r_0 ~\approx~  \Big[4\, \sum_I \, d^I\Big]^{-1}
 \big[ \, j_L ~-~ d^I \widehat Q_I \, \big] \,,}
 \ee
where $j_L$ the angular momentum of the supertube  (\ref{Langmomdecomp}).
In making the comparison to the results of  Section \ref{OffCenter}  recall that for a black ring
with a black hole  exactly in the center, the embedding radius in
the standard, flat $\IR^4$ metric is given by:
\be \label{embedR}{   R^2 ~=~{ l_p^6  \over L^4 }\,
\,\Big[\sum d^I \Big]^{-1} \,  \Big(J_T ~-~ d^I \widehat Q_I  \Big) \,.}
\ee
The transformation between a flat $\IR^4$ and the GH metric
with $V ={1 \over r}$ involves setting  $r = {1 \over 4} \rho^2$,   and this
leads to the relation  $R^2 = 4 R_T$.  We therefore have complete consistency
with the classical merger result.

Note that the classical merger condition is simply $r_0 \to 0$, which is, of course,
very natural.  This might, at first, seem to fall outside the validity of our
approximation, however we will see in the next section that for irreversible mergers
one does indeed maintain $\Delta, r_j \ll r_0$ in the limit  $r_0 \to 0$.
Reversible mergers cannot however be described in this approximation, and have to be analyzed numerically.

\subsection{Irreversible mergers and scaling solutions}
\label{Scaling}

All the results we have obtained in  Sections \ref{ExactRes} and  \ref{Approximations}
apply  equally to
reversible and irreversible mergers. However, since our main purpose is to obtain
microstates of a BPS back hole with classically large horizon area, we now focus
 on irreversible mergers.

We will show that an irreversible merger occurs in such a manner
that the ring radius, $r_0$, the ring width, $\Delta$, and a typical separation
of points within the black-hole blob all limit to zero while their ratios all limit to
finite values.  We will call these scaling solutions, or scaling mergers. As the
merger progresses, the throat of the solution becomes deeper and deeper, and
corresponding redshift becomes larger and larger. The resulting microstates
have a very deep throat, and will be called ``deep microstates.''

Using the solution constructed in the previous  sections, we begin decreasing
the radius of the bubbled ring, $r_0$,  by decreasing some of its flux parameters.
We take all the flux parameters of the  $(N-2)$ points in the blob to be parallel:
\be \label{kpara}
{k_j^I ~=~ \hat k_0^I  ~=~ k^I\,, \qquad j~=~1, \dots, N-2 \,,}
\ee
Further assume that all the GH charges in the black-hole blob obey
$q_j = (-1)^{j+1}$, $j=1,\dots,N-2$.  We therefore have
\bea \label{QJhat}
\widehat Q_I &~=~& 2\,(N-1)(N-3)\, C_{IJK}\, k^J k^K \,, \\
\widehat J_R&~=~& \coeff{8}{3}\,(N-1)(N-2)(N-3)\, C_{IJK}\, k^I k^J k^K  \,.
\nonumber
\eea
Define:
\be \label{muidefn}
{\mu_i ~\equiv~  \coeff{1}{6}\, (N-2 - q_i)^{-1} \,
C_{IJK} \, \sum_{{\scriptstyle j=1} \atop {\scriptstyle j \ne
i}}^{N-2} \,  \,  \Pi^{(I)}_{ij} \,   \Pi^{(J)}_{ij} \,
\Pi^{(K)}_{ij} \ {q_j \over r_{ij} }  \,, }
\ee
then the bubble equations for this blob in isolation ({\it i.e.} with
no additional bubbles, black holes or rings) are simply:
\be \label{bhbubbles}{\mu_i ~=~   \sum_{I=1}^3 \, k^I \,,}
\ee
More generally,  in any solution satisfying (\ref{kpara}), if one finds a blob in which
the $\mu_i$ are all equal to the same constant, $\mu_0$, then the
GH points in the blob must all be arranged in the same way
as an isolated black hole, but with all the positions scaled by
$\mu_0^{-1}  \big(\sum_{I=1}^3 \, k^I \big)$.

Now  consider the full set of $N$ points with $\Delta, r_j \ll r_0$.  In
Section \ref{Approximations} we solved the last two bubble equations and
determined $\Delta$ and $r_0$ in terms of the flux parameters.
The remaining bubble equations are then:
\be \label{newbubblei}{ (N-2 - q_i)\, \mu_i  +  {\alpha_i \over | \vec r_N
- \vec r_i |} - {\beta_i \over | \vec r_{(N-1)} -\vec r_i |} ~=~
\sum_{I=1}^3 \, \big((N-2 - q_i)\,k^I  + {d^I \over 2} \big) \,, }
\ee
for $i =1,\dots, N-2$.  Once again we use the multipole expansion in these equations:
\be \label{ithmultipole}{ (N-2 - q_i)\, \mu_i  ~+~  {(\alpha_i -\beta_i)  \over r_0} ~-~
 {\alpha_i \,  \Delta  \over r_0^2}~=~
\sum_{I=1}^3 \, \big((N-2 - q_i)\,k^I  + {d^I \over 2} \big) \,, }
\ee

It is elementary to show that:
\be \label{ABdiff}
{ \alpha_i - \beta_i ~=~ \coeff{1}{8}\, (j_L - d^I \,
\widehat Q_I ) ~+~  \gamma  ~-~
 \coeff{1}{8}\,  (N-2 - q_i)\, C_{IJK} \,  d^I\, k^J\, X^K \,,}
\ee
where $X^I $ is defined in (\ref{Xdefn}).
If one now uses this identity, along with (\ref{deltares}) and (\ref{rradangmom}) in
 (\ref{ithmultipole}) one obtains:
\bea
(N-2 - q_i)\, \mu_i  &~-~&  {1  \over  r_0} \,C_{IJK} \,\Big[
\coeff{1}{8}\,  (N-2 - q_i)\,  d^I\, k^J\, X^K  ~-~  \Big(1-{\alpha_i \over \alpha_0}\Big)\,
\gamma \Big]  \nonumber \\
 & ~\approx~ & (N-2 - q_i)\, \sum_{I=1}^3 \, k^I   \,.
\label{newith}
\eea
Finally, note that:
\be \label{AzeroAi}{\alpha_0 -  \alpha_i ~=~ Q \,(N-2 - q_i)\,C_{IJK} \, \big[ \coeff{1}{32}
(X^I - \coeff{1}{Q} d^I)\, (X^J - \coeff{1}{Q} d^J)\,k^K ~+~ \coeff{1}{6}\,
k^I\, k^J\,k^K \big]\,,}
\ee
and so the bubble equations (\ref{newbubblei}) reduce to:
\bea
\label{unieqns}
\mu_i & \approx &
\Big(\sum_{I=1}^3 \, k^I \Big)  ~+~ {1  \over  r_0} \,C_{IJK} \,
\Big[ \coeff{1}{8}\,  d^I\, k^J\, X^K \nonumber \\
&~& \qquad \qquad ~-~
 \alpha_0^{-1}\,Q\, \gamma\, \big( \coeff{1}{32}\,
(X^I - \coeff{1}{Q} d^I)\, (X^J - \coeff{1}{Q} d^J)\,k^K ~+~ \coeff{1}{6}\,
k^I\, k^J\,k^K \big) \,   \Big]    \nonumber \\
& \approx &
\Big(\sum_{I=1}^3 \, k^I \Big)  ~+~ {1  \over  r_0} \,C_{IJK} \,
\Big[ \coeff{1}{8}\,  d^I\, k^J\, X^K  \\
&~& \qquad \qquad \qquad \qquad \qquad  -~ \alpha_0^{-1}\,Q\, \gamma\, \big(
\coeff{1}{32}\, X^I  \,  X^J  \,k^K +  \coeff{1}{6}\, k^I\, k^J\,k^K \big) \,   \Big]   \,,
\eea
since we are assuming $X^I$ is large compared to $Q^{-1} d^I$.

Observe that the right-hand side of (\ref{unieqns}) is independent of
$i$, which means that the first $(N-2)$ GH points satisfy a scaled
version of the equations (\ref{bhbubbles}) for a isolated, bubbled
black hole.  Indeed, if $\vec r_i^{BH}$ are the positions of a set of
GH points satisfying (\ref{bhbubbles}) then we can solve
(\ref{unieqns}) by scaling the black hole solution, $\vec r_i =
\lambda^{-1} \vec r_i^{BH}$, where the scale factor is given by:
\bea
  \lambda &~\approx~&
1~+~  {1  \over  r_0} \,\Big(\sum_{I=1}^3 \, k^I \Big)^{-1}\,C_{IJK} \,
\Big[ \coeff{1}{8}\,  d^I\, k^J\, X^K  \nonumber \\
&~& \qquad \qquad  ~-~ \alpha_0^{-1}\,Q\, \gamma\, \big(
\coeff{1}{32}\, X^I  \,  X^J  \,k^K ~+~ \coeff{1}{6}\, k^I\, k^J\,k^K \big) \,   \Big]
 \label{scalefac} \,.
\eea

Notice that as one approaches the critical ``merger'' value, at which
$\Omega = j_L - d^I \, \widehat Q_I =0$, (\ref{scalefac}) implies that
the distance, $r_0$, must also scale as $\lambda^{-1}$.  Therefore the
merger process will typically involve sending $r_0 \to 0$ while
respecting the assumptions made in our approximations ($\Delta, r_i
\ll r_0$).  The result will be a ``scaling solution'' in which all
distances in the GH base are vanishing while preserving their relative
sizes.

In \cite{deep} this picture of the generic merger process was verified
by making quite a number of numerical computations\footnote{A merger
  was tracked through a range where the scale factor, $\lambda$,
  varied from about $4$ to well over $600$.  It was also verified that
  this scaling behavior is not an artefact of axial symmetry.
  Moreover, in several numerical simulations the GH points of the
  black-hole blob were arranged along a symmetry axis but the bubbled
  ring approached the black-hole blob at various angles to this axis;
  the scaling behavior was essentially unmodified by varying the angle
  of approach.  }; we urge the curious reader to refer to that paper
for more details. In  Section \ref{Numerics} we will only present one very
simple scaling solution, which illustrates the physics of these
mergers.

An important exception to the foregoing analysis arises when the term
proportional to $r_0^{-1}$ in (\ref{unieqns}) vanishes to leading
order.  In particular, this happens if we violate one of the
assumptions of our analysis, namely, if one has:
\be
 \label{XIzero}{X^I ~\equiv~  2\,f^I - d^I - 4\,(N-2)\,k^I ~\approx~0\,,}
\ee
to leading order order in $Q^{-1} d^I$.  If $X^I$ vanishes one can see
that, to leading order, the merger condition is satisfied:
\bea
 \Omega &~\equiv~& j_L - d^I \, \widehat Q_I \nonumber \\
 &~=~ &  \coeff{1}{8}  \, C_{IJK} \, d^I\, \big[X^J X^K - \coeff{1}{3}\, Q^{-2}\,
 (4\,Q -1) \, d^J d^K - 16  \,k^J k^K \big]  \label{Omform}\\
&~\approx~& 0  \nonumber \,,
\eea
and so one must have $r_0 \to 0$.  However, the foregoing analysis
is no longer valid, and so the merger will not necessarily result in a
scaling solution.

An important example of this occurs when  $k^I$, $d^I$ and $f^I$ are all parallel:
\be \label{kdfpara}{k^I ~=~ k\, u^I\,, \qquad d^I ~=~ d\, u^I\,, \qquad f^I ~=~ f\, u^I\,,}
\ee
for some fixed $u^I$.  Then the merger condition (\ref{Omform}) is
satisfied to leading order, only when $X \equiv (2\, f - d -4\,
(N-2)\,k)$ vanishes.

For non-parallel fluxes it is possible to satisfy the merger
condition, (\ref{Omform}), while keeping $X^I$ large, and the result
is a scaling solution.

Even if it looks like irreversible mergers progress until the final
size on the base vanishes, this is an artifact of working in a
classical limit an ignoring the quantization of the fluxes. After
taking this into account we can see from (\ref{unieqns}) that $r_0$
cannot be taken continuously to zero because the $d^I, f^I, X^I$ and
$k^I$ are integers of half-integers.  Hence, the final result of an
irreversible merger is a microstate of a high, but finite, redshift
and whose throat only becomes infinite in the classical limit.

In order to find the maximum depth of the throat, one has to find the
smallest allowed value for the size of the ensemble of GH points in the $\IR^3$
base of the GH space.
During the irreversible merger all the distances scale, the size of
the ensemble of points will be approximately equal to the distance
between the ring blob and the black hole blob, which is given by
(\ref{rradangmom}). Since $j_L - d^I \widehat Q_I$ is quantized, the minimal
size of the ensemble of GH points is given by:
\be
\label{rminbase}{r|_{\rm min} \approx {1 \over d^1+d^2+d^3} \,.}
\ee
More generally, in the scaling limit, the GH size of a solution with left-moving
angular momentum $J_L$ is
\be
\label{rminbas}{r|_{\rm min} \approx {J_L \over d^1+d^2+d^3} \,.}
\ee

Since the $d^I$ scale like the square-roots of the ring charges, we
can see that in the classical limit, $r|_{\rm min}$ becomes zero and the
throat becomes infinite.

\subsection{Numerical results for a simple merger}
\label{Numerics}

Given that most of the numerical investigations and most of the
derivations we have discussed above use black hole microstate made
from a very large number of points, it is quite hard to illustrate
explicitly the details of a microstate merger.

To do this, it is much more pedagogical to investigate a black hole
microstate that is made from three points, of GH charges $-n$, $2n+1$,
and $-n$, and its merger with the black ring microstate of GH charges
$-Q$ and $+Q$. This black-hole microstate can be obtained by
redistributing the position of the GH points inside the black-hole blob
considered in Section \ref{Scaling},  putting all the $+1$ charges
together and putting half of the $-1$ charges together on one side of
the positive center and the other half on the other
side\footnote{Since the $k$ parameters on the black-hole points are the same,
 the bubble equations give no obstruction to moving black-hole centers of the
  same GH charge on top of each other.}

We consider a configuration with 5 GH centers of charges
\be \label{ghch}{q_1=-12,~q_2=25,~q_3=-12,~q_4=-20,~q_5=20~.}
\ee
The first three points give the black-hole ``blob,'' which can be
thought as coming from a blob of $N-2 = 49$ points upon redistributing
the GH points as described above; the $k^I$ parameters of the black
hole points are
\be
\label{ghch2}
{k_1^I = q_1 \hat k_0^I,~k_2^I = q_2 \hat k_0^I,~k_3^I = q_3  \hat
k_0^I~,}
\ee
where $ \hat k_0^I $ is the average of the $k^I$ over the black-hole points,
defined in (\ref{kIavg}). To merge the ring and the black hole microstates we
have varied $ \hat k_0^2 $ keeping $ \hat k_0^1 $ and $ \hat k_0^3 $
fixed:
\be
 \label{khatnum}{ \hat k_0^1 ={5 \over 2},~ \hat k_0^3 ={1\over 3},~  }
\ee
We have also kept fixed the ring parameters $f^I $ and $d^I$:
\be
\label{fdnum}{d^1=100,~d^2=130,~d^3=80,~f^1=f^2=160,~f^3=350  }
\ee
The relation between these parameters and the $k^I$ of the ring is
given in (\ref{ringchgs}), where $N-2$ (the sum of $|q_i|$ for the
black hole points) is now $|q_1| +|q_2|+ |q_3| = 49$~.


\def\nicespacea#1{{~~#1~~}}
\goodbreak
{\vbox{
{
$$
\vbox{\offinterlineskip\tabskip=0pt
\halign{\strut\vrule#
&\vrule\hfil #\hfil
&\vrule\hfil #\hfil
&\vrule\hfil #\hfil
&\vrule\hfil #\hfil
&\vrule\hfil #\hfil
&\vrule\hfil #\hfil
&\vrule\hfil #\hfil
&\vrule\hfil #\hfil
&\vrule\hfil #\hfil
\cr
\noalign{\hrule}
& &\nicespacea{$ \hat k^2_0$}&\nicespacea{$x_4-x_3$}&
\nicespacea{$\displaystyle{x_4-x_3 \over x_2-x_1}$}&
\nicespacea{$\displaystyle{x_2-x_1\over  x_3-x_{2}}$}
&\nicespacea{$\displaystyle{x_2-x_1\over x_5-x_4}$}&\nicespacea{$J_L$}&
\nicespacea{${\cal{H}}$}&
\cr
\noalign{\hrule height1pt}
&~0~& 3.0833 &~175.5~ & 2225 & 1.001 & 2.987 &~215983~&.275
&
\cr
\noalign{\hrule }
&~1~& 3.1667 &23.8 & 2069 & 1.001 & 3.215 &29316&.278
&
\cr
\noalign{\hrule }
&~2~& 3.175 &8.65 & 2054 & 1.001 & 3.239 &10650&.279
&
\cr
\noalign{\hrule }
&~3~& 3.1775 &4.10 & 2049 & 1.001 & 3.246 &5050&.279
&
\cr
\noalign{\hrule }
&~4~& 3.178 &3.19 & 2048 & 1.001 & 3.248 &3930&.279
&
\cr
\noalign{\hrule }
&~5~& 3.17833 &2.59 & 2048 & 1.001 & 3.249 &3183&.279
&
\cr
\noalign{\hrule }
&~6~& 3.17867 &1.98 & 2047 & 1.001 & 3.250 &2437&.279
&
\cr
\noalign{\hrule }
&~7~& 3.1795 &.463 & 2046 & 1.001 & 3.252 &570&.279
&
\cr
\noalign{\hrule }
&~8~& ~3.17967~ &.160 & 2045 & 1.001 & 3.253 &197&.279
&
\cr
\noalign{\hrule}}
\hrule}$$
\leftskip 2pc
\rightskip 2pc\noindent{
  \sl \baselineskip=8pt {\bf Table 1}: Distances between points in the
  scaling regime. The parameter ${\cal H} \equiv {Q_1 Q_2 Q_3-J_R^2 /4
    \over Q_1 Q_2 Q_3}$ measures how far away the angular momentum of the
resulting solution  is from the angular momentum of the maximally-spinning
black hole with identical charges.  The value of $\hat
  k_0^2$ is varied to produce the merger, and the other parameters of
  the configuration are kept fixed: $ Q=20,~q_1=q_3=-12,~q_2=25,~\hat
  k_0^1={5 \over 2},~\hat k_0^3={1\over
    3},~d^1=100,~d^2=130,~d^3=80,~f^1=f^2=160,~f^3=350$. Both the
  charges and $J_R$ remain approximately constant, with $J_R \approx 3.53
  \times 10^7$
}
}
\vskip7pt}}

The charges and $J_R $ angular momentum of the solutions are approximately
\be \label{qjnum}{Q_1 \approx 68.4 \times 10^3,~Q_2 \approx 55.8 \times 10^3,~
Q_3 \approx 112.8 \times 10^3,~J_R \approx 3.53 \times 10^7,}
\ee
while $J_L$ goes to zero as the solution becomes deeper and deeper.

Solving the bubble equations (\ref{BubbleEqns}) numerically,  one
obtains the positions $x_i$ of the five points as a function of $\hat
k^2_0$. As we can see from the table above, a very small increase in
the value of $\hat k^2_0$ causes a huge change in the positions of the
points on the base. If we were merging real black holes and real black
rings, this increase would correspond to the black hole and the black
ring merging. For the microstates, this results in the scaling
described above: all the distances on the base become smaller, but
their ratios remain fixed.

Checking analytically that these solutions have no closed timelike
curves is not that straightforward, since the quantities in
(\ref{noCTCs}) have several hundred terms. However, in \cite{deep} it
was found numerically that such closed timelike curves are absent, and
that the equations (\ref{noCTCs}) are satisfied throughout the scaling
solution.

\subsection{The metric structure of the deep microstates}
\label{MetricStruc}

The physical metric is given by (\ref{elevenmetric})  and (\ref{fivemetric}) and the
physical distances are related to the coordinate distances on the
the $\IR^3$ base of the GH space, $d \vec y \cdot d \vec y$  via:
\be  \label{physlen}{ds^2 ~=~ (Z_1 Z_2 Z_3)^{1/3} \, V \, d \vec y \cdot d \vec y \,.}
\ee
The physical lengths are thus determined by the functions,
$Z_I V$, and if one has:
\be  \label{AdSthroat}{(Z_1 Z_2 Z_3)^{1/3} \, V  ~\sim~ {1 \over r^2} \,,}
\ee
then the solution looks is an $AdS_2 \times S^3$ black hole throat. In
the region where the constants in the harmonic functions become
important, this throat turns into an asymptotically flat $\IR^{(4,1)}$
region.  Near the GH centers that give the black-hole bubbles, the
function $Z_1 Z_2 Z_3$ becomes constant.  This corresponds to the
black-hole throat ``capping off''.  As the GH points get closer in the
base, the region where (\ref{AdSthroat}) is valid becomes larger, and hence
the throat becomes longer.

As one may intuitively expect, in a scaling solution the ring is
always in the throat of the black hole. Indeed, the term ``1'' on the
right hand side of (\ref{scalefac}) originates from the constant terms in
$L_I$ and $M$, defined in (\ref{LMexp}). In the scaling regime this term is
sub-leading, which implies the ring is in a region where the $1$ in the
$L_I$ (and hence the $Z_I$) is also sub-leading.  Hence, the ring lies
in the $AdS$ throat of the black-hole blob.

Increasing the scale factor, $\lambda$, in (\ref{scalefac}) means that
the bubbles localize in a smaller and smaller region of the GH base,
which means that the throat is getting longer and longer.  The
physical circumference of the throat is fixed by the charges and the
angular momentum, and remains finite even though the blob is shrinking
on the GH base. Throughout the scaling the throat becomes deeper and
deeper; the ring remains in the throat, and also descends deeper and
deeper into it, in direct proportion to the overall depth of the
throat.

On a more  mechanistic level, the physical distance through the blob
and the physical distance from the blob to the ring are
controlled by  integrals of the form:
\be  \label{physsize}
{ \int \, (Z_1 Z_2 Z_3 \, V^3 )^{1/6}  \, d \ell \,.}
\ee
In the throat the behavior of this function is given by (\ref{AdSthroat})
and this integral is logarithmically divergent as $r \to 0$.  However,
the $Z_I$ limit to finite values at $\vec r = \vec r_j$  and between two very
close, neighboring GH points in the blob, the integral has a dominant
contribution  of the form
\be  \label{lowerbd}
{ C_0 \ \int \,|(x-x_i)(x-x_{j})|^{-1/2}   \, d x \,,}
\ee
for some constant, $C_0$, determined by the flux parameters.  This
integral is finite and indeed is equal to $C_0 \, \pi$.   Thus we see
that the throat gets very long but then caps off with bubbles of finite
physical size.

\subsection{Are deep microstates dual to typical boundary microstates?}
\label{DeepTypical}

As we have seen in   Section \ref{MetricStruc}, the throats of the deep
microstates become infinite in the classical limit. Nevertheless,
taking into account flux quantization one can find that the GH radius
of microstates does not go all the way to zero, but to a finite value
(\ref{rminbase}), which corresponds to setting $J_L=1$.

One can estimate the energy gap of the solution by considering the
lightest possible state at the bottom of the throat, and estimating
its energy as seen from infinity.  The lightest massive particle one
can put on the bottom of the throat is not a Planck-mass object, but a
Kaluza-Klein mode on the $S^3$. Its mass is
\be  \label{massKK}
{m_{KK} = {1 \over R_{S^3}} = {1 \over (Q_1 Q_2 Q_3)^{1\over 6} }}
\ee
and therefore the mass gap in a microstate of size $r_{\rm min}$ in the GH base is:
\be  \label{gapmicro}
{\Delta E_{r_0} = m_{KK} \sqrt{g_{00}}|_{r=r_{\rm min}}
=  m_{KK} (Z_1 Z_2 Z_3)^{-1/3}|_{r=r_{\rm min}} = {r_{\rm min} \over (Q_1 Q_5 Q_P)^{1/2}}.}
\ee
For a ring-hole merger, $r_{\rm min}$ depends on the
sum of the $d^I$, and so its relation with the total charges of the system is
not straightforward. Nevertheless, we can consider a regime where
$Q_1\sim Q_5 > Q_P$, and in this regime the dipole charge that
dominates the sum in (\ref{rminbas})
is $d^3 \approx \sqrt{ Q_1 Q_5 \over Q_P}$. Hence
\be  \label{rminlim}
{r_{\rm min} = {J_L \over d^3} \approx J_L \sqrt{Q_P \over Q_1 Q_5}~, }
\ee
%

\exbox{Show that the mass gap for a KK mode sitting on the bottom
of the throat at
$r\sim r_{\rm min}$ is
\be  \label{gap}
{\Delta E_{r_{\rm min}} \approx {J_L \over Q_1 Q_5}~.}
\ee
}

This M-theory frame calculation is done in the limit $Q_1 \sim Q_5 >
Q_P$, which is the limit in which the solution, when put into the
D1-D5-P duality frame, becomes asymptotically $AdS_3 \times S^3 \times
T^4$. As shown in \cite{Bena:2004tk}, in this limit $d^1+d^2+d^3 \approx d^3$,
which justifies going from (\ref{rminbas}) to (\ref{rminlim}).

For $J_L=1$, the mass gap computed in the bulk (\ref{gap}) matches the
charge dependence of the mass gap of the black hole \cite{MaldacenaDS}.
Moreover, this mass gap should also match the
mass gap of the dual microstate in the D1-D5 CFT.

As it is well known (see \cite{DavidWN,magoo} for reviews)
the states of this CFT can be
characterized by various ways of breaking a long effective string of
length $N_1 N_5$ into component strings.  BPS momentum modes on these
component strings carry $J_R$. The fermion zero modes of each
component string allow it in addition to carry one unit of $J_L$.
The typical CFT microstates that contribute to the entropy of the
three-charge black hole have one component string
\cite{bmpv}; microstates dual to objects that have a
macroscopically large $J_L$ have the effective string broken into many
component strings \cite{LuninJY,lmm,Bena:2004tk}.

Hence, the only way a system can have a large $J_L$ is to be have many
component strings.  The CFT mass gap corresponds to exciting the longest
component string, and is proportional to the inverse of its length.

The formula (\ref{gap}) immediately suggests what the dual of a deep
microstate should be.  Consider a long effective string of length $N_1
N_5$ broken into $J_L$ component strings of equal length.  Each
component string can carry one unit of left-moving angular momentum,
totaling up to $J_L$. The length of each component string is
\be  \label{lbit}
{l_{\rm component} = {N_1 N_5 \over J_L}~,}
\ee
and hence the CFT mass gap is
\be  \label{cftgap}
{\Delta E_{CFT} \approx {J_L \over N_1 N_5}~.}
\ee
This agrees with {\it both} the $J_L$ dependence and the dependence on
the charges of the gap computed in the bulk.  While we have been
cavalier about various numerical factors of order one, the agreement
that we have found suggests that deep microstates of angular momentum
$J_L$ are dual to CFT states with $J_L$ component strings.  If this is
true, then the deepest microstates, which have $J_L = 1$, correspond
to states that have only one component string, of length $N_1 N_5$.
This is a feature that typical microstates of the three-charge black
hole have, and the fact that deep microstates share this feature is
quite remarkable.

Our analysis here has been rather heuristic.  It would be very
interesting to examine this issue in greater depth by finding, at
least approximate solutions to the wave equation in these backgrounds,
and performing an analysis along the lines of \cite{LuninJY,lmm}.

\section{Implications for black-hole physics}
\label{Conclusions}

\subsection{Microstate geometries}
\label{MicrostateGeometries}

As we have seen, string theory contains a huge number of smooth
configurations that have the same charges and asymptotics as the
three-charge BPS black hole in five dimensions. Counting these
configurations, or relating them to the states of the boundary CFT
will allow one to prove or disprove the claim that black holes in
string theory are not fundamental objects, but rather a statistical
way to describe an ensemble of black-hole-sized configurations with no
horizon and with unitary scattering.  This will help in establishing
the answer to the key question {\it ``What is the $AdS$-CFT dual of
  the states of the D1-D5-P system?''}  Nevertheless, even if a
definitive answer may be hard to establish and prove, it is well worth
exploring in more detail the three (or four) possible answers to this
question, particularly in light of our current understanding of black
hole microstates:

\medskip

\noindent  {\bf Possibility  1: One bulk solution dual to
many boundary microstates}  \\

\noindent
It is possible that some of the states of the CFT, and in particular
the typical ones (whose counting gives the black hole entropy) do not
have individual bulk duals, while some other states do.  However this
runs counter to all our experience with the $AdS$-CFT correspondence:
In all the examples that have been extensively studied and
well-understood (like the D1-D5 system, Polchinski-Strassler
\cite{pol-str}, giant gravitons and LLM \cite{giant,Lin:2004nb}, the
D4-NS5 system \cite{d4ns5}) the $AdS$-CFT correspondence relates
boundary states to bulk states and boundary vacua to bulk vacua.

\begin{figure}
\centering
\includegraphics[height=4cm]{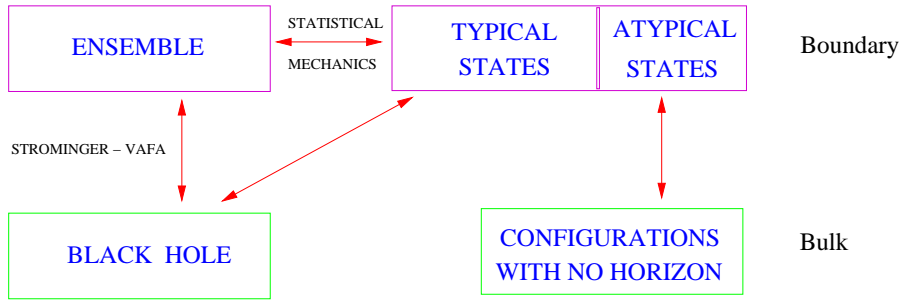}
\caption{An schematic description of Possibility 1.
}
\label{non-Mathur1}
\end{figure}

It is logically possible that, for the D1-D5-P system only, the path
integrals in the bulk and on the boundary are related in the standard
way via the $AdS$-CFT correspondence, and yet all the boundary states
that give the CFT entropy are mapped into one black hole solution in
the bulk.  This possibility is depicted in Fig. \ref{non-Mathur1}.
However, this possibility raises a lot of questions.  First, why would
the D1-D5-P system be different from all the other systems mentioned
above.  Moreover, from the microscopic (or CFT) perspective, there is
nothing special about having three charges: One can map the boundary
states in the D1-D5-KKM system in {\it four} dimensions to the
corresponding bulk microstates
\cite{Bena:2005ay,Srivastava:2006pa,3charge4d}.  The only reason for
which the D1-D5-P system would be different from all the other systems
would be the fact that it has the right amount of charges to create a
macroscopically-large event horizon in five dimensions.  To have such
divergently different behaviour for the D1-D5-P system in five
dimensions would be, depending on one's taste, either very deep or,
more probably, very bizarre.

Even if the typical states of the three-charge system
correspond to one single black hole, we have seen that besides this
black hole there exists a huge number of smooth solutions that also
are dual to individual states of this CFT. Hence, according to Possibility 1,
some states of the CFT would have individual bulk duals and
some others would not (they would be dual as an ensemble to the black
hole).  This distinction is very unnatural.   One might explain this if
the states dual to the black hole and the ones dual to
microstate geometries are in different sectors of the CFT, but this is
simply not the case. We have seen in Section
\ref{DeepTypical} that the deep bulk microstates correspond to
boundary states that have one (or several) long component string(s).
Hence, they belong to the same CFT sector as the typical microstates. If
typical microstates did not have individual bulk duals, then in the same
sector of the CFT we would have both states with a bulk dual and
states without one.  While not obviously wrong, this appears, at least,
dubious and unjustifiable from the point of view of the CFT.

\medskip

\noindent {\bf Possibility 2: Typical bulk microstate very similar to
black hole.} \\

\begin{figure}
\centering
\includegraphics[height=4cm]{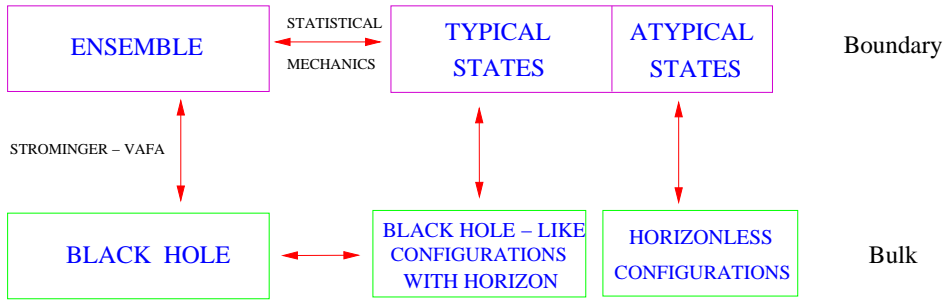}
\caption{An schematic description of Possibility 2.
}
\label{non-Mathur2}
\end{figure}

\noindent
It is possible that all the states of the CFT are
dual to geometries in the bulk, but the typical states are dual to
geometries that have a horizon, and that only differ from the
classical black hole by some Planck-sized fuzz near the singularity.
This situation is depicted in Fig. \ref{non-Mathur2}.

This also has a few problems. First, there are arguments,
\cite{mathur-rev}, that if the microstates of the black hole only differ
from the classical geometry near the singularity, this does not solve
the information paradox.  Putting such arguments on one side, there is
a more obvious objection: Possibility 2 means that typical microstates
would have horizons, and so it would seem that one would have to ascribe
an entropy to each microstate, which violates one of the principles of
statistical mechanics.  A counterargument here is to observe that one
can always ascribe an ad hoc entropy to a microstate of any system
simply by counting the number of states with the same macroscopic
properties.  What really distinguishes a microstate from an ensemble
is that one has complete knowledge of the state of the former and that
one has lost some knowledge of the state in the latter.  The
counterargument asserts that the presence of the horizon does not
necessarily indicate information loss, and that the complete
information might ultimately be extracted from something like the
Hawking radiation.  Thus microstates could have a horizon if
information is somehow stored and not lost in the black hole.  This is
a tenable viewpoint and it is favored by a number of relativists but
it defers the issue of how one decodes the microstate information to
some unknown future physics whereas string theory appears to be
pointing to a very interesting answer in the present.

There is also one of the objections raised in possibility 1: We have
seen that some CFT states corresponding to long component strings are
dual to deep microstates, that have no horizon.  If the second
possibility is correct, then other states in {\it the same sector} of
the CFT would be dual to geometries that have a horizon and a
singularity, and are therefore drastically different.  Moreover, for
extremal black holes, the distance to the horizon is infinite, while
the distance to the cap of the microstates is finite (though divergent
in the classical limit). Hence, in the same sector of the CFT,
some states would be dual to supergravity solutions with an infinite
throat, while others would be dual to solutions with a finite throat.
This again appears quite dubious from the point of view of the CFT.

One can also think about obtaining the bulk microstate geometries
by starting from a weak-coupling microstate (which is a certain
configuration of strings and branes) and increasing the string
coupling. During this process, we can imagine measuring the distance
to the configuration. If a horizon forms, then this distance would
jump from being finite to being infinite.  However, for the
smooth microstates, this distance is always a continuous function of
the string coupling, and never becomes infinite. While the infinite
jump of the length of the throat is a puzzling phenomenon, equally
puzzling is the fact that only some microstates would have this
feature, while some very similar ones would not\footnote{We thank
  Samir Mathur for pointing out this argument to us.}.

\medskip

\noindent {\bf Possibility  3:  Typical bulk microstate differs from black hole
at the scale of the horizon.} \\

\begin{figure}
\centering
\includegraphics[height=4cm]{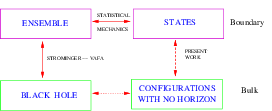}
\caption{An schematic description of Possibility 3.
}
\label{Mathur-end}
\end{figure}

\noindent
It is possible that all boundary microstates are dual to
horizonless configurations. The classical black hole geometry is only
a thermodynamic description of the physics, which stops being valid at
the scale of the horizon, much like fluid mechanics stops being a good
description of a gas at scales of order the mean free path. For
physics at the horizon scale, one cannot rely on the thermodynamic
description, and has to use a ``statistical'' description in terms of
a large number of microstates.  This possibility is depicted in
Fig. \ref{Mathur-end}.

Since these microstates have no horizon, they have unitary scattering but
it takes a test particle a very long time to escape from this
microstate. Hence, if this possibility is correct, the information
paradox is reduced to nothing but an artifact of using a thermodynamic
description beyond its regime of validity.  This possibility now
splits into two options, having to do with the appropriate description
of the typical black hole microstates: \medskip

\noindent {\bf Possibility 3A:  Typical microstates
cannot be described in supergravity and require the full
force of string theory.}

 \medskip

\noindent {\bf Possibility 3B: Typical microstates
can be described in supergravity. }

\medskip

As we cannot, yet, explore or count large strongly-interacting
horizon-sized configurations of branes and strings using our current
string theory technology,  Possibility 3A would be more challenging to
establish or analyze.
We therefore need to examine
Possibility 3B in great detail to see if it is true, or at least
determine the extent to which supergravity can be used.  One way to do
this is by counting the microstates, using for example counting
techniques of the type used in \cite{counting}.  Another approach is
to find the exact (or even approximate) dictionary between the states
of the CFT and the bubbled geometries in the bulk.  Anticipating (or
perhaps speculating) a bit, one could imagine that, as a result of
this investigation, one could relate the number of bubbles of a deep
microstate to the distribution of the momentum on the long component
string of the dual CFT state. Such a relation (which could in
principle be obtained using scattering experiments as in
\cite{lmm,GiustoIP}) would indicate whether typical bulk microstates
have large bubbles or Planck-sized bubbles, and would help distinguish
between Possibilities 3A and 3B.

One of the interesting questions that needs to be addressed here is:
{\it What about non-extremal black holes?}.  All the arguments
presented in this review in favor of the third possibility have been
based on supersymmetric black holes, and one can legitimately argue
that even if these black holes describe an ensemble of smooth
horizonless configurations, it may be that non-supersymmetric black
holes (like the ones we have in the real world) are fundamental
objects, and not ensembles.  The arguments put forth to support
Possibility 3 for non-extremal black holes are rather more limited. Indeed,
on a technical level, it is much more difficult to find
non-supersymmetric, smooth microstate geometries, but some progress
has been made \cite{ross}.  There are nevertheless some interesting physical
arguments based primarily on the phenomenon of charge fractionation.

The idea of charge fractionation \cite{Das:1996ug,Maldacena:1996ds}
is most simply illustrated by the
fact that when you put $N_1$ D1 branes (or strings) in a periodic box
of length $L$, then the lowest mass excitation carried by this system
is not of order $L^{-1}$, but of order $(N_1 L)^{-1}$.  The
explanation is that the branes develop multi-wound states with the
longest effective length being of order $N_1 L$.  Similarly, but via a
rather more complex mechanism, the lowest mass excitations of the
D1-D5 system vary as $(N_1 N_5)^{-1}$.  This is called charge
fractionation.  It is this phenomenon that leads to the CFT mass gap
given in (\ref{cftgap}).  The other important consequence of
fractionation is that the corresponding ``largest'' natural physical length
scale of the system grows as $N_1 N_5$.  One of the crucial physical
questions is how does the ``typical'' length scale grow with charge.
That is, what is the physical scale of the most likely (or typical)
configuration.  It is believed that this will grow as some positive
power of the underlying charges, and this is the fundamental reason
why it is expected that microstate geometries are ``large'' compared
to the Planck scale and that microstate geometries are not just
relevant within a few Planck units of the singularity, but extend to
the location of the classical horizon.

This argument can be extended to non-BPS systems.  Configurations of
multiple species of branes also exhibit fractionation.  For this
reason, it is believed that, given a certain energy budget, the way to
get most entropy is to make brane-antibrane pairs of different
sorts\footnote{This idea has been used in formulating microscopic
  brane-antibrane models for near-BPS black holes \cite{nearBPS} and
  for black branes \cite{Danielsson:2001xe}, and has recently received
  a beautiful confirmation in the microscopic calculation of the
  entropy of extremal non-BPS black holes \cite{gary-roberto}. It has
  also been applied to cosmology \cite{mathur-cosmo} and to
  understanding the Gregory Laflamme instability \cite{Gregory:1993vy}
  microscopically \cite{greglaf,Harmark:2006df}.}.  Putting together
these different kinds of branes creates a system with very light
(fractionated) modes, whose mass is much much lower than the Planck
scale.  These modes can then ``extend'' all the way to the horizon,
and have to be taken into account when discussing physics at this
scale.

One of the counterarguments to the third possibility is that one can
collapse a shell of dust and create a horizon at very weak curvatures,
long before the black hole singularity forms. Moreover, the larger the
mass, the longer will be the time elapsed between the formation of the
horizon and the singularity. Hence, it naively appears that the
horizon cannot possibly be destroyed by effects coming from a
singularity that is so far away.  Nevertheless, if fractionation gives
the correct physics, then one can argue that as the mass of the
incoming shell increases, the number of brane-antibrane pairs that are
created becomes larger, and hence the mass of the ``fractionated''
modes becomes smaller; these modes will then affect the physics at
larger and larger scales, which can be argued to be of order the
horizon size.  In this picture the collapsing shell would reach a
region where a whole new set of very light degrees of freedom exist.
Since these ``fractionated'' degrees of freedom have a much larger
entropy, the shell will dump all its energy into these modes, which
would then expand to the horizon and destroy the classical geometry up
to this scale.  More details in support of these arguments can be
found in \cite{mathur-rev}.

On the other hand, one may hope to preserve the status quo for
non-extremal black holes by arguing that fractionation is a phenomenon
that is based on weakly coupled D-brane physics, and is not
necessarily valid in the range of parameters where the black hole
exists.  This, however, leaves one with the problem of explaining why
fractionation appears to be occurring in extremal black holes and why
non-BPS black holes should be any different.  Indeed, if the classical
solution for the extremal black hole is proven to give an incorrect
description of the physics at the horizon when embedded into a quantum
theory of gravity, it is hard to believe that other similar,
non-extremal solutions will give a correct description of the physics
at the horizon.  It will be much more reasonable to accept that all
the classical black hole solutions are thermodynamic descriptions of
the physics, which break down at the scale of the horizon.

The most direct support for the smooth microstate structure of
non-extremal black holes would be the construction and counting of
smooth, non-extremal geometries generalizing those presented here,
like those constructed in \cite{ross}. Such constructions are
notoriously difficult and, barring a technical miracle in the
construction of non-BPS solutions, it is hard to hope that there will
be a complete classification of such geometries in the near future.
On the other hand, it is instructive and encouraging to recall the
developments that happened shortly after the original state counting
arguments of Strominger and Vafa for BPS black holes: There was a lot
of analysis of near-BPS configurations and confirmation that the
results could be generalized perturbatively to near-BPS states with
small numbers of anti-branes.  This might prove fruitful here and
would certainly be very useful in showing that generic smooth
microstate geometries are not special properties of BPS objects.  It
would thus be interesting to try, either perturbatively, or perhaps
through microstate mergers, to create near-BPS geometries.

Finally, the fact that the classical black hole solution does not
describe the physics at the scale of the horizon seems to contradict
the expectation that this solution should be valid there since its
curvature is very small.  There are, however, circumstances in which
this this expectation can prove wrong.  First, if a solution has a
singularity, it oftentimes does not give the correct physics even at
very large distance away from this singularity because the boundary
conditions at the singularity generate incorrect physics even in
regions where the curvature is very low.  Such solutions therefore
have to be discarded.  A few examples of such solutions are the
Polchinski-Strassler flow \cite{pol-str} without brane
polarization \cite{gppz-pw},
or the singular KK giant graviton \cite{giant,Bena:2004qv}.  The
reason why we do not automatically discard black hole solutions is
that their singularities are hidden behind horizons and sensible
boundary conditions can be imposed at the horizon.  However, this does
not imply that all solutions with singularities behind horizons must
be good: It only shows that they should not be discarded {\it a
  priori}, without further investigation.  What we have tried to show
is that if the third possibility is correct then the investigation
indicates that the classical BPS black-hole solution should not be
trusted to give a good description of the physics at the scale of the
horizon.

\subsection{A simple analogy}
\label{Analogy}

To understand Possibility 3 a little better, it is
instructive to recall the physics of a gas, and to propose an analogy
between the various descriptions of a black hole and the various
descriptions of this gas.

For scales larger than the mean free path, a gas can be described by
thermodynamics, or by fluid mechanics. At scales below the mean free
path, the thermodynamic description breaks down, and one has to use a
classical statistical description, in which one assumes all the
molecules behave like small colliding balls. When the molecules are
very close to each other, this classical statistical description
breaks down, and we have to describe the states of this gas quantum
mechanically. Moreover, when the temperature becomes too high, the
internal degrees of freedom of the molecules become excited, and they
cannot be treated as small balls. There are many features, such as
shot noise or Brownian motion, that are not seen by the thermodynamic
description, but can be read off from the classical statistical
description. There are also features that can only be seen in the full
quantum statistical description, such as Bose-Einstein condensation.

For black holes, if Possibilities 3A or 3B are correct, then the
$AdS$-CFT correspondence relates quantum states to quantum states, and
we expect the bulk dual of a given boundary state to be some
complicated quantum superposition of horizonless configurations.
Unfortunately, studying complicated superpositions of geometries is
almost impossible, so one might be tempted to conclude that even if
Possibilities 3A or 3B are correct, there is probably no new physics
one can learn from it, except for an abstract paradigm for a solution
to the information paradox. Nevertheless, we can argue by analogy to a
gas of particles that this is not the case.

Consider a basis for the Hilbert space of the bulk configurations.  If
this basis is made of coherent states, some of the states in this
basis will have a semiclassical description in terms of a supergravity
background. This would be very similar to the situation explored in
\cite{Lin:2004nb}, where bubbled geometries correspond to coherent CFT
states. The supergravity solutions we have discussed in these notes
are examples of such coherent states.  The main difference between the
Possibility 3A and 3B has to do with whether the coherent states that
form a basis of the Hilbert space can be described using supergravity
or whether one has to use string theory to describe them. By analogy
with the gas, this is the difference between the regime where the
simple ``colliding ball'' model is valid, and the regime where one
excites internal degrees of freedom of the molecules.

If supergravity is a good description of most of the coherent states,
we can argue that we have constructed the black hole analogue of the
classical statistical description of an ideal gas. Even if most of the
coherent states can only be described in a full string-theoretic
framework, one can still hope that this will give the analogue of an,
albeit more complicated, classical statistical description of the gas.
Both these descriptions are more complete than the thermodynamic
description, and for the gas they capture physics that the
thermodynamic description overlooks.  Apart from solving the
information problem, it would be very interesting to identify
precisely what this physics is for a black hole.  Indeed, as we will
explain below, it might lead to some testable signature of string
theory.

On the other hand, the black-hole analogue of the quantum statistical
description involves a complicated and hard-to-study quantum
superposition of microstates, and is therefore outside our present
theoretical grasp. One can speculate, again in analogy with the ideal
gas, that there are probably interesting physical phenomena that can
only be captured by this description, and not by the classical
statistical description.

We should also note that in \cite{babel} it has been argued that from
the point of view of the dual CFT, the difference between the typical
microstates and the classical black hole solution can only be
discerned by doing a very atypical measurement, or waiting for a very
long time\footnote{See \cite{thermal} for other interesting work in
  this direction.}.  This is analogous to the case of a gas, where if
one waits for a very long time, of order the Poincar\'e recurrence
time, one will observe spikes in the pressure coming from very
unlikely events, such as a very large number of molecules hitting the
wall at the same time. In the thermodynamic approximation one ignores
the small energy gap between microstates, and such phenomena are not
visible.  The fact that the classical black hole geometry has an
infinite throat and no mass gap implies that this geometry will not
display such fluctuations at very large time-scales.  Since the CFT
does have a mass gap, and fluctuations at large scales occur, one can
argue \cite{Maldacena:2001kr} that the black hole gives a
thermodynamic description of the physics, and not a microscopic one.

Since, by standard $AdS$-CFT arguments, a long time on the boundary
corresponds to a large distance into the bulk, one can argue that
atypical CFT measurements involving very long times correspond in the
bulk to propagators that reach very close to the black hole horizon
\cite{babel}. Hence, this supports the intuition that one can
distinguish between different microstates by making experiments at the
scale of the horizon. Moreover, in a gas one can distinguish between
the ensemble and the microstates by making experiments at scales
smaller than the mean free path. At this scale the thermodynamic
description breaks down, and new phenomena that cannot be captured by
thermodynamics appear.  By analogy, for the black hole we have argued
that the scale where thermodynamics breaks down is that of the
horizon. Therefore, both our arguments and the arguments of
\cite{babel} indicate that experiments made {\it at the scale of the
  horizon} should distinguish between a microstate and the classical
solution. While from the point of the dual CFT these experiments
appear to be very atypical, they might not be so atypical from the
point of view of the dual bulk.  It would certainly be very
interesting to propose and analyze in more detail such gedanken
experiments, and explore more thoroughly the implications of this
fact.

The whole problem with finding experimental or observational tests of
string theory is that the string scale and the Planck scale are so far
out of reach of present accelerations.  However, the ideas of
fractionation and the present ideas about the microstate structure of
black holes show us that we can get stringy effects on very large
length scales.  It would obviously be very exciting if we could make
black holes at the LHC and thereby test these ideas, but even if this
were not to happen, we may still be able to see some signature of stringy
black holes within the next decade. Indeed, the gravitational wave
detectors LIGO and LISA are very likely to detect the gravitational
``ring-down'' of merging black holes within the next few years and,
while the underlying computations will be extremely difficult, one
might reasonably hope that the microstate structure arising from
string theory could lead to a new, detectable and recognizable
signature in the LIGO or LISA data.

\bigskip
\leftline{\bf Acknowledgements}
\smallskip
\noindent We would like to thank our collaborators, Per Kraus and
Chih-Wei Wang, who played instrumental roles in much of the research
reviewed here.  We would also like to thank Samir Mathur, Gary Gibbons,
Thibault Damour, Eric Gimon, Gary Horowitz, Don Marolf and Simon Ross,
for interesting discussions
that have helped us clarify many of the things presented here. IB
would also like to thank Stefano Bellucci and Sergio Ferrara, for
organizing the excellent Winter School on the Attractor Mechanism in
Frascati, where the lectures on which these notes are based were given.
NW and IB would like to thank CERN, the Institute for Advanced Study, and the
Aspen Center for Physics where they were affiliated in the early stage of
the writing of these notes.  The work of NW was supported in part by
funds provided by the DOE under grant DE-FG03-84ER-40168. The work of
IB was supported in part by the NSF grant PHY-0503584, and by the
Dir\'ection des Sciences de la Mati\`ere of the Commissariat \`a
L'En\'ergie Atomique of France.

%
%

%
%

\end{document}